\documentclass[a4paper,11pt]{article}

\usepackage{stmaryrd}
\usepackage{bm}
\usepackage{cite}
\usepackage{color}
\usepackage{dsfont}
\usepackage{nicefrac}
\usepackage{multirow}
\usepackage{graphics}
\usepackage{graphicx}
\usepackage{epsfig}
\usepackage{subfigure}
\usepackage{pgf}
\usepackage{amsmath,amssymb,amsfonts}
\usepackage{wasysym}
\usepackage{yhmath}
\usepackage{booktabs}

\begin{document}

\title{Topology optimization on two-dimensional manifolds}

\author{Yongbo Deng$^{1,2}$\footnote{dengyb@ciomp.ac.cn (Y.~Deng)}, Zhenyu Liu$^{3}$, Jan G. Korvink$^{1}$\footnote{jan.korvink@kit.edu (J.~G.~Korvink)} \\
1 Institute of Microstructure Technology (IMT), \\
Karlsruhe Institute of Technology (KIT), \\
Hermann-von-Helmholtzplatz 1, Eggenstein-Leopoldshafen 76344, Germany; \\
2 State Key Laboratory of Applied Optics (SKLAO),\\
Changchun Institute of Optics, Fine Mechanics and Physics (CIOMP),\\
Chinese Academy of Sciences, Dongnanhu Road 3888, Changchun 130033, China; \\
3 Changchun Institute of Optics, Fine Mechanics and Physics (CIOMP),\\
Chinese Academy of Sciences, Dongnanhu Road 3888, Changchun 130033, China.}

\maketitle

\abstract{This paper implements topology optimization on two-dimensional manifolds. In this paper, the material interpolation is implemented on a material parameter in the partial differential equation used to describe a physical field, when this physical field is defined on a two-dimensional manifold; the material density is used to formulate a mixed boundary condition of the physical field and implement the penalization between two different types of boundary conditions, when this physical field is defined on a three-dimensional domain with its boundary conditions defined on the two-dimensional manifold corresponding a surface or an interface of this three-dimensional domain. Based on the homeomorphic property of two-dimensional manifolds, typical two-dimensional manifolds, e.g., sphere, torus, M\"{o}bius strip and Klein bottle, are included in the numerical tests, which are provided for the problems on fluidic mechanics, heat transfer and electromagnetics.

\textbf{Keywords:} Topology optimization; two-dimensional manifold (2-manifold); material distribution method; tangential gradient operator; mixed boundary condition.}


\section{Introduction} \label{sec:IntroductionManifold}

Topology optimization is a robust method used to determine the structural configuration, which corresponds to the material distribution in a structure \cite{Bendsoe2003}. In contrast to designing devices by tuning a handful of structural parameters in size and shape optimization, topology optimization utilizes the full-parameter space to design structures based on the user-desired performance, and it is more flexible and robust, because of its low dependence on initial guess and implicit expression of the structures.

Optimization of structural topology was investigated as early as 1904 for trusses by Michell \cite{Michell1904}. Material distribution method for topology optimization was pioneered by Bends{\o}e and Kikuchi for elasticity \cite{Bendsoe1988}, and this method was extended to a variety of areas, e.g. acoustics, electromagnetics, fluid dynamics and thermodynamics \cite{Sigmund2001,Sigmund1997,Rozvany2001,Andkjaer2011,Bendsoe1999,Andkjaer2012,Fujii2013,Zhou2011,Diaz2010,Zhou2010Meta,Andkjaer20101,Zhou20102,Feichtner2012,Deng2011,Hassan2014,Shim2009,Borrvall2003,Gersborg-Hansen2006,Nomura2007,Sigmund2008,Duhring2008,Akl2008,Kreissl2011,Zhou2008,Guest2006,Takezawa2014}. Several other methods have also been proposed and developed for the implementation of topology optimization, e.g. the level set method \cite{Wang2003,Allaire2004,Liu2008,Xing2010}, the evolutionary techniques \cite{XieStevenSpringer1997,StevenComputMech2000,TanskanenCMAME2002}, the moving morphable components \cite{GuoJAM2014,GuoCMAME2016} and the phase field method \cite{TakezawaJCP2010}.

In topology optimization, structures were usually defined on a three-dimensional (3D) domain in $\mathbb{R}^3$ or reduced two-dimensional (2D) plane on $\mathbb{R}^2$.
The current development of additive manufacturing, e.g., 3D-printing, has effectively enlarged the structural-design space. Implementing topology optimization on 2-manifolds can be helpful to enhance the freedom of structural design, where a 2-manifold is a topological space with its arbitrary interior point having a neighborhood homeomorphic to a sub-region of $\mathbb{R}^2$ and it can be used to describe a structural surface or a material interface. Related researches have been implemented for the structural design on 2-manifolds based on the conformal geometry theory \cite{QianYe2019,PanagiotisVogiatzis2018}, layouts on surfaces for shell structures \cite{KrogOlhoffComputersStructures1996,AnsolaComputersStructures2002,HassaniSMO2013,LochnerAldingerSchumacher2014,ClausenActaMechanicaSinica2017}, and material interfaces for stiffness and multi-material structures \cite{AllaireESAIM2014,VermaakSMO2014,SigmundJMPS1997,GibianskyJMPS2000,GaoZhangIJNME2011,
LuoAiaaJ2012,YinSMO2001,WangCMAME2004,ZhouWangSMO2007}, fluid-structure interaction \cite{YoonFSIIJNME2010,LundgaardFSISMO2018,AndreasenFSISMO2019}, energy absorption \cite{AuligUlm2012}, cohesion \cite{MauteIJNME2017} and actuation \cite{MauteSMO2005}.



In this paper, topology optimization on 2-manifolds is implemented by using the material distribution method and this implementation includes the following two cases. When a physical field is defined on a 2-manifold, the material interpolation of topology optimization is implemented on a material parameter in the partial differential equation (PDE) with the tangential gradient operator used to describe the physical field. When the physical field is defined on a 3D domain and its boundary conditions are defined on a 2-manifold of codimension one in a 3D Euclidian space, the material density representing the structural configuration is used to implement the penalization between two different types of boundary conditions and formulate a mixed boundary condition of the physical field. In the second case, the 2-manifold can be an exterior surface or an interface of the 3D domain. When the 2-manifold is an exterior surface, it can mix the Dirichlet and Neumann types of boundary conditions; when the 2-manifold is an interface, it can mix the no-jump and Dirichlet types of boundary conditions.

For the two cases of topology optimization on 2-manifolds, a monolithic description of the corresponding optimization problems is presented in Section \ref{sec:MethodologyManifold}, including the adjoint analysis and numerical implementation; test problems on fluidic mechanics, heat transfer and electromagnetics are implemented in Section \ref{sec:TestManifold}. In the following, all the mathematical descriptions are performed in the Cartesian system.

\section{Methodology} \label{sec:MethodologyManifold}

Details for topology optimization implemented on 2-manifolds are provided as follows.

\subsection{2-manifold} \label{sec:2Manifolds}

\begin{figure}[!htbp]
  \centering
  \subfigure[Sphere]
  {\includegraphics[height=0.22\columnwidth]{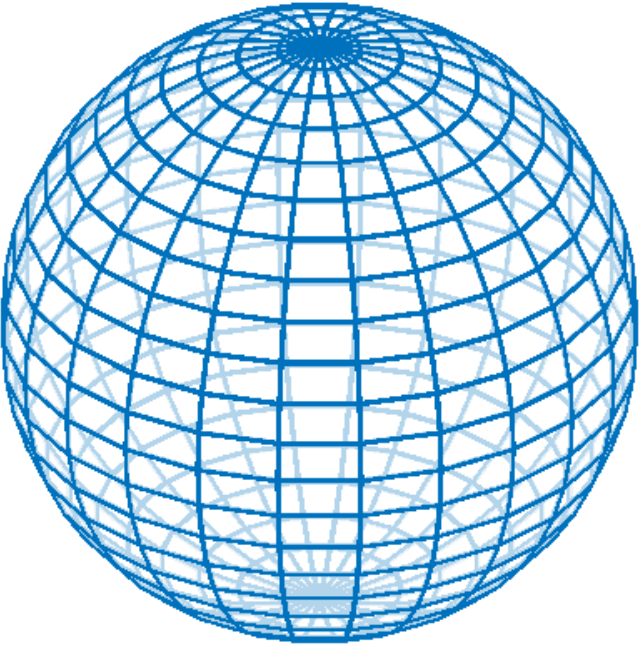}}\hspace{0.5em}
  \subfigure[Torus]
  {\includegraphics[height=0.22\columnwidth]{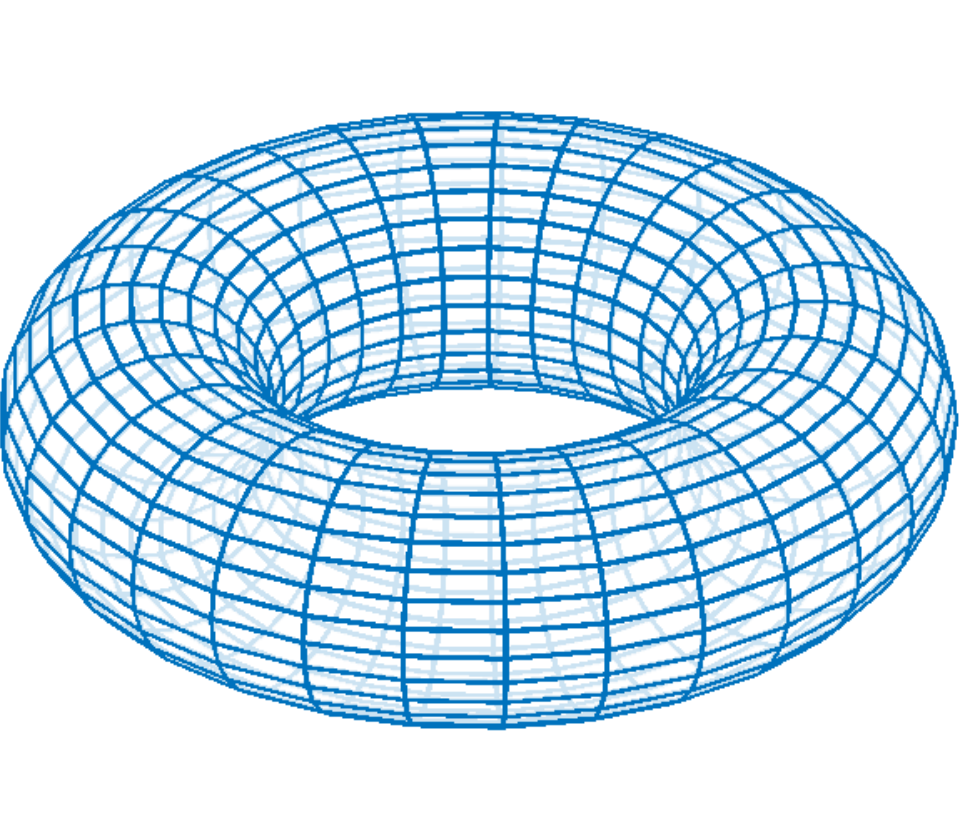}}\hspace{0.5em}
  \subfigure[M\"{o}bius strip]
  {\includegraphics[height=0.22\columnwidth]{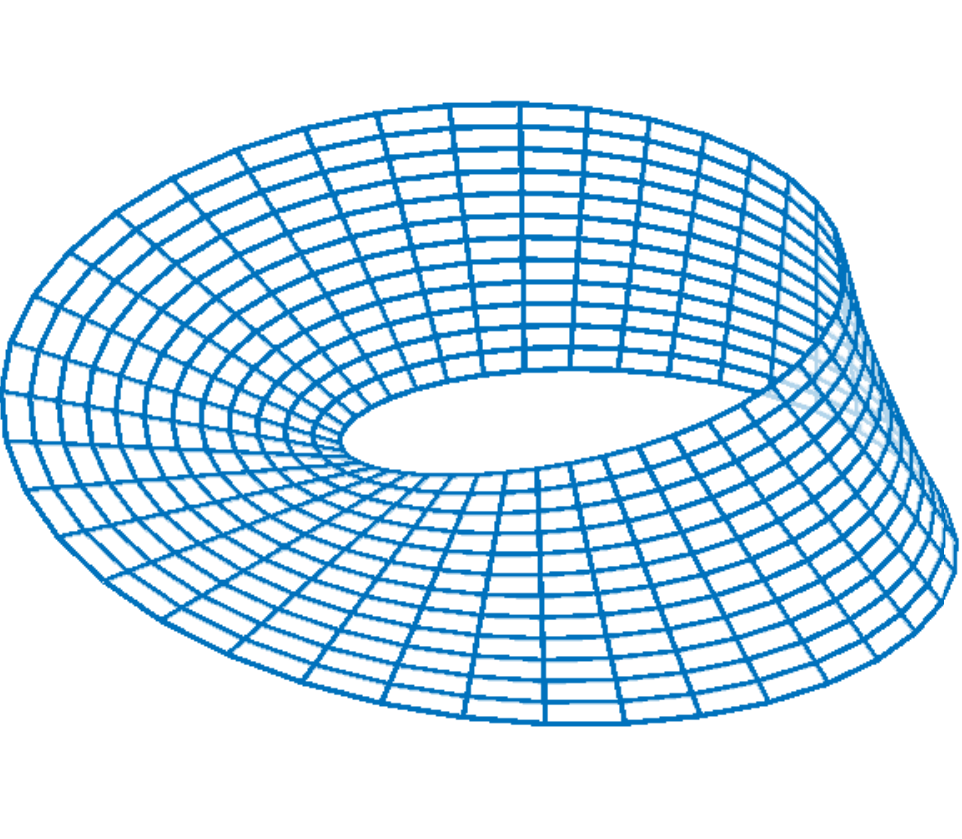}}\hspace{0.5em}
  \subfigure[Klein bottle]
  {\includegraphics[height=0.25\columnwidth]{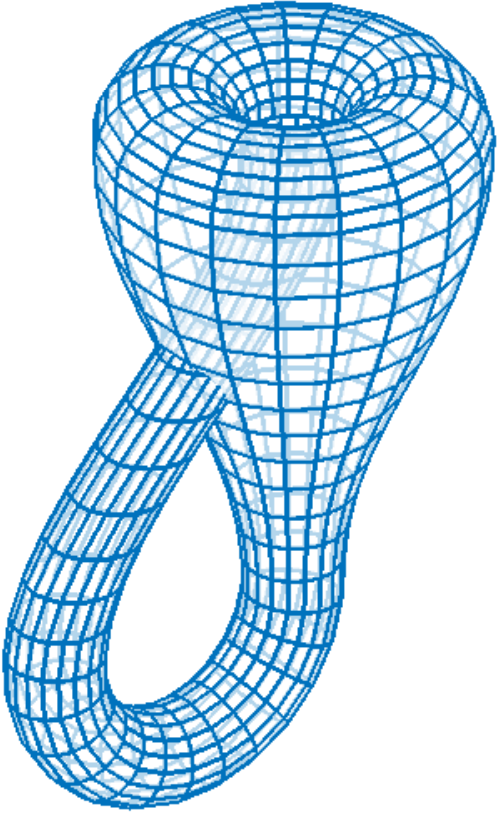}}
  \caption{Typical examples of 2-manifolds.}\label{fig:ManifoldDemon}
\end{figure}

According to the classification theorem \cite{Reshetnyak2Manifolds1993}, a 2-manifold without boundary is compact if every open cover of it has a finite subcover; its family can be exhausted by the two infinite families $\{$$\mathbb{S}^2$ (sphere), $\mathbb{T}^2$ (torus), $\mathbb{T}^2\#\mathbb{T}^2$ (double torus), $\cdots$$\}$ and $\{$$\mathbb{P}^2$ (projective plane), $\mathbb{P}^2\#\mathbb{P}^2$ (Klein bottle), $\cdots$$\}$, where $\#$ denotes the connected sum of two manifolds. A 2-manifold with boundary can be derived by removing an open disk from a 2-manifold without boundary. All 2-manifolds without boundary can be derived by gluing the basic 2-manifolds with boundaries, e.g., the disk, cylinder, M\"{o}bius strip. Structural surfaces can be described as the orientable 2-manifolds, e.g., sphere, on which the normal vector can be defined globally. Structural interfaces can include both the orientable and non-orientable 2-manifolds, e.g. M\"{o}bius strip is a non-orientable 2-manifold, on which the normal vector can be defined locally instead of globally. Therefore, without loosing generality, topology optimization in this paper is implemented on the 2-manifolds homeomorphic to sphere, torus, M\"{o}bius strip, and their connected sum or glued manifolds (Figure \ref{fig:ManifoldDemon}).

\subsection{Physical PDEs} \label{sec:PhysicalPDEs}

To implement topology optimization on a 2-manifold, a design variable, which is a relaxed binary distribution, is defined on this 2-manifold to represent the structural configuration; and it is bounded in the typical interval $\left[0,1\right]$, with $0$ and $1$ representing two different material phases, respectively. An optimization problem can be formulated by minimizing or maximizing a design objective used to evaluate a desired performance of the structure implicitly expressed on the 2-manifold. The physical field related with this structure can be described by using a corresponding PDE.
Then, this optimization problem is a PDE constrained optimization problem. It is nonlinear and challenging to be solved directly. Therefore, the iterative solution procedure is widely utilized. To ensure the monolithic convergence of the iterative procedure, regularization based on a PDE filter and threshold projection is imposed on the design variable, with the projected design variable referred to as the material density. This iterative procedure including the PDE filter and the threshold projection can control the feature size of the structure and remove the gray regions from the derived structural pattern.

For the case with the physical field defined on a 2-manifold, the physical PDE, used to describe the distribution of the physical field, can be expressed in a typical abstract form with the material parameter interpolated by the material density:
\begin{equation}\label{equ:GeneralPhysicalPDE2Manifold}
\begin{split}
  & \nabla_s \cdot \left[ p\left(\bar{\gamma}\right) \mathbf{g}\left(\nabla_s u, u \right) \right] = c_s,~\mathrm{in}~\Sigma_S \\
  & \left[ p\left(\bar{\gamma}\right) \mathbf{g}\left(\nabla_s u, u \right)\right]\cdot \boldsymbol{\tau} = c_b,~\mathrm{on}~\partial\Sigma_S \\
  & u = u_0,~\mathrm{at}~\mathcal{P}\subset\Sigma_S \\
\end{split}
\end{equation}
where $u$ is the physical variable; $\Sigma_S$ is the 2-manifold; $c_s$ and $c_b$ are known distributions defined on $\Sigma_S$ and $\partial\Sigma_S$, respectively; if $\Sigma_S$ is treated as the 2-manifold of codimension one in an 3D Euclidian space, $\nabla_s= \nabla-\left(\mathbf{n}\cdot\nabla\right)\mathbf{n}$ is the tangential gradient operator defined on $\Sigma_S$, with $\mathbf{n}$ denoting the unitary normal vector of $\Sigma_S$; $\boldsymbol\tau$ perpendicular to $\partial \Sigma_S$ is the unitary outward tangential vector of $\Sigma_S$; $\mathcal{P}\subset\Sigma_S$ is a finite point set with known physical variable $u_0$; $p\left(\bar{\gamma}\right)$ is the material parameter interpolated by the material density $\bar{\gamma}\in\left[0,1\right]$, derived from sequential filtering and projecting the design variable defined on $\Sigma_S$ \cite{LazarovIntJNumerMethodsEng2011,WangStructMultidiscipOptim2011,GuestIntJNumerMethodsEng2004}; $\mathbf{g}\left(\nabla_s u, u \right)$ is a vector functional of $\nabla_s u$ and $u$. Based on the Green's formula including the tangential gradient operator \cite{GilbargTrudingerSpringer1988}, the physical PDE in equation \ref{equ:GeneralPhysicalPDE2Manifold} can be transformed into the following variational formulation:
\begin{itemize}
  \item find $u \in \mathcal{H} \left(\Sigma_S\right)$ with $u=u_0$ at $\mathcal{P}$, satisfying
\begin{equation}\label{equ:WeakGeneralPhysicalPDE2Manifold}
\begin{split}
& e\left(u;\bar{\gamma}\right):= \int_{\Sigma_S} - p\left(\bar{\gamma}\right) \mathbf{g}\left(\nabla_s u, u \right) \cdot \nabla_s \hat{u} - c_s \hat{u} \, \mathrm{d}s + \int_{\partial\Sigma_s} c_b \hat{u} \, \mathrm{d}l = 0,\\
& \forall \hat{u} \in \mathcal{H} \left(\Sigma_S\right)
\end{split}
\end{equation}
\end{itemize}
where $\hat{u}$ is the test function of $u$; $\mathcal{H}\left(\Sigma_S\right) = \big\{ u\in \mathcal{L}^2\left(\Sigma_S\right) \big| \nabla_s u \in \left(\mathcal{L}^2\left(\Sigma_S\right)\right)^3 \big\} $ is the first-order Sobolev space on $\Sigma_S$, with $\mathcal{L}^2\left(\Sigma_S\right)$ denoting the second-order Lebesgue integrable functional space on $\Sigma_S$; $\mathrm{d}s$ and $\mathrm{d}l$ are the corresponding Riemann metrics on $\Sigma_S$ and $\partial\Sigma_S$, respectively.

For the case with the physical field defined on a volume domain and design variable defined on the 2-manifold corresponding to a boundary surface of this volume domain, the physical PDE can be expressed in a typical abstract form as
\begin{equation}\label{equ:GeneralPhysicalPDEVolumeSurface}
\begin{split}
  & \nabla \cdot \left[ p \mathbf{g}\left(\nabla u, u \right) \right] = c_s,~\mathrm{in}~\Omega \\
  & \left[ p \mathbf{g}\left(\nabla_s u, u \right)\right] \cdot \mathbf{n} = \alpha\left(\bar{\gamma}\right)\left(u - u_d\right),~\mathrm{on}~\Sigma_S = \partial \Omega \\
  & u = u_0,~\mathrm{at}~\mathcal{P}\subset\Omega \\
\end{split}
\end{equation}
where $u_d$ is the known physical variable on the structure defined on $\Sigma_S$; $\Omega$ is the open and bounded volume domain with the boundary of Lipschitz type; $\alpha\left(\bar{\gamma}\right)$ is the material interpolation used to implement the penalization between the Neumann and Dirichlet types of boundary conditions. Based on this material interpolation, the mixed boundary condition is formulated as that in equation \ref{equ:GeneralPhysicalPDEVolumeSurface}. When the material density $\bar{\gamma}$ takes the value of $0$, $\alpha$ is valued to be large enough to ensure the dominance of the Dirichlet term $\left(u-u_d\right)$. The mixed boundary condition degenerates into the Neumann type, when $\alpha$ is valued as $0$ with $\bar{\gamma}$ taking the value of $1$. Sequentially, the structural pattern with the known physical variable $u_d$ can be determined. The physical PDE in equation \ref{equ:GeneralPhysicalPDEVolumeSurface} can be transformed into the variational formulation as
\begin{itemize}
  \item find $u \in \mathcal{H} \left(\Omega\right)$ with $u=u_0$ at $\mathcal{P}$, satisfying
\begin{equation}\label{equ:WeakGeneralPhysicalPDEVolumeSurface}
\begin{split}
e\left(u;\bar{\gamma}\right):= \int_\Omega - p \mathbf{g}\left(\nabla u, u \right) \cdot \nabla \hat{u} \, \mathrm{d}\Omega + \int_{\Sigma_S} \alpha\left(\bar{\gamma}\right)\left(u - u_d\right) \hat{u} \, \mathrm{d}s = 0,~\forall \hat{u} \in \mathcal{H} \left(\Omega\right)
\end{split}
\end{equation}
\end{itemize}
where $\mathcal{H}\left(\Omega\right) = \big\{ u\in \mathcal{L}^2\left(\Omega\right) \big| \nabla u \in \left(\mathcal{L}^2\left(\Omega\right)\right)^3 \big\} $ is the first-order Sobolev space defined on $\Omega$. In this case, when the design variable is defined on an interface of the volume domain, the typical abstract form of the physical PDE can be
\begin{equation}\label{equ:GeneralPhysicalPDEVolumeInterface}
\begin{split}
  & \nabla \cdot \left[ p \mathbf{g}\left(\nabla u, u \right) \right] = c_s,~\mathrm{in}~\Omega \\
  & \llbracket p \mathbf{g}\left(\nabla_s u, u \right) \rrbracket \cdot \mathbf{n} = \alpha\left(\bar{\gamma}\right)\left(u - u_d\right),~\mathrm{on}~\Sigma_S \hookrightarrow \Omega \\
  & u = u_d,~\mathrm{at}~\mathcal{P} \subset \Sigma_S \\
  & u = u_0,~\mathrm{on}~\partial\Omega \\
\end{split}
\end{equation}
where $\llbracket\cdot\rrbracket$ denotes the local jump of a variable across $\Sigma_S$ embedded in $\Omega$. Correspondingly, the variational formulation of this physical PDE is
\begin{itemize}
  \item find $u \in \mathcal{H} \left(\Omega\right)$ with $u=u_d$ at $\mathcal{P}$ and $u=u_0$ on $\partial\Omega$, satisfying
\begin{equation}\label{equ:WeakGeneralPhysicalPDEVolumeSurface}
\begin{split}
e\left(u;\bar{\gamma}\right):= \int_\Omega - p \mathbf{g}\left(\nabla u, u \right) \cdot \nabla \hat{u} \, \mathrm{d}\Omega + \int_{\Sigma_S} \alpha\left(\bar{\gamma}\right)\left(u - u_d\right) \hat{u} \, \mathrm{d}s = 0,~\forall \hat{u} \in \mathcal{H} \left(\Omega\right)
\end{split}
\end{equation}
\end{itemize}

The above abstract forms of physical PDEs are expressed with divergence operator. They can be directly transformed into the forms with curl operator in the following numerical examples for electromagnetics.

\subsection{Regularization} \label{sec:RegularizationDesignVariable}

The material density is derived by sequentially implementing the filter and projection operations on the design variable. The filter operation of the design variable is implemented by solving the PDE defined on the 2-manifold \cite{LazarovIntJNumerMethodsEng2011}:
\begin{equation}\label{equ:PDEFilterDesignVarManifold}
\begin{split}
& \nabla_s \cdot \left( - r^2 \nabla_s \tilde{\gamma} \right) + \tilde{\gamma} = \gamma,~\mathrm{on}~ \Sigma_S \\
& - r^2 \nabla_s \tilde{\gamma} \cdot \boldsymbol\tau = 0,~\mathrm{at}~ \partial \Sigma_S \\
\end{split}
\end{equation}
where $r$ is the filter radius; $\gamma\in \mathcal{L}^2\left(\Sigma_S\right)$ is the design variable; $\tilde{\gamma}$ is the filtered design variable; $\hat{\tilde{\gamma}}$ is the test function for $\tilde{\gamma}$. The variational formulation of the PDE filter in equation \ref{equ:PDEFilterDesignVarManifold} is
\begin{itemize}
  \item find $\tilde{\gamma} \in \mathcal{H} \left(\Omega\right)$ satisfying
\begin{equation}\label{equ:WeakPDEFilterDesignVarManifold}
\begin{split}
f\left(\tilde{\gamma};\gamma\right):= \int_{\Sigma_S} r^2 \nabla_s \tilde{\gamma}\cdot\nabla_s\hat{\tilde{\gamma}} + \tilde{\gamma} \hat{\tilde{\gamma}} - \gamma \hat{\tilde{\gamma}} \, \mathrm{d}s = 0,~\forall \hat{\tilde{\gamma}} \in \mathcal{H} \left(\Sigma_S\right)
\end{split}
\end{equation}
\end{itemize}
The filtered design variable is projected by the threshold projection\cite{WangStructMultidiscipOptim2011,GuestIntJNumerMethodsEng2004}, to derive the material density:
\begin{equation}\label{equ:ProjectionDesignVarManifold}
  \bar{\gamma} = { \tanh\left(\beta \xi\right) + \tanh\left(\beta \left(\tilde{\gamma}-\xi\right)\right) \over \tanh\left(\beta \xi\right) + \tanh\left(\beta \left(1-\xi\right)\right)}
\end{equation}
where $\beta$ and $\xi$ are the projection parameters and their values are chosen based on numerical experiments \cite{GuestIntJNumerMethodsEng2004}.



\subsection{Topology optimization problem}\label{sec:VarProBulkFieldManifold}

The 2-manifold and the material density together composes a fiber bundle \cite{ChernWorldScientificPublishing1999}, where $\Sigma_S$ is its base manifold and $\bar{\gamma}: \Sigma_S \rightarrow \left[0,1\right]$ is its fibers. This fiber bundle can be expressed as $\left(\Sigma_S \times \bar{\gamma}\left(\Sigma_S\right), \Sigma_S, proj_1, \bar{\gamma}\left(\Sigma_S\right) \right)$, with the natural projection $proj_1: \Sigma_S \times \bar{\gamma}\left(\Sigma_S\right) \rightarrow \Sigma_S$ satisfying $ proj_1\left(\mathbf{x},\bar{\gamma}\left(\mathbf{x}\right)\right)=\mathbf{x}$ for $\forall \mathbf{x}\in\Sigma_S $.
Then the topology optimization problem in this paper can be formulated as an optimization problem constrained by the PDEs defined on the 2-manifold or a volume domain with this 2-manifold as the boundary or interface. The topology optimization problem can be described in the following form:
\begin{equation}\label{equ:VarProBulkField}
\begin{split}
& \mathrm{find}~\gamma\left(\Sigma_S\right)\in\left[0,1\right]~\mathrm{for}~\left(\Sigma_S \times \gamma\left(\Sigma_S\right), \Sigma_S, proj_1, \gamma\left(\Sigma_S\right) \right), \\
& \mathrm{to~minimize}~J\left(u;\bar{\gamma}\right)~\mathrm{constrained~by} \\
& \left\{\begin{split}
& e\left(u;\bar{\gamma}\right) = 0 ~~
 (\mathrm{Physical}~\mathrm{PDE}) \\
& f\left(\tilde{\gamma};\gamma\right) = 0 ~~(\mathrm{PDE}~\mathrm{filter}) \\
& \bar{\gamma} = { \tanh\left(\beta \xi\right) + \tanh\left(\beta \left(\tilde{\gamma}-\xi\right)\right) \over \tanh\left(\beta \xi\right) + \tanh\left(\beta \left(1-\xi\right)\right)} ~~ (\mathrm{Threshold}~\mathrm{projection}) \\
& \left|{1\over \left|\Sigma_S\right| } \int_{\Sigma_S} \bar{\gamma} \,\mathrm{d}s - V_f \right| \leq 10^{-3} ~~ (\mathrm{Area} ~\mathrm{constraint}) \\
\end{split}\right.
\end{split}
\end{equation}
where $J: \mathcal{H}\left(\Sigma_S\right) \times \mathcal{H}\left(\Sigma_S\right) \rightarrow \mathbb{R}$ or $\mathcal{H}\left(\Omega\right) \times \mathcal{H}\left(\Sigma_S\right) \rightarrow \mathbb{R}$ is a bounded continuous mapping operator; $V_f\in\left(0,1\right)$ is the area fraction of the structure on the 2-manifold with an admissible tolerance of $10^{-3}$; $\left|\Sigma_S\right|$ is the area of the 2-manifold $\Sigma_S$.

\subsection{Adjoint analysis}\label{sec:AdjointAnalysisBulkFieldManifold}

The adjoint analysis can be implemented to derive the adjoint derivative of the topology optimization problem and define the descent direction of the design objective for the iterative solution procedure. For the PDE used to describe a physical field, the usual situation is that $e$ is continuously Fr\'{e}chet-differentiable and  $e_u\left(u;\bar{\gamma}\right)$ is a linear operator with bounded inverse. According to the implicit function theorem \cite{ZeidlerSpringer1986}, $e\left(u;\bar{\gamma}\right) = 0$ locally defines a continuously Fr\'{e}chet-differentiable $\bar{\gamma} \mapsto u\left(\bar{\gamma}\right)$ with the following Fr\'{e}chet derivative:
\begin{equation}\label{equ:FDerBulkAbstract}
  u_{\bar{\gamma}}\left(\bar{\gamma}\right) = -e_u^{-1} \left(u;\bar{\gamma}\right) e_{\bar{\gamma}} \left(u;\bar{\gamma}\right).
\end{equation}
The PDE filter defines a continuously Fr\'{e}chet-differentiable $\gamma \mapsto \tilde{\gamma}\left(\gamma\right)$ with the following Fr\'{e}chet derivative:
\begin{equation}\label{equ:FDerHelmholtzFilterAbstract}
  \tilde{\gamma}_\gamma\left(\gamma\right) = -f_{\tilde{\gamma}}^{-1} \left(\tilde{\gamma};\gamma\right) f_\gamma\left(\tilde{\gamma};\gamma\right).
\end{equation}
Then, the G\^{a}teaux derivative of $J$ is
\begin{equation}\label{equ:FDrvBulkField}
\begin{split}
  & \left\langle J', t \right\rangle_{\mathcal{L}^2\left(\Sigma_S\right),\mathcal{L}^2\left(\Sigma_S\right)} \\
= & \left\langle J_u, u_{\bar{\gamma}} \bar{\gamma}_{\tilde{\gamma}} \tilde{\gamma}_\gamma t \right\rangle_{\mathcal{U}^*\left(\Omega\right),\mathcal{U}\left(\Omega\right)} + \left\langle J_{\bar{\gamma}}, \bar{\gamma}_{\tilde{\gamma}} \tilde{\gamma}_\gamma t \right\rangle_{\mathcal{H}\left(\Sigma_S\right),\mathcal{H}\left(\Sigma_S\right)} \\
= & \left\langle \tilde{\gamma}_\gamma^* \bar{\gamma}_{\tilde{\gamma}} \left( u_{\bar{\gamma}}^* J_u + J_{\bar{\gamma}} \right), t \right\rangle_{\mathcal{L}^2\left(\Sigma_S\right),\mathcal{L}^2\left(\Sigma_S\right)} \\
= & \left\langle -f_\gamma^* \left(f_{\tilde{\gamma}}^{-1}\right)^* \bar{\gamma}_{\tilde{\gamma}} \left( -e_{\bar{\gamma}}^* \left(e_u^{-1}\right)^* J_u + J_{\bar{\gamma}} \right), t \right\rangle_{\mathcal{L}^2\left(\Sigma_S\right),\mathcal{L}^2\left(\Sigma_S\right)},~~\forall t \in \mathcal{L}^2\left(\Sigma_S\right) \\
\end{split}
\end{equation}
where $*$ denotes the adjoint of an operator; $\left\langle\cdot,\cdot \right\rangle$ denotes the pairing between a functional space and its dual. Under the precondition that $J$ is Fr\'{e}chet-differentiable, the adjoint equations and the adjoint derivative can be derived from the G\^{a}teaux derivative, according to the Kurash-Kuhn-Tucker condition \cite{HinzeSpringer2009}.

When the physical field is defined on the 2-manifold $\Sigma_S$, $e\left(u;\bar{\gamma}\right)$ can include the terms of surface-integrals and curve-integral; then, it is expressed to be the sum of the corresponding terms as
\begin{equation}\label{equ:eSum2manifoldSur}
  e\left(u;\bar{\gamma}\right) = e^{\Sigma_S}\left(u;\bar{\gamma}\right) + e^{\partial\Sigma_S}\left(u\right).
\end{equation}
By setting
\begin{equation}
\begin{split}
\mu & := -\left(e_u^{-1}\right)^* J_u = \left(e_u^{\Sigma_S*} + e_u^{\partial\Sigma_S*} \right)^{-1} J_u, \\
\nu & := -\left(f_{\tilde{\gamma}}^{-1}\right)^* \bar{\gamma}_{\tilde{\gamma}} J_{\bar{\gamma}} = \left(f_{\tilde{\gamma}}^*\right)^{-1} \bar{\gamma}_{\tilde{\gamma}} J_{\bar{\gamma}},
\end{split}
\end{equation}
the adjoint equations can be obtained as
\begin{itemize}
  \item find $\mu\in\mathcal{H}\left(\Sigma_S\right)$ and $\nu\in\mathcal{H}\left(\Sigma_S\right)$, satisfying
\begin{equation}\label{equ:WeakAdjEquSurField}
\begin{split}
  & \left\langle \left(e_u^{\Sigma_S*} + e_u^{\partial\Sigma_S*} \right) \mu, v \right\rangle_{\mathcal{H}\left(\Sigma_S\right),\mathcal{H}\left(\Sigma_S\right)} = \left\langle -J_u, v \right\rangle_{\mathcal{H}\left(\Sigma_S\right),\mathcal{H}\left(\Sigma_S\right)},~\forall v \in \mathcal{H}\left(\Sigma_S\right) \\
  & \left\langle f_{\tilde{\gamma}}^* \nu, y \right\rangle_{\mathcal{H}\left(\Sigma_S\right),\mathcal{H}\left(\Sigma_S\right)} = \left\langle - J_{\bar{\gamma}}\bar{\gamma}_{\tilde{\gamma}}, y \right\rangle_{\mathcal{H}\left(\Sigma_S\right),\mathcal{H}\left(\Sigma_S\right)},~\forall y \in \mathcal{H}\left(\Sigma_S\right) \\
\end{split}
\end{equation}
\end{itemize}
where $\mu$ and $\nu$ are the adjoint variables of $u$ and $\tilde{\gamma}$, respectively. The adjoint derivative of $J$ can be derived as
\begin{equation}\label{equ:ReducedFDrvSurField}
\begin{split}
\left\langle J', t \right\rangle_{\mathcal{L}^2\left(\Sigma_S\right),\mathcal{L}^2\left(\Sigma_S\right)} = \left\langle f_\gamma^* \nu J_{\bar{\gamma}}^{-1} \left(e^{\Sigma_S}_{\bar{\gamma}}\right)^* \mu + f_\gamma^* \nu, t \right\rangle_{\mathcal{L}^2\left(\Sigma_S\right),\mathcal{L}^2\left(\Sigma_S\right)},~\forall t \in \mathcal{L}^2\left(\Sigma_S\right) \\
\end{split}
\end{equation}
where the adjoint variables $\mu$ and $\nu$ are derived by solving equation \ref{equ:WeakAdjEquSurField}.

When the physical field is defined on the volume domain $\Omega$, $e\left(u;\bar{\gamma}\right)$ includes the terms of volume-integral and surface-integrals; then, it is expressed to be the sum of the corresponding terms as
\begin{equation}\label{equ:eSum2manifoldVolume}
  e\left(u;\bar{\gamma}\right) = e^{\Omega}\left(u\right) + e^{\Sigma_S}\left(u;\bar{\gamma}\right).
\end{equation}
By setting $\mu = -\left(e_u^{-1}\right)^* J_u = \left(e_u^{\Omega*} + e_u^{\Sigma_S*} \right)^{-1} J_u$ and $\nu = -\left(f_{\tilde{\gamma}}^{-1}\right)^* \bar{\gamma}_{\tilde{\gamma}} J_{\bar{\gamma}} = \left(f_{\tilde{\gamma}}^*\right)^{-1} \bar{\gamma}_{\tilde{\gamma}} J_{\bar{\gamma}}$, the adjoint equations can be obtained as
\begin{itemize}
  \item find $\mu\in\mathcal{H}\left(\Omega\right)$ and $\nu\in\mathcal{H}\left(\Sigma_S\right)$, satisfying
\begin{equation}\label{equ:WeakAdjEquBulkField}
\begin{split}
  & \left\langle \left(e_u^{\Omega*} + e_u^{\Sigma_S*} \right) \mu, v \right\rangle_{\mathcal{H}\left(\Omega\right),\mathcal{H}\left(\Omega\right)} = \left\langle -J_u, v \right\rangle_{\mathcal{H}\left(\Omega\right),\mathcal{H}\left(\Omega\right)},~\forall v \in \mathcal{H}\left(\Omega\right) \\
  & \left\langle f_{\tilde{\gamma}}^* \nu, y \right\rangle_{\mathcal{H}\left(\Sigma_S\right),\mathcal{H}\left(\Sigma_S\right)} = \left\langle - J_{\bar{\gamma}}\bar{\gamma}_{\tilde{\gamma}}, y \right\rangle_{\mathcal{H}\left(\Sigma_S\right),\mathcal{H}\left(\Sigma_S\right)},~\forall y \in \mathcal{H}\left(\Sigma_S\right) \\
\end{split}
\end{equation}
\end{itemize}
where $\mu$ and $\nu$ are the adjoint variables of $u$ and $\tilde{\gamma}$, respectively. The adjoint derivative of $J$ can be derived in the same form as that in equation \ref{equ:ReducedFDrvSurField}, with the adjoint variables $\mu$ and $\nu$ derived by solving equation \ref{equ:WeakAdjEquBulkField}.

Similarly, for $V= {1\over \left|\Sigma_S\right| } \int_{\Sigma_S} \bar{\gamma} \,\mathrm{d}s$ in the area constraint, the adjoint derivative of $V$ can be derived as
\begin{equation}\label{equ:WeakAdjDrvVolConstr}
\begin{split}
\left\langle V', t \right\rangle_{\mathcal{L}^2\left(\Sigma_S\right),\mathcal{L}^2\left(\Sigma_S\right)} = \left\langle f_\gamma^* \nu, t \right\rangle_{\mathcal{L}^2\left(\Sigma_S\right),\mathcal{L}^2\left(\Sigma_S\right)},~\forall t \in \mathcal{L}^2\left(\Sigma_S\right) \\
\end{split}
\end{equation}
where the adjoint variable $\nu$ is derived by solving the following adjoint equation:
\begin{itemize}
  \item find $\nu\in\mathcal{H}\left(\Sigma_S\right)$ satisfying
\begin{equation}\label{equ:WeakAdjEquVolConstr}
\begin{split}
  \left\langle f_{\tilde{\gamma}}^* \nu, y \right\rangle_{\mathcal{H}\left(\Sigma_S\right),\mathcal{H}\left(\Sigma_S\right)} = \left\langle - V_{\bar{\gamma}}\bar{\gamma}_{\tilde{\gamma}}, y \right\rangle_{\mathcal{H}\left(\Sigma_S\right),\mathcal{H}\left(\Sigma_S\right)},~\forall y \in \mathcal{H}\left(\Sigma_S\right).
\end{split}
\end{equation}
\end{itemize}

Based on the derived adjoint derivatives with adjoint variables solved from the corresponding adjoint equations, the design variable can be iteratively updated.

\subsection{Numerical implementation} \label{sec:NumericalImplementation}

By using an iterative procedure, the design variable is penalized and evolved into a binary distribution on the 2-manifold. The iterative procedure is implemented as outlined by the pseudocode in Table \ref{tab:IterativeProcedure}. The surface finite element method is utilized to solve the relevant PDEs and the adjoint equations defined on the 2-manifold \cite{DziukActaNumerica2013}. And it can be implemented by choosing a software-package including a surface finite element solver. In this iterative procedure, the projection parameter $\beta$ with initial value $1$ is doubled after every $30$ iterations and $\xi$ is set to be $0.5$; the loop for solving the optimization problem in equation \ref{equ:VarProBulkField} is stopped when the maximal iteration number is reached, or the averaged variation of the design objective over continuous 5 iterations and the residual of the area constraint are simultaneously less than the specified tolerance of $10^{-3}$; and the design variable is updated using the method of moving asymptotes \cite{SvanbergIJNME1987}, which has the merits on dealing with multiple integral constraints and bound constraint of the design variable.

\begin{table}[!htbp]
\centering
\begin{tabular}{l}
  \hline
  Choose $r$, $V_f$ and $n^{sub}_{max}$;\\
  Set $i \leftarrow 1$, $\gamma \leftarrow V_f$, $n^{sub} \leftarrow 1$, $\xi \leftarrow 0.5$ and $\beta \leftarrow 1$; \\
  \textbf{loop} \\
          \hspace{1em} Derive $\bar{\gamma}$ by filtering and projecting $\gamma$, and evaluate $V$; \\
          \hspace{1em} Solve $u$ from the physical PDE, evaluate $J$, and set $J_{n^{sub}}\leftarrow J$; \\
          \hspace{1em} Solve $\mu$ and $\nu$ from the adjoint equations for $J$, and evaluate $J'$; \\
          \hspace{1em} Solve $\nu$ from the adjoint equation for $V$, and evaluate $V'$; \\
          \hspace{1em} Update $\gamma$ based on $J'$ and $V'$; \\
          \hspace{1em} \textbf{if}$\mod\left(n^{sub},n^{upt}\right)==0$ \\
          \hspace{2em} $\beta \leftarrow 2\beta$; \\
          \hspace{1em} \textbf{end}(\textbf{if})\\
          \hspace{1em} \textbf{if} $\left(n^{sub}==n^{sub}_{max}\right)$ or \\
          \hspace{2em} $\left(\beta == 2^{10}, {1\over5}\sum_{m=0}^4 \left| J_{n^{sub}} - J_{n^{sub}-m} \right|\Big/J_0 \leq 10^{-3},~\mathrm{and}~\left|V-V_f\right| \leq 10^{-3}\right)$ \\
          \hspace{2em} break; \\
          \hspace{1em} \textbf{end}(\textbf{if})\\
          \hspace{1em} $n^{sub} \leftarrow n^{sub}+1$ \\
      \textbf{end}(\textbf{loop})\\
  \hline
\end{tabular}
\caption{Pseudocode for the iterative solution of topology optimization on 2-manifolds. In the loop, $n^{sub}$ is the loop-index, $n^{sub}_{max}$ is the maximal value of $n^{sub}$, $n^{upt}$ is the updating interval of the projection parameter $\beta$, and $J_{n^{sub}}$ is the objective value corresponding to the loop-index $n^{sub}$.}\label{tab:IterativeProcedure}
\end{table}

\section{Test problems} \label{sec:TestManifold}

In this section, the first case of topology optimization implemented on 2-manifolds is demonstrated by the optimization of the micro-textures for wetting behaviors on the solid surfaces in the form of 2-manifolds; and the second case is demonstrated by topology optimization of patterns of heat sink for heat transfer and patterns of the perfect conductor for electromagnetics.

\subsection{Microtextures for wetting behaviors in Cassie-Baxter mode} \label{subsubsec:MicrotextureInWettingBehavior}

Wettability is an important aspect of fluidic mechanics \cite{FengAdvMater2006}. Constructing artificial micro-textures on a solid surface with complicated geometries can be interesting and attractive for stable hydrophobic wettability. Recently, topology optimization has been implemented to design micro-textures on flat solid surfaces \cite{DengCMAME2018}. On solid surfaces, the micro-textures can support two modes of hydrophobic performance, i.e. the Wenzel mode with the liquid completely filling the micro-textures and the Cassie-Baxter mode with vapour pockets trapped in the micro-textures. The Cassie-Baxter mode can be transitioned into the Wenzel mode, when the liquid is pressurized \cite{BicoEurophysLett1999,LafumaNatMater2003}. Therefore, it is desired to use reasonable micro-textures to keep the Cassie-Baxter mode from transition and enhance the stability of hydrophobic wettability. On a solid surface with complicated geometry, the pattern of the micro-textures is defined on a 2-manifold used to describe the solid surface. In this case, the mean curvature of the lquid/vapor interface supported by the micro-textures can be regarded to be a physical field defined on the 2-manifold.

The mean curvature of the lquid/vapor interface supported by the micro-textures on the 2-manifold can be described by the dimensionless Young-Laplace equation \cite{YoungPhilTrans1805,LaplaceMechaniqueCeleste1806}:
\begin{equation}\label{equ:ConstSurf1}
\begin{split}
  & \nabla_s \cdot \left( \bar{\sigma} { \nabla_s \bar{z} \over \sqrt { \left(L/z_0\right)^2 + \left|\nabla_s \bar{z} \right|^2}} \right) = 1 - 2 \bar{\sigma}_l \kappa,~\mathrm{on}~ \Sigma_S \\
  & \bar{\sigma} { \nabla_s \bar{z} \over \sqrt { \left(L/z_0\right)^2 + \left|\nabla_s \bar{z} \right|^2} } \cdot \boldsymbol \tau = 0,~\mathrm{at}~ \partial\Sigma_S \\
  & \bar{z} = 0,~\mathrm{at}~ \mathcal{P}
\end{split}
\end{equation}
where $\bar{z} = z/z_0$ is the normalized displacement of the liquid/vapor interface relative to the 2-manifold $\Sigma_S$, with $z_0$ denoting the magnitude of the original displacement $z$; $L$ is the characteristic size of the solid surface; $\bar{\sigma} = \sigma/\left(L P\right)$ is the dimensionless surface tension, with $\sigma$ denoting the surface tension and $P$ denoting the static pressure at the liquid/vapor interface; $\bar{\sigma}_l$ is the dimensionless surface tension of the liquid; $\kappa$ is the mean curvature of the 2-manifold $\Sigma_S$. To ensure the uniqueness of the solution, the liquid/vapor interface is constrained at a specified point set $\mathcal{P}$ localized on the 2-manifold.

For the wetting behavior in the Cassie-Baxter mode, the performance of the micro-textures can be measured by the volume of the liquid bulges enclosed by the convex liquid/vapor interface and untextured solid surface. Therefore, the topology optimization of the micro-textures is implemented by
minimizing the normalized volume of the liquid bulges expressed as
\begin{equation}\label{equ:MicrotextureObj}
  J = {1\over\left|\Sigma_S\right|^2 } \int_{\Sigma_S} \bar{z}^2 \,\mathrm{d}s
\end{equation}
which is constrained by an area constraint with the specified area fraction of $V_f$. In topology optimization of the micro-textures, the material density is used to interpolate the dimensionless surface tension as
\begin{equation}\label{equ:MicrotextureInterpolation}
  \bar{\sigma} = \bar{\sigma}_l + \left( \bar{\sigma}_s - \bar{\sigma}_l \right) { 1 - \bar{\gamma} \over 1 + 10^4\bar{\gamma} }
\end{equation}
where $\bar{\sigma}_s$ is set as $10^5 \sup_{\mathbf{x}\in\Sigma_S} \left|1 - 2 \bar{\sigma}_l \kappa\left(\mathbf{x}\right)\right|$ to approximate the liquid/solid interface.

Based on the adjoint analysis introduced in Section \ref{sec:AdjointAnalysisBulkFieldManifold}, the adjoint derivative of $J$ is derived as
\begin{equation}\label{equ:AdjDrvHeatTransfer2ManifoldJ}
  \left\langle J',t\right\rangle_{\mathcal{L}^2\left(\Sigma_S\right),\mathcal{L}^2\left(\Sigma_S\right)} = - \int_{\Sigma_S} \tilde{\gamma}_a t \, \mathrm{d}s,~\forall t \in \mathcal{L}^2\left(\Sigma_S\right).
\end{equation}
The adjoint variable $\tilde{\gamma}_a$ in equation \ref{equ:AdjDrvHeatTransfer2ManifoldJ} can be derived by solving the following adjoint equations:
\begin{itemize}
  \item find $\bar{z}_a\in \mathcal{H}\left(\Sigma_S\right)$ with $\bar{z}_a=0$ at $\mathcal{P}$, satisfying
  \begin{equation}\label{equ:AdjEquHeatTransfer2ManifoldT}
  \begin{split}
  & \int_{\Sigma_S} {2 \over \left|\Sigma_S\right|^2 } \bar{z} \hat{\bar{z}}_a - \bar{\sigma} {\nabla_s \bar{z}_a \cdot \nabla_s \hat{\bar{z}}_a \over \sqrt { \left(L/z_0\right)^2 + \left|\nabla_s \bar{z} \right|^2} } + \bar{\sigma} { \left(\nabla_s \bar{z} \cdot \nabla_s \bar{z}_a \right) \left(\nabla_s \bar{z} \cdot \nabla_s \hat{\bar{z}}_a \right) \over \left( \sqrt { \left(L/z_0\right)^2 + \left|\nabla_s \bar{z} \right|^2} \right)^3 }  \,\mathrm{d}s = 0,\\
  & \forall \hat{\bar{z}}_a \in \mathcal{H}\left(\Sigma_S\right)
  \end{split}
  \end{equation}
  \item find $\tilde{\gamma}_a \in \mathcal{H}\left(\Sigma_S\right)$, satisfying
  \begin{equation}\label{equ:HelmholtzFilterAdjEqu}
  \begin{split}
  & \int_{\Sigma_S} - r^2 \nabla_s \tilde{\gamma}_a \cdot \nabla_s \hat{\tilde{\gamma}}_a + \tilde{\gamma}_a \hat{\tilde{\gamma}}_a - {\partial \bar{\sigma} \over \partial \bar{\gamma}} {\partial \bar{\gamma} \over \partial \tilde{\gamma}} {\nabla_s \bar{z} \cdot \nabla_s \bar{z}_a \over \sqrt { \left(L/z_0\right)^2 + \left|\nabla_s \bar{z} \right|^2}} \hat{\tilde{\gamma}}_a \,\mathrm{d}s = 0,\\
  & \forall \hat{\tilde{\gamma}}_a \in \mathcal{H}\left(\Sigma_S\right)
  \end{split}
  \end{equation}
\end{itemize}
where $\bar{z}_a$ is the adjoint variable of $\bar{z}$. For $V= {1\over \left|\Sigma_S\right| } \int_{\Sigma_S} \bar{\gamma} \,\mathrm{d}s$ in the area constraint, the adjoint derivative is
\begin{equation}\label{equ:AdjDrvHeatTransfer2ManifoldVolConstr}
\begin{split}
\left\langle V', t \right\rangle_{\mathcal{L}^2\left(\Sigma_S\right),\mathcal{L}^2\left(\Sigma_S\right)} = - \int_{\Sigma_S} \tilde{\gamma}_a t \, \mathrm{d}s,~\forall t \in \mathcal{L}^2\left(\Sigma_S\right)
\end{split}
\end{equation}
where the adjoint variable $\tilde{\gamma}_a$ is derived by solving the following adjoint equation:
\begin{itemize}
  \item find $\tilde{\gamma}_a \in \mathcal{H}\left(\Sigma_S\right)$, satisfying
\begin{equation}\label{equ:WeakAdjEquHeatTransfer2ManifoldVolConstr}
\begin{split}
  \int_{\Sigma_S} {1\over \left|\Sigma_S\right| } {\partial \bar{\gamma} \over \partial \tilde{\gamma}} \hat{\tilde{\gamma}}_a \,\mathrm{d}s + r^2\nabla_s \tilde{\gamma}_a \cdot \nabla_s \hat{\tilde{\gamma}}_a + \tilde{\gamma}_a \hat{\tilde{\gamma}}_a \, \mathrm{d}s = 0,~\forall\hat{\tilde{\gamma}}_a\in \mathcal{H}\left(\Sigma_S\right).
\end{split}
\end{equation}
\end{itemize}
After adjoint analysis, the topology optimization problem is solved by the numerical implementation procedure introduced in Section \ref{sec:NumericalImplementation}.

\begin{figure}[!htbp]
  \centering
  \includegraphics[width=0.9\columnwidth]{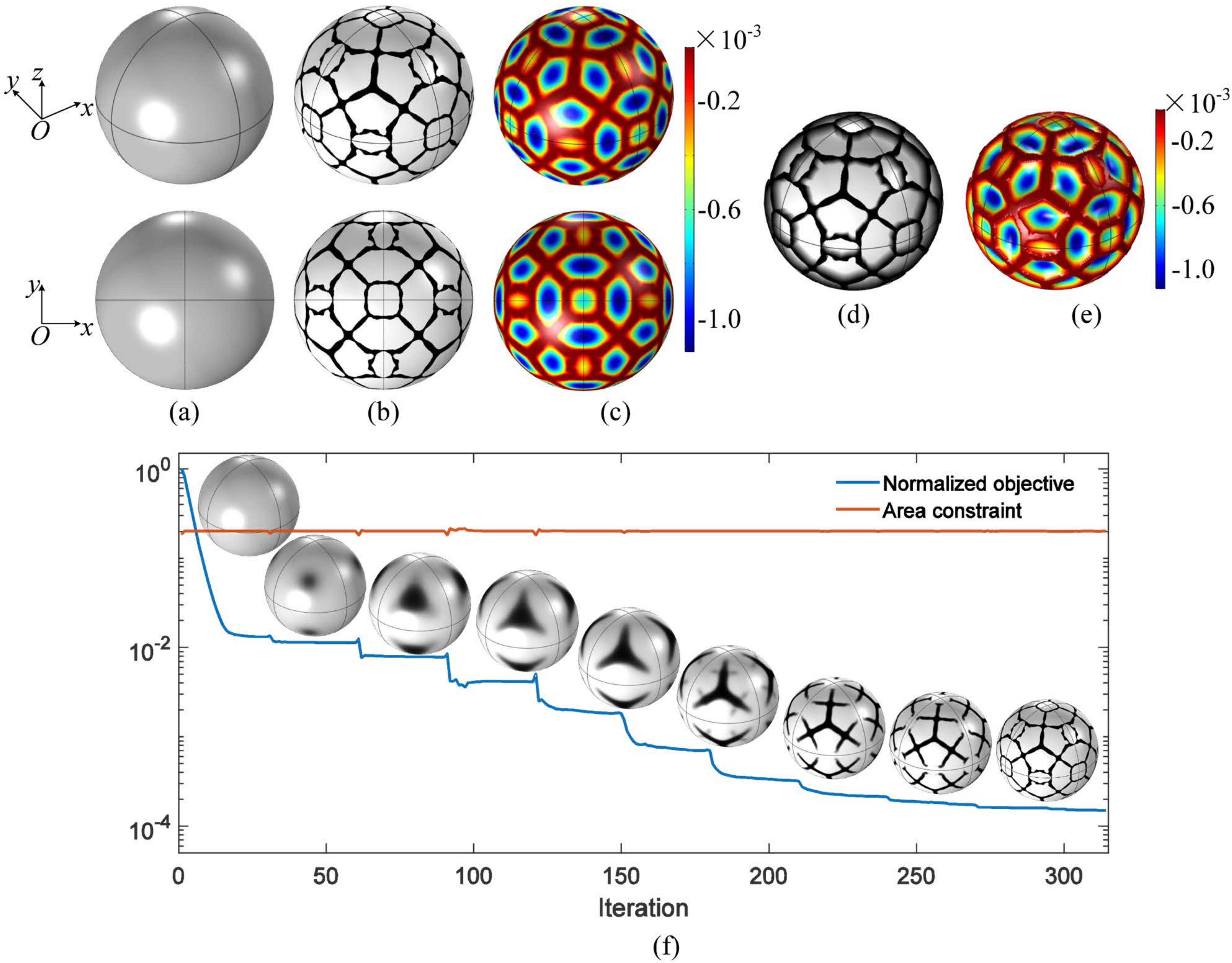}
  \caption{(a-c) Stereo and vertical views of the sphere surface divided by regular spherical-triangles, the derived pattern of the micro-textures, and the normalized displacement of the liquid/vapor interface supported on the micro-textures with the derived pattern; (d-e) deformation views of the micro-textures and the liquid/vapor interfaces; (f) convergent histories of the optimization objective and the area constraint, including the snapshots for the evolution of the material density.}\label{fig:MicrotexturesOnSphereWithTriangle}
\end{figure}

\begin{figure}[!htbp]
  \centering
  \includegraphics[width=0.9\columnwidth]{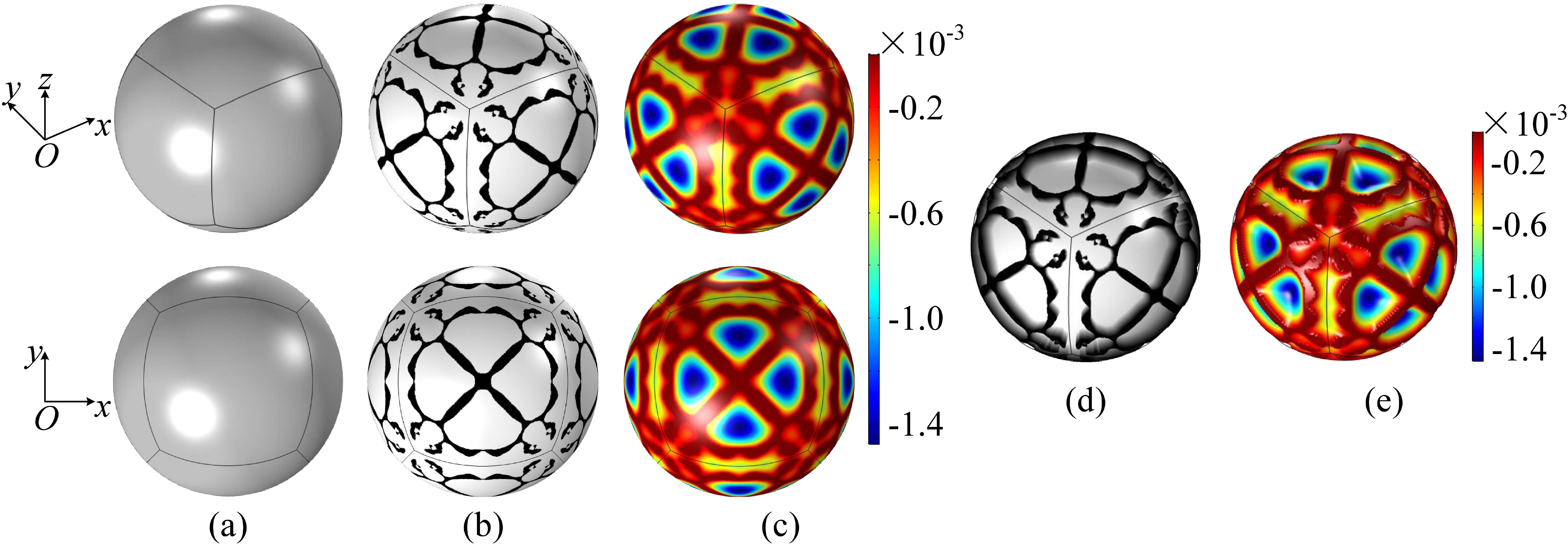}
  \caption{(a-c) Stereo and vertical views of the sphere surface divided by regular spherical-quadrangles, the derived pattern of the micro-textures, and the normalized displacement of the liquid/vapor interface supported on the micro-textures with the derived pattern; (d-e) deformation views of the micro-textures and the liquid/vapor interfaces.}\label{fig:MicrotexturesOnSphereWithQuadrangle}
\end{figure}

\begin{figure}[!htbp]
  \centering
  \includegraphics[width=0.95\columnwidth]{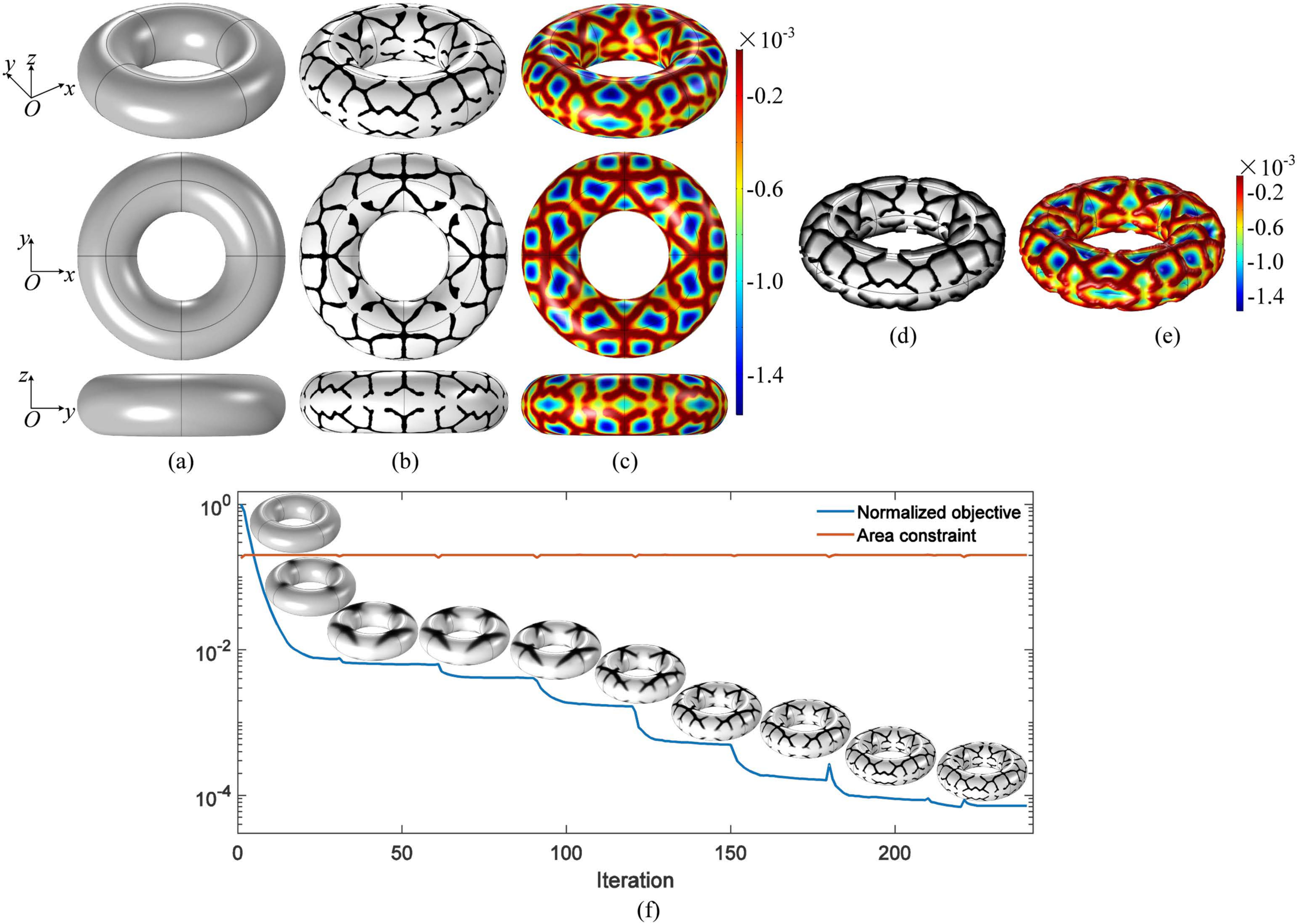}
  \caption{(a-c) Stereo, vertical and front views of the torus surface divided by the illustrated circles, the derived pattern of the micro-textures, and the normalized displacement of the liquid/vapor interface supporting on the micro-textures with the derived pattern; (d-e) deformation views of the micro-textures and the liquid/vapor interfaces; (f) convergent histories of the optimization objective and the area constraint, including the snapshots for the evolution of the material density.}\label{fig:MicrotexturesOnTorusCstP1}
\end{figure}

\begin{figure}[!htbp]
  \centering
  \includegraphics[width=0.95\columnwidth]{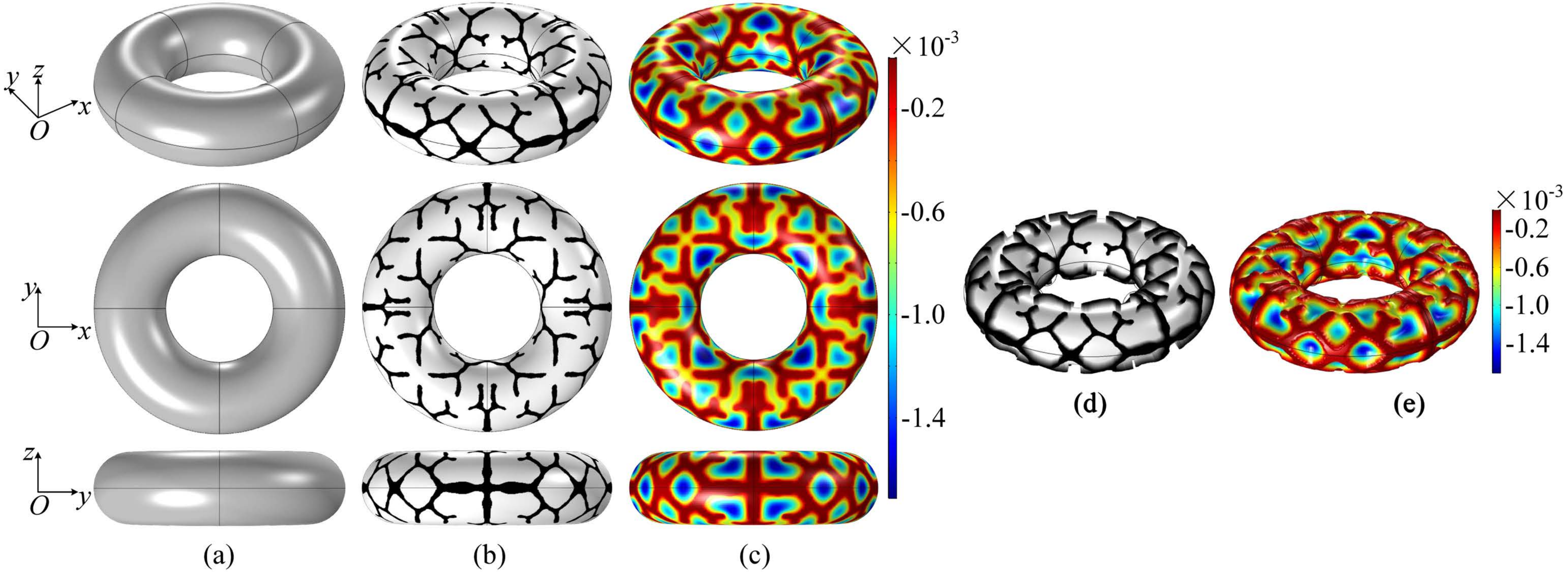}
  \caption{(a-c) Stereo, vertical and front views of the torus surface divided by the illustrated circles, the derived pattern of the micro-textures, and the normalized displacement of the liquid/vapor interface supporting by the micro-textures with the derived pattern; (d-e) deformation views of the micro-textures and the liquid/vapor interfaces.}\label{fig:MicrotexturesOnTorusCstP2}
\end{figure}

\begin{figure}[!htbp]
  \centering
  \includegraphics[width=0.95\columnwidth]{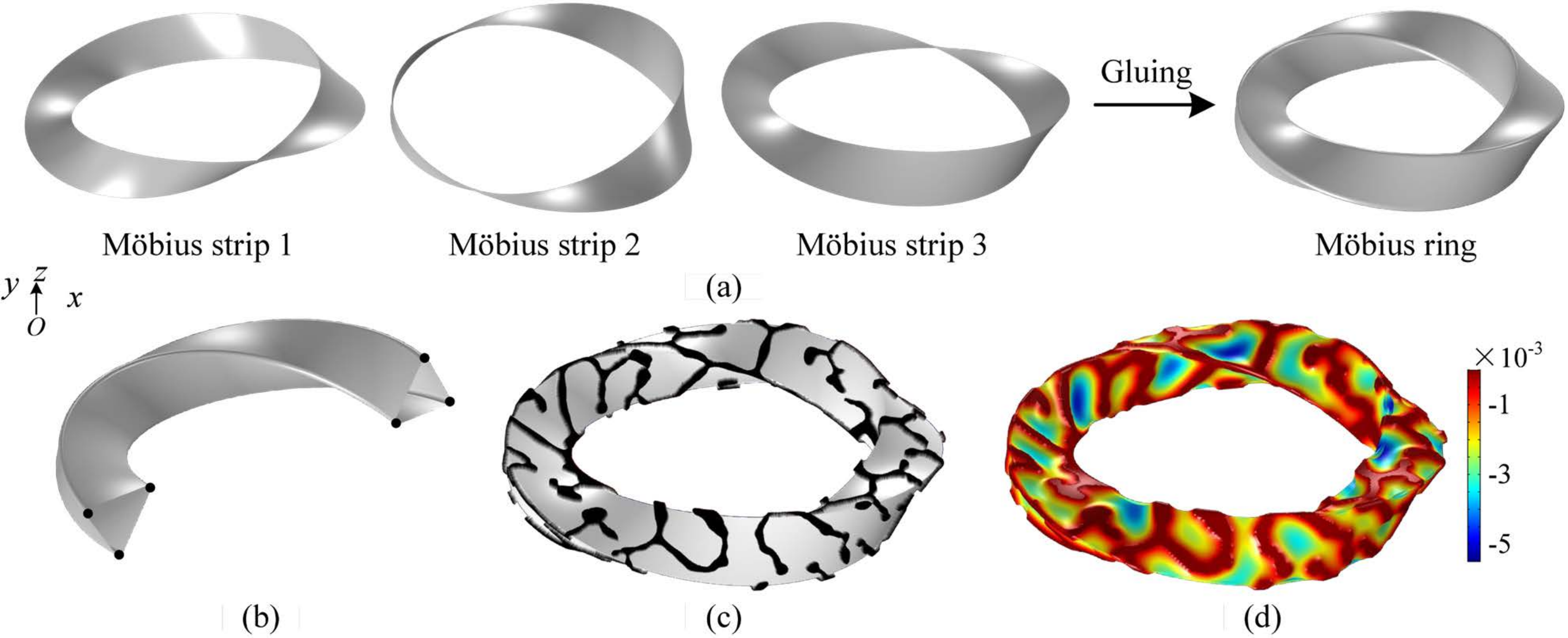}
  \caption{(a) Sketch for gluing three M\"{o}bius strips to derive a M\"{o}bius ring; (b) cutaway view of the M\"{o}bius ring used to sketch the points of the point set $\mathcal{P}$; (c-d) deformation views of the micro-textures and the liquid/vapor interfaces.}\label{fig:MoebiusRing}
\end{figure}

Topology optimization of the micro-textures is implemented for the sphere- and torus-shaped surfaces. By setting the parameters as listed in Table \ref{fig:ParametersMicrotextures}, the patterns of the micro-textures are derived for different choices of the points in the point set $\mathcal{P}$. For the sphere, the surface is divided into regular spherical-triangles (Figure \ref{fig:MicrotexturesOnSphereWithTriangle}a) and spherical-quadrangles (Figure \ref{fig:MicrotexturesOnSphereWithQuadrangle}a), respectively. The point set $\mathcal{P}$ is set as the central points of these spherical-polygons. For the torus, the surface is divided by two sets of circles (Figure \ref{fig:MicrotexturesOnTorusCstP1}a and \ref{fig:MicrotexturesOnTorusCstP2}a), where $\mathcal{P}$ is set as the  intersection points of the circles. Then, the patterns of the micro-textures are derived as shown in Figure \ref{fig:MicrotexturesOnSphereWithTriangle}b and \ref{fig:MicrotexturesOnTorusCstP2}b, with the corresponding liquid/vapor interfaces shown in Figure \ref{fig:MicrotexturesOnSphereWithTriangle}c and \ref{fig:MicrotexturesOnTorusCstP2}c. In Figure \ref{fig:MicrotexturesOnSphereWithTriangle}d and \ref{fig:MicrotexturesOnTorusCstP2}d, \ref{fig:MicrotexturesOnSphereWithTriangle}e and \ref{fig:MicrotexturesOnTorusCstP2}e, the derived patterns and the corresponding liquid/vapor interfaces are shown in the deformation views. The convergent histories of the normalized optimization objective and the area constraints are shown in Figure \ref{fig:MicrotexturesOnSphereWithTriangle}f and \ref{fig:MicrotexturesOnTorusCstP1}f, including the snapshots for the evolution of the material density. From the convergent histories and the snapshots, the convergent performance of the topology optimization procedures can be confirmed. The topology optimization of the micro-textures is further implemented on a M\"{o}bius ring derived by gluing three M\"{o}bius strips shown in Figure \ref{fig:MoebiusRing}a, where the point set $\mathcal{P}$ is set as the six points sketched in the cutaway view shown in Figure \ref{fig:MoebiusRing}b. The micro-textures are derived as shown in Figure \ref{fig:MoebiusRing}c with the corresponding liquid/vapor interface shown in Figure \ref{fig:MoebiusRing}d.

\begin{table}
\centering
\begin{tabular}{c|c|c|c|c|c}
  $L$ & $\bar{\sigma}_l$ & $z_0$ & $V_f$ & $n^{sub}_{max}$ & $n^{upt}$ \\
  \hline
  $10\mu\mathrm{m}$ & $10^2$ & $1\mu\mathrm{m}$ & $0.2$ & $315$ & $30$ \\
\end{tabular}
\caption{Parameters used to implement topology optimization of the micro-textures for the wetting behaviors in the Cassie-Baxter mode.}\label{fig:ParametersMicrotextures}
\end{table}

In Table \ref{tab:TextureOptimalityConfirmation}, the volumes of the liquid-bulges are cross compared for the derived micro-textures on the sphere and torus for the cases with two different dimensionless surface tension of $10^2$ and $10^3$, respectively. The point set $\mathcal{P}$ is set to be the same as that in Figure \ref{fig:MicrotexturesOnSphereWithTriangle} and \ref{fig:MicrotexturesOnTorusCstP1}. From a cross
comparison of the values in every row of the sub-tables in Table \ref{tab:TextureOptimalityConfirmation}, the improved performance of the micro-textures corresponding to the derived patterns can be confirmed.

\begin{table}[!htbp]
\centering
\subtable[Sphere]{
\begin{tabular}{c|cc}
  \toprule
        & \includegraphics[height=0.08\columnwidth]{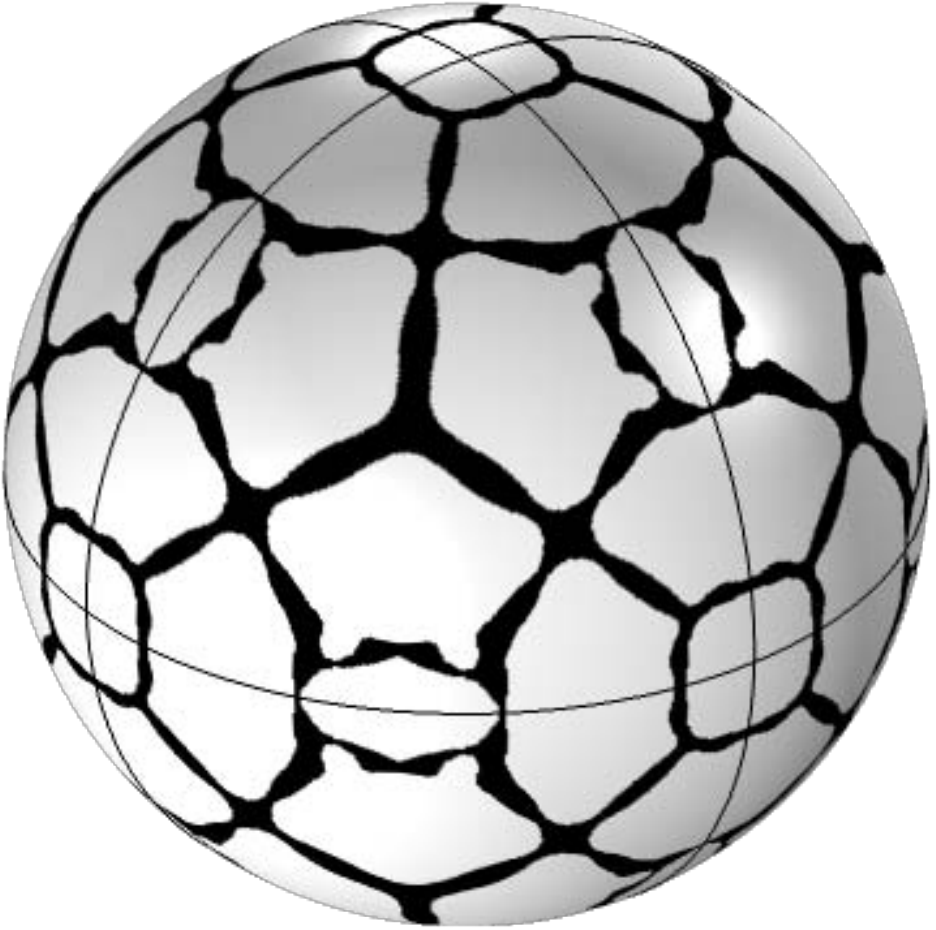} & \includegraphics[height=0.08\columnwidth]{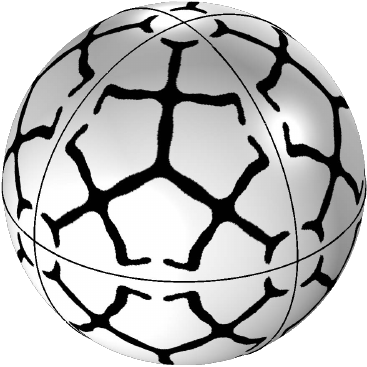} \\
  \midrule
  $\bar{\sigma}_l=10^2$ & $\mathbf{2.97\times10^{-6}}$ & $1.27\times10^{-5}$ \\
  \midrule
  $\bar{\sigma}_l=10^3$ & $9.43\times10^{-8}$ & $\mathbf{4.55\times10^{-8}}$ \\
  \bottomrule
\end{tabular}}\hspace{2ex}
\subtable[Torus]{
\begin{tabular}{c|cc}
  \toprule
        & \includegraphics[height=0.08\columnwidth]{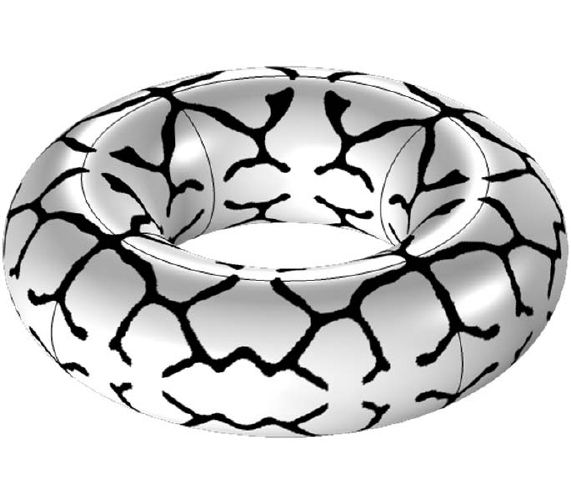} & \includegraphics[height=0.08\columnwidth]{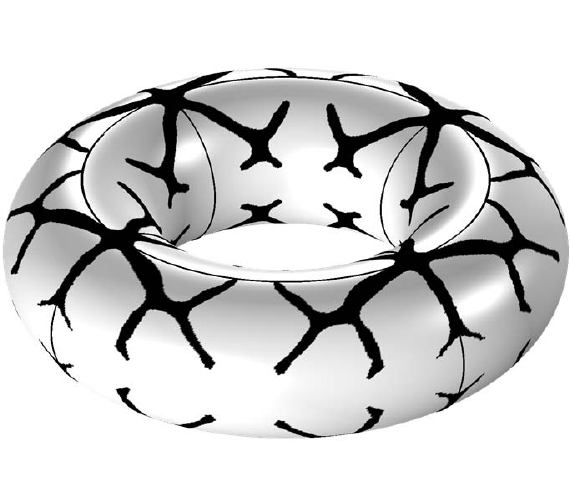} \\
  \midrule
  $\bar{\sigma}_l=10^2$ & $\mathbf{9.78\times10^{-6}}$ & $1.13\times10^{-4}$ \\
  \midrule
  $\bar{\sigma}_l=10^3$ & $2.04\times10^{-6}$ & $\mathbf{1.10\times10^{-6}}$ \\
  \bottomrule
\end{tabular}}
\caption{Volume of the liquid bulges supported by the micro-textures with the derived patterns on the sphere and the torus for the dimensionless surface tension $10^2$ and $10^3$, respectively. The optimized entries are noted in bold.}\label{tab:TextureOptimalityConfirmation}
\end{table}

When the solid objects are textured using the derived patterns, the three-phase contact lines of the liquid/vapour interfaces can be anchored at the geometrically singular corners formed by the top and side walls of the micro-textures. This anchoring effect is caused by the hysteresis of contact angle. The wetting behaviors are thus in the Cassie-Baxter mode.

If the difference between the static pressure imposed on the liquid is large enough to make the contact angle between the liquid/vapour interface and side wall of the micro-textures reach the critical advancing angle, the liquid/vapour interface can burst and transition can occur from the Cassie-Baxter mode to the Wenzel mode. As the contact angle evolves towards the critical advancing angle, the robustness of the Cassie-Baxter mode decreases; simultaneously, the liquid/vapour interface supported on the micro-textures becomes more convex, and the volume of the liquid bulges suspended at the liquid/vapour interface increases. Therefore, reasonable micro-textures on a solid surface can keep the Cassie-Baxter mode from transition by keeping the contact angle aloof of the crucial advancing angle.

\subsection{Heat sink for heat transfer problem} \label{subsubsec:HeatSinkInHeatTransfer}

The heat transfer problem has been investigated by using topology optimization to find the layout of the heat conductive materials \cite{LiIJHMT2004,GersborgHansenSMO2006,ZhuangCMAME2007} and optimal match of the structural topology and heat source \cite{DengJCP2014}.
This section implements topology optimization on 2-manifolds for heat transfer problems to determine the patterns of heat sinks corresponding to the minimized thermal compliance. The temperature distribution in a 3D domain can be described as
\begin{equation}\label{equ:HeatTransPDE2Manifold}
\begin{split}
& - \nabla \cdot \left(k \nabla T \right) = Q, ~ \mathrm{in} ~ \Omega\\
& T = T_d, ~ \mathrm{on} ~ \Gamma_d \\
& - k \nabla T \cdot \mathbf{n} = 0, ~ \mathrm{on} ~ \Gamma_n \\
& T = T_0, ~ \mathrm{at} ~ \mathcal{P}
\end{split}
\end{equation}
where $k$ is the thermal conductivity; $T$ is temperature; $Q$ is the heat source; $\Gamma_d$ is heat sink with known temperature $T_d$ and it is attributed to the Dirichlet type; $\Gamma_n$ is the insulation boundary, attributed to the Neumann type; $\Gamma_d$ and $\Gamma_n$ satisfy $\Gamma_d\cup\Gamma_n=\Sigma_M$; $\Omega$ is the domain enclosed by $\Sigma_M$; $T_0$ is the known temperature at the points in the point set $\mathcal{P}$. By setting the design domain as $\Sigma_S=\Sigma_M$, topology optimization can be implemented to determine the crosswise distribution of the heat sink $\Gamma_d\left(\bar{\gamma}=0\right)$ and the insulation $\Gamma_n\left(\bar{\gamma}=1\right)$ on $\Sigma_S$. Then, the material interpolation is implemented on $\Sigma_S$ to define the mixed boundary condition:
\begin{equation}\label{equ:InterpolatedBNDHeatTransfer2Manifold}
 -k \nabla T \cdot \mathbf{n} = \alpha\left(\bar{\gamma}\right) \left(T - T_d\right),~\mathrm{on}~\Sigma_S
\end{equation}
where $\alpha\left(\bar{\gamma}\right)$ is the penalization function of the material density. The penalization function $\alpha\left(\bar{\gamma}\right)$ is expressed as
\begin{equation}\label{equ:MaterialInterpolationHeatTransfer2Manifold}
  \alpha\left(\bar{\gamma}\right) = \alpha_{\max} {q\left(1-\bar{\gamma}\right)\over q+\bar{\gamma}}
\end{equation}
where $\alpha_{\max}$ and $q$ are the parameters used to implement the penalization and to tune the convexity of the penalization function, respectively. The value of $\alpha_{\max}$ should be chosen to be large enough to ensure the domination of the term $\left(T - T_d\right)$ in equation \ref{equ:InterpolatedBNDHeatTransfer2Manifold}, when the material density takes on the value of $0$; equation \ref{equ:InterpolatedBNDHeatTransfer2Manifold} degenerates into the insulation boundary condition, when the material density takes on the value of $1$. Based on numerical experiments, $\alpha_{\max}$ and $q$ are set as $10^4$ and $10^{-6}$, respectively. The variational formulation of equation \ref{equ:VarProBulkField} is
\begin{itemize}
  \item find $T\in\mathcal{H}\left(\Omega\right)$ with $T=T_0$ at $\mathcal{P}$, satisfying
\begin{equation}\label{equ:PhysicalPDEHeatTransfer2Manifold}
  \int_\Omega k \nabla T \cdot \nabla \hat{T} - Q \hat{T} \,\mathrm{d}v + \int_{\Sigma_S} \alpha \left(T - T_d\right) \hat{T} \,\mathrm{d}s=0,~\forall \hat{T} \in \mathcal{H}\left(\Omega\right)
\end{equation}
\end{itemize}
where $\hat{T}$ is the test function of $T$; $\mathcal{H}\left(\Omega\right)$ is the first order Sobolev space on $\Omega$.

The distribution of the heat sink on $\Sigma_S$ is determined to minimize the thermal compliance:
\begin{equation}\label{equ:ThermalCompliance}
  J = \int_\Omega k \nabla T \cdot \nabla T \,\mathrm{d}\Omega.
\end{equation}
Based on the adjoint analysis introduced in Section \ref{sec:AdjointAnalysisBulkFieldManifold}, the adjoint derivative of $J$ is derived as
\begin{equation}\label{equ:AdjDrvHeatTransfer2ManifoldJ}
  \left\langle J',t\right\rangle_{\mathcal{L}^2\left(\Sigma_S\right),\mathcal{L}^2\left(\Sigma_S\right)} = - \int_{\Sigma_S} \tilde{\gamma}_a t \, \mathrm{d}s,~\forall t \in \mathcal{L}^2\left(\Sigma_S\right).
\end{equation}
The adjoint variable $\tilde{\gamma}_a$ in equation \ref{equ:AdjDrvHeatTransfer2ManifoldJ} can be derived by sequentially solving the following adjoint equations:
\begin{itemize}
  \item find $T_a\in\mathcal{H}\left(\Omega\right)$ with $T_a = 0$ at $\mathcal{P}$, satisfying
  \begin{equation}\label{equ:AdjEquHeatTransfer2ManifoldT}
  \begin{split}
    \int_\Omega 2 k \nabla T \cdot \nabla \hat{T}_a + k \nabla T_a \cdot \nabla \hat{T}_a \,\mathrm{d}v + \int_{\Sigma_S} \alpha T_a \hat{T}_a \,\mathrm{d}s = 0,~\forall \hat{T}_a \in \mathcal{H}\left(\Omega\right)
  \end{split}
  \end{equation}
  \item find $\tilde{\gamma}_a\in\mathcal{H}\left(\Sigma_S\right)$ satisfying
  \begin{equation}\label{equ:AdjEquHeatTransfer2ManifoldGamma}
  \begin{split}
    & \int_{\Sigma_S} {\partial \alpha \over \partial \bar{\gamma}} {\partial \bar{\gamma} \over \partial \tilde{\gamma}} \left(T - T_d\right) T_a \hat{\tilde{\gamma}}_a \,\mathrm{d}s + r^2\nabla_s \tilde{\gamma}_a \cdot \nabla_s \hat{\tilde{\gamma}}_a + \tilde{\gamma}_a \hat{\tilde{\gamma}}_a \, \mathrm{d}s = 0,~\forall\hat{\tilde{\gamma}}_a\in \mathcal{H}\left(\Sigma_S\right)
  \end{split}
  \end{equation}
\end{itemize}
where $T_a$ is the adjoint variables of $T$.
For the area constraint, the adjoint derivative and the adjoint equation are the same as those in equations \ref{equ:AdjDrvHeatTransfer2ManifoldVolConstr} and \ref{equ:WeakAdjEquHeatTransfer2ManifoldVolConstr}.

\begin{figure}[!htbp]
  \centering
  \includegraphics[width=0.68\columnwidth]{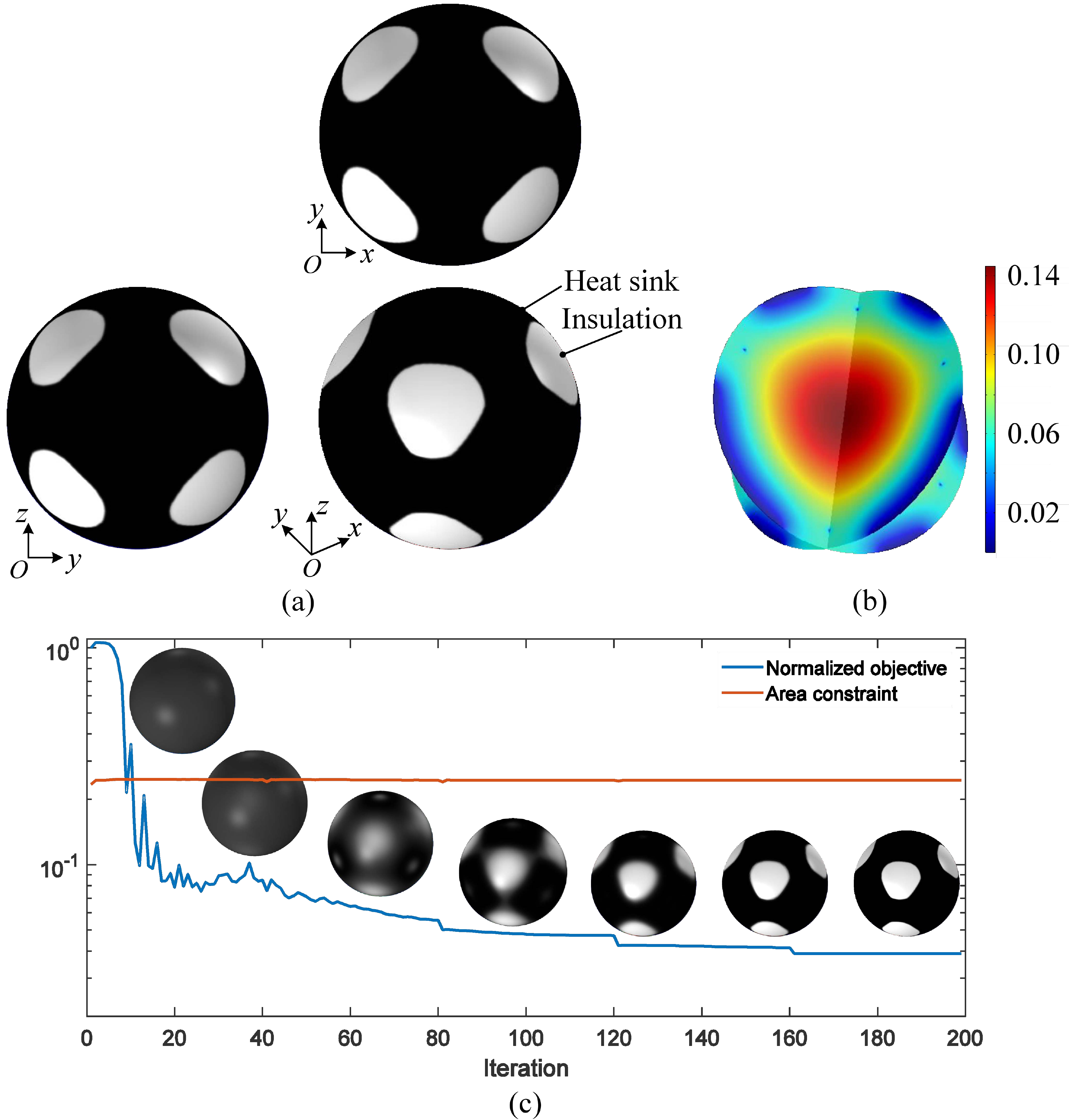}
\caption{(a) Stereo, vertical and front views of the derived pattern of the heat sink on the sphere; (b) temperature distribution in the cross-sections of the spherical domain; (c) convergent histories of the optimization objective and the area constraint, including the snapshots for the evolution of the material density.}\label{fig:Heat_Sphere_Topology_T0_0}
\end{figure}

\begin{figure}[!htbp]
  \centering
  {\includegraphics[width=0.95\columnwidth]{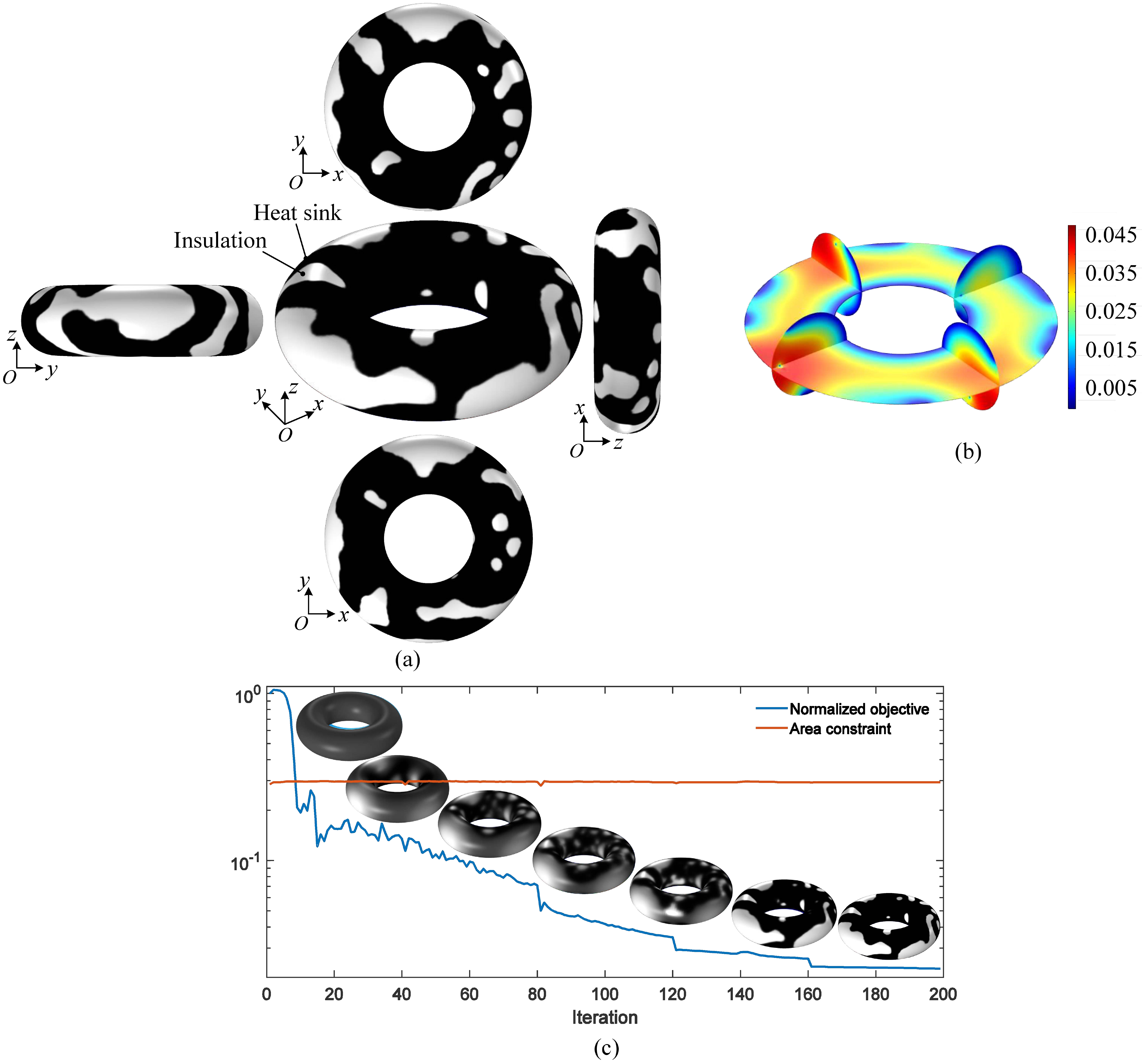}}
  \caption{(a) Stereo and lateral views of the derived pattern of the heat sink on the torus; (b) temperature distribution in the cross-sections of the torus domain; (c) convergent histories of the optimization objective and the area constraint, including the snapshots for the evolution of the material density.}\label{fig:Heat_Torus_Topology_T0_0}
\end{figure}

Based on the numerical implementation introduced in Section \ref{sec:NumericalImplementation}, the computational domain $\Omega$ is set to be the volume domains with the genus 0 and 1, respectively. The corresponding typical 2-manifolds are sphere and torus (Figure \ref{fig:ManifoldDemon}a and \ref{fig:ManifoldDemon}b). The coordinate origin is set as the centers of the sphere and torus; the radius of the sphere is $1$; the inner radius and outer radii of the torus are $3/4$ and $7/4$, respectively. For these two 2-manifolds, the thermal conductivity is set as $1$; the temperature of the heat sink $T_d$ and known point-temperature $T_0$ are set as $0$; the heat source is set as $Q= 1/\left(1+\mathbf{x}^2\right)$; the area fractions are set as $0.75$ and $0.7$, respectively. For the sphere, the point set $\mathcal{P}$ is set as the one composed of the vertexes of a cube with the center and one of the vertexes localized at the sphere center and $\left(0.5, 0.5, 0.5\right)$, respectively. For the torus, $\mathcal{P}$ is set as $\left\{\left(-1.65, 0, 0\right),\left(0, 1.25, 0.4\right),\left(0.85, 0, 0\right),\left(0, -1.25, -0.4\right)\right\}$. The maximal iteration number and the updating interval are set as  $n^{sub}_{max}=200$ and $n^{upt}=40$, respectively. The patterns of the heat sinks are derived as shown in Figure \ref{fig:Heat_Sphere_Topology_T0_0} and \ref{fig:Heat_Torus_Topology_T0_0}, including the corresponding convergent histories and snapshots for the evolution of the material density. From the convergent histories, the convergent performance of the topology optimization procedure can be confirmed for this heat transfer problem. The temperature distributions in the cross-sections of the volume domains are shown in Figure \ref{fig:Heat_Sphere_Topology_T0_0}b and \ref{fig:Heat_Torus_Topology_T0_0}b. The results show that the heat insulations are localized at the parts of the manifolds nearest the zero-temperature points and the heat sinks distribute around the heat insulations. Such distributions of the heat sinks and insulations can preserve the thermal energy in the regions around the zero-temperature points and reduce the temperature gradient to minimize the thermal compliance.

\begin{figure}[!htbp]
  \centering
  \includegraphics[width=0.7\columnwidth]{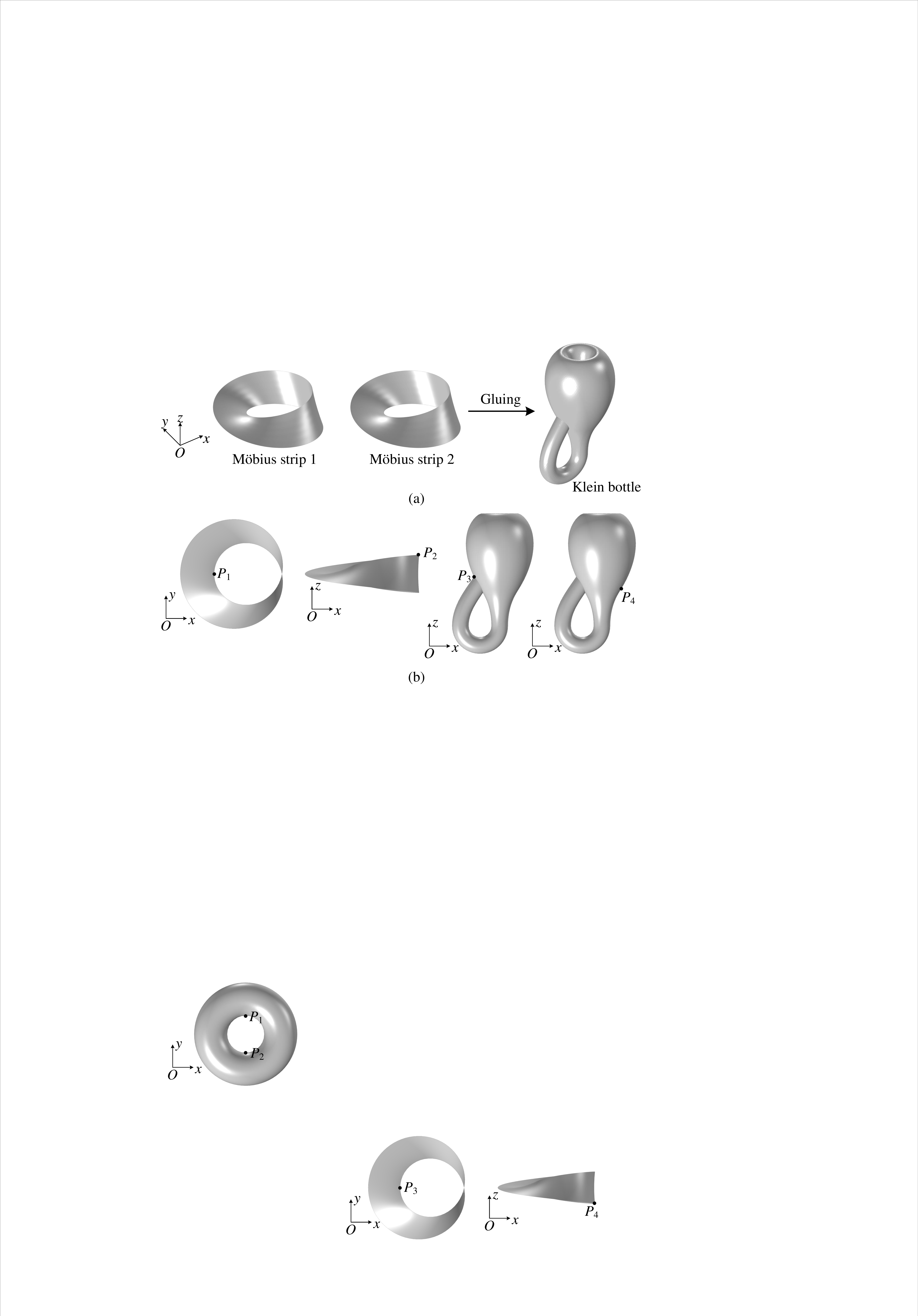}
  \caption{(a) Sketch for the Klein bottle, derived by gluing two M\"{o}bius strips; (b) sketches for the points of the point set $\mathcal{P}$.}\label{fig:MoebiusKleinBottleHeat}
\end{figure}

\begin{figure}[!htbp]
  \centering
  \includegraphics[width=0.62\columnwidth]{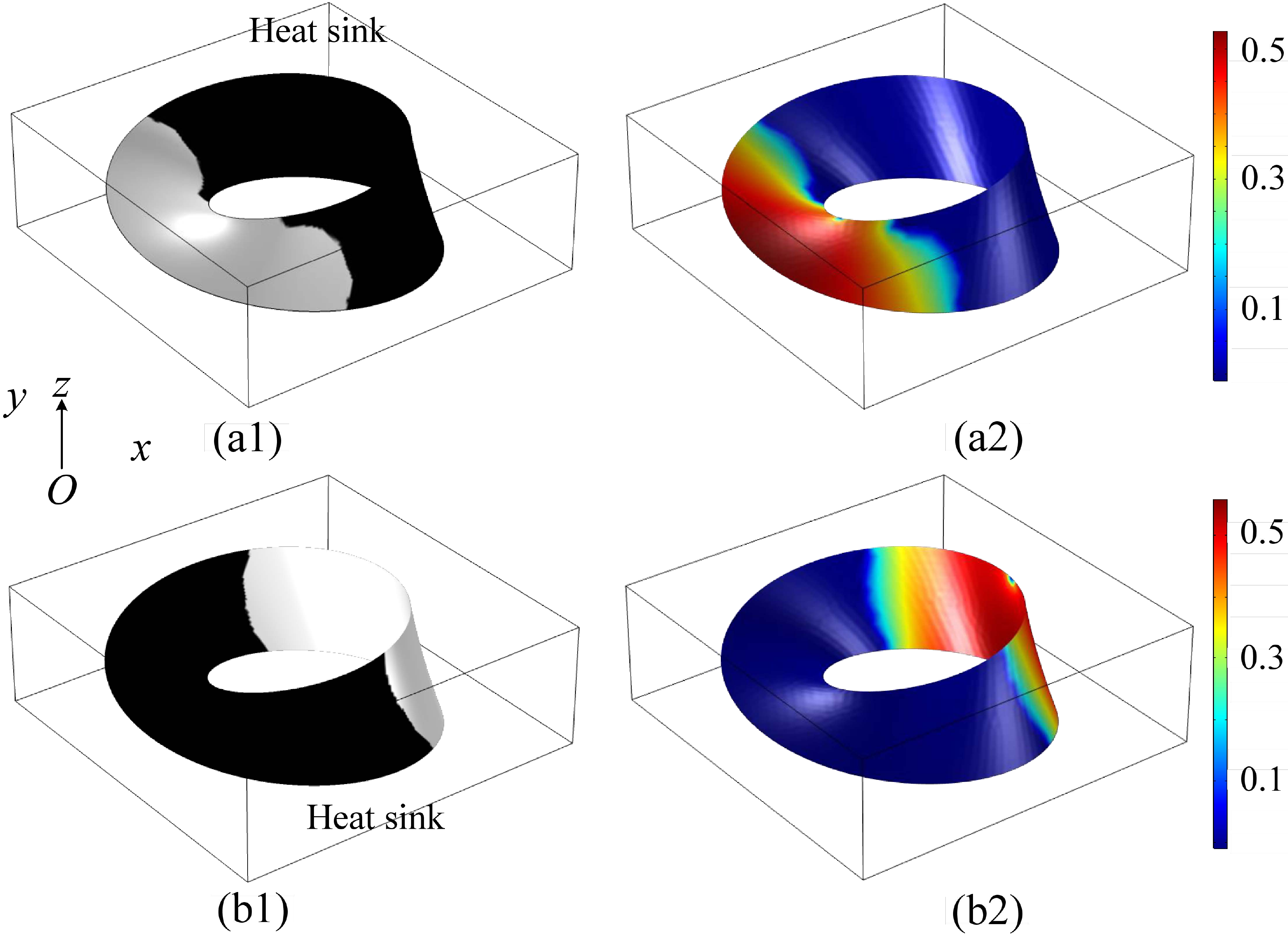}
  \caption{Derived patterns of the heat sinks and the corresponding temperature distributions on the M\"{o}bius strip: (a1-a2) for the case with $\mathcal{P}=\left\{P_1\right\}$; (b1-b2) for the case with $\mathcal{P}=\left\{P_2\right\}$. $P_1$ and $P_2$ has been sketched in Figure \ref{fig:MoebiusKleinBottleHeat}b.}\label{fig:MoebiusHeat}
\end{figure}

\begin{figure}[!htbp]
  \centering
  \includegraphics[width=0.85\columnwidth]{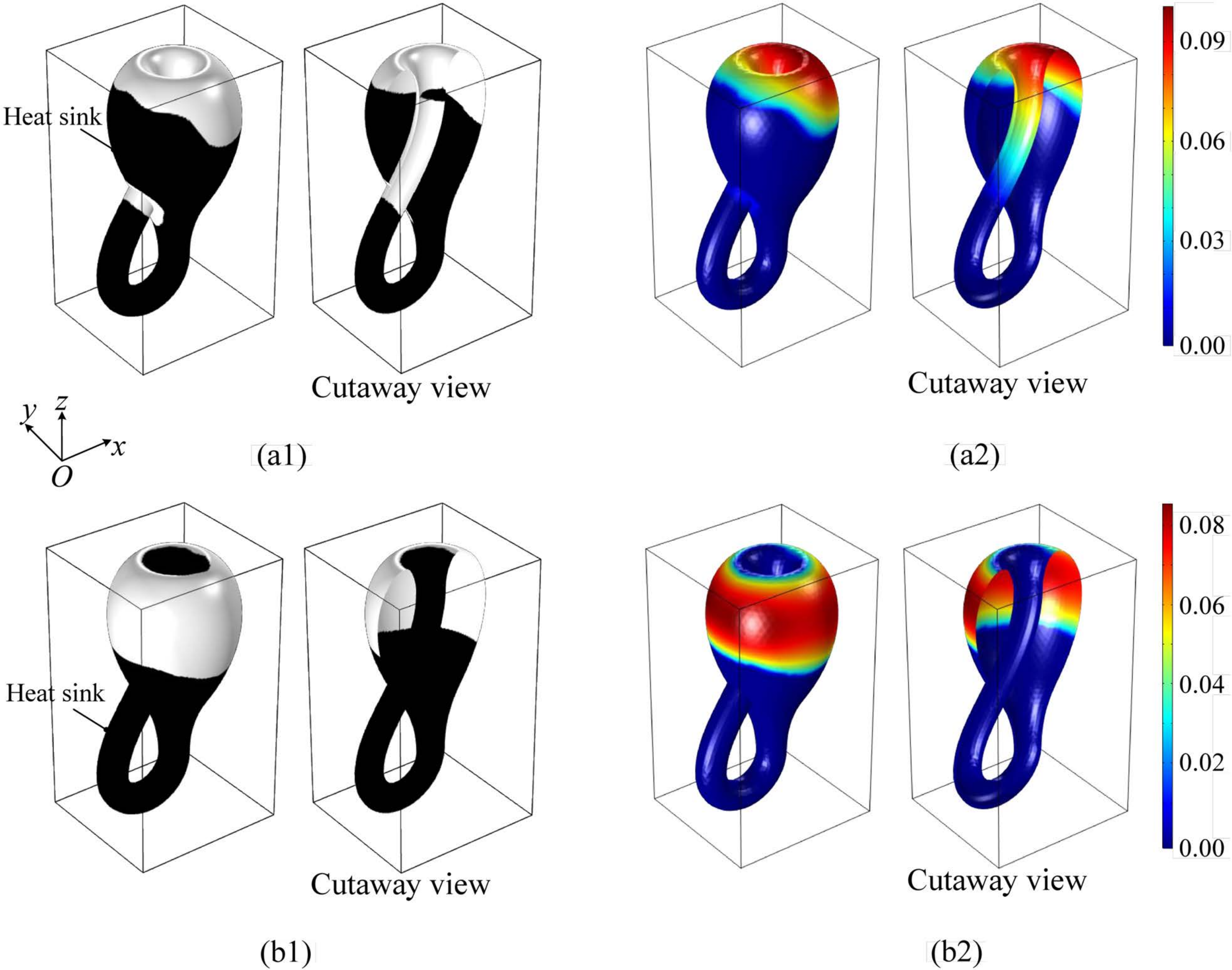}
  \caption{Stereo and cutaway views of the derived pattern of the heat sink and the corresponding temperature distribution on the Klein bottle: (a1-a2) for the case with $\mathcal{P}=\left\{P_3\right\}$; (b1-b2) the case with $\mathcal{P}=\left\{P_4\right\}$. $P_3$ and $P_4$ has been sketched in Figure \ref{fig:MoebiusKleinBottleHeat}b.}\label{fig:KleinBottleHeat}
\end{figure}

The thermal compliance on 2-manifolds can also be minimized by using topology optimization, where the 2-manifolds are imbedded in a block-shaped 3D domain $\Omega$ enclosed by the insulation boundaries and the center of $\Omega$ is localized at the coordinate origin. In this case, the optimization objective is
\begin{equation}\label{equ:ThermalCompliance}
  J = \int_{\Sigma_S} k \nabla_s T \cdot \nabla_s T \,\mathrm{d}s,
\end{equation}
where the 2-manifold $\Sigma_S$ is an interface of $\Omega$. The material interpolation implemented on $\Sigma_S$ is the mixture of the Dirichlet ($T = T_d$) and no-jump ($-k \left \llbracket \nabla T \right \rrbracket \cdot \mathbf{n}=0$) boundary conditions:
\begin{equation}\label{equ:InterpolatedBNDHeatTransfer2ManifoldNoJump}
 -k \left \llbracket \nabla T \right \rrbracket \cdot \mathbf{n} = \alpha\left(\bar{\gamma}\right) \left(T - T_d\right),~\mathrm{on}~\Sigma_S.
\end{equation}
The adjoint equation in equation \ref{equ:AdjEquHeatTransfer2ManifoldT} is changed as
\begin{itemize}
  \item find $T_a\in\mathcal{H}\left(\Omega\right)$ with $T_a = 0$ at $\mathcal{P}$, satisfying
  \begin{equation}\label{equ:AdjEquHeatTransfer2ManifoldTExt}
  \begin{split}
    \int_\Omega k \nabla T_a \cdot \nabla \hat{T}_a \,\mathrm{d}v + \int_{\Sigma_S} 2 k \nabla_s T \cdot \nabla_s \hat{T}_a + \alpha T_a \hat{T}_a \,\mathrm{d}s = 0,~\forall \hat{T}_a \in \mathcal{H}\left(\Omega\right).
  \end{split}
  \end{equation}
\end{itemize}
By setting the area fraction as $0.6$ and keeping the other parameters without change, topology optimization for the patterns of the heat sinks is implemented on the M\"{o}bius strip and Klein bottle which can be derived by gluing two M\"{o}bius strips (Figure \ref{fig:MoebiusKleinBottleHeat}a). Because these two 2-manifolds are non-orientable, the untary normal vector $\mathbf{n}$ is defined locally. The patterns of the heat sinks corresponding to different choices of the point set $\mathcal{P}$ (Figure \ref{fig:MoebiusKleinBottleHeat}b) are derived as shown in Figure \ref{fig:MoebiusHeat} and \ref{fig:KleinBottleHeat}, where the temperature distributions on these 2-manifolds are included.  The derived patterns of the heat sinks change the temperature distribution in the volume domain $\Omega$ with the tendency to reduce or eliminate the temperature gradient on the remained part of the interface $\Sigma_S$. From the temperature distributions, the performance of the derived patterns of the heat sinks on minimizing the thermal compliance can be confirmed.

To check the optimality, the thermal compliance is cross-compared as listed in Table \ref{tab:HeatsinkPatternOptimalityConfirmation} for the derived patterns of the heat sinks in Figure \ref{fig:MoebiusHeat}a1, \ref{fig:MoebiusHeat}b1, \ref{fig:KleinBottleHeat}a1 and \ref{fig:KleinBottleHeat}b1. From a cross comparison of the values in every row of the sub-tables in Table \ref{tab:HeatsinkPatternOptimalityConfirmation}, the optimized performance of the derived patterns can be confirmed.

\begin{table}[!htbp]
\centering
\subtable[M\"{o}bius strip]{
\begin{tabular}{c|cc}
  \toprule
        & Figure \ref{fig:MoebiusHeat}(a1) & Figure \ref{fig:MoebiusHeat}(b1) \\
  \midrule
  $P_1$ & $\mathbf{1.31}$ & $2.58$ \\
  \midrule
  $P_2$ & $2.23$ & $\mathbf{1.21}$ \\
  \bottomrule
\end{tabular}}\hspace{2ex}
\subtable[Klein bottle]{
\begin{tabular}{c|cc}
  \toprule
        & Figure \ref{fig:KleinBottleHeat}(a1) & Figure \ref{fig:KleinBottleHeat}(b1) \\
  \midrule
  $P_3$ & $\mathbf{0.09}$ & $0.34$ \\
  \midrule
  $P_4$ & $0.20$ & $\mathbf{0.14}$ \\
  \bottomrule
\end{tabular}}
\caption{Values of the thermal compliance corresponding to the patterns of the heat sinks on the M\"{o}bius strip and the Klein bottle for the different choices of the point set $\mathcal{P}$. The optimized entries are noted in bold.}\label{tab:HeatsinkPatternOptimalityConfirmation}
\end{table}

\subsection{Perfect conductor for electromagnetics} \label{subsubsec:ElectrodesInElectromagnetics}

In electromagnetics, perfect conductor condition is widely used to approximate a completely conductive and thin metal layer \cite{KrausCarverElectromagnetics1973}. Topology optimization on 2-manifolds can be used to determine the patterns of the perfect conductor by maximizing the scattering energy of an incident field. The electric field scattered by the perfect conductor can be described by the wave equation, where the tangential component of the electric field on the 2-manifold is zero. The scattering field of the perfect conductor is described as
\begin{equation}\label{WaveEquE3D}
\begin{split}
    & \nabla \times \left[ \mu_r^{-1} \nabla \times \left(\mathbf{E}_s+\mathbf{E}_i\right)\right] - k_0^2 \epsilon_r \left( \mathbf{E}_s+\mathbf{E}_i \right) = \mathbf{0},~\mathrm{in}~\Omega \\
    & \nabla \cdot \left[ \epsilon_r \left( \mathbf{E}_s+\mathbf{E}_i \right) \right] = 0,~\mathrm{in}~\Omega
\end{split}
\end{equation}
where $\mathbf{E}_s$ and $\mathbf{E}_i$ are the scattering and incident fields, respectively; the electric field $\mathbf{E}=\mathbf{E}_s+\mathbf{E}_i$ is the total field; the second equation is the divergence-free condition of the electric displacement; $\Omega$ is a cuboid-shaped domain. The infinite computational space is truncated by the perfectly matched layers (PMLs). In the PMLs, the wave equations with the complex-valued coordinate transformation is described as \cite{JinWiley2002,BerengerJCP1994}
\begin{equation}\label{WaveEquE3DPML}
\begin{split}
    & \nabla_\mathbf{x'} \times \left( \mu_r^{-1} \nabla_\mathbf{x'} \times \mathbf{E}_s \right) - k_0^2 \epsilon_r \mathbf{E}_s = \mathbf{0},~\mathrm{in}~\Omega_P\\
    & \nabla \cdot \mathbf{E}_s = 0,~\mathrm{in}~\Omega_P
\end{split}
\end{equation}
where $\mathbf{x}'$ is the complex-valued coordinate transformed from the original Cartesian coordinate in $\Omega_P$; $\mathbf{E}_s$ in the PMLs satisfy the divergence-free condition described in the original Cartesian coordinate system, because the fields are source-free in the PMLs; $\nabla_\mathbf{x'}$ is the gradient operator in the PMLs with transformed coordinates; $\Omega_P$ is the union of the PML domains. The transformed coordinates and the original Cartesian coordinates satisfy the following transformation:
\begin{equation}\label{CoordTransf}
\begin{split}
    \mathbf{x}' = \mathbf{T} \mathbf{x},~\forall\mathbf{x}\in\Omega_P
\end{split}
\end{equation}
where $\mathbf{T}$ and $\mathbf{x}$ are the transformation matrix and the original Cartesian coordinate, respectively. The transformation matrix is \cite{JinWiley2002}
\begin{equation}\label{TransformationMatrix}
\begin{split}
    \mathbf{T} = \left\{
    \begin{split}
    & \mathrm{diag}\left(\left(1-j\right)\lambda/d,1,1\right),~\mathrm{in}~\Omega_{fb}\\
    & \mathrm{diag}\left(1,\left(1-j\right)\lambda/d,1\right),~\mathrm{in}~\Omega_{lr}\\
    & \mathrm{diag}\left(1,1,\left(1-j\right)\lambda/d\right),~\mathrm{in}~\Omega_{td}\\
    & \mathrm{diag}\left(1,\left(1-j\right)\lambda/d,\left(1-j\right)\lambda/d\right), ~\mathrm{in}~\Omega_{fbe} \\
    & \mathrm{diag}\left(\left(1-j\right)\lambda/d,1,\left(1-j\right)\lambda/d\right), ~\mathrm{in}~\Omega_{lre}\\
    & \mathrm{diag}\left(\left(1-j\right)\lambda/d,\left(1-j\right)\lambda/d,1\right), ~\mathrm{in}~\Omega_{tde}\\
    & \mathrm{diag}\left(\left(1-j\right)\lambda/d,\left(1-j\right)\lambda/d, \left(1-j\right)\lambda/d\right), ~\mathrm{in}~\Omega_c\\
    \end{split}
    \right.
\end{split}
\end{equation}
where $\lambda$ the wavelength of the incident wave; $d$ the thickness of the PMLs; $\Omega_{fb}$, $\Omega_{lr}$ and $\Omega_{td}$ are the PMLs attached on the surfaces of $\Omega$ with normal vector parallel to $x$, $y$ and $z$ axes, respectively; $\Omega_{fbe}$, $\Omega_{lre}$ and $\Omega_{tde}$ are the PMLs attached on the edges of $\Omega$ with tangential vectors perpendicular to $yOz$, $zOx$ and $xOy$ planes, respectively; $\Omega_{c}$ are the PMLs attached on the vertexes of $\Omega$. The no-jump boundary condition for the scattering field is imposed on the interface $\partial\Omega$ between $\Omega_P$ and $\Omega$:
\begin{equation}\label{JumpBNDPML}
\begin{split}
     & \mu_r^{-1} \left(\nabla \times \mathbf{E}_s - \nabla_\mathbf{x'} \times \mathbf{E}_s\right) \times \mathbf{n} = \mathbf{0},~\mathrm{on}~\partial\Omega.
\end{split}
\end{equation}
The perfect electric conductor condition $\mathbf{n}\times\mathbf{E}_s = \mathbf{0}$ is imposed on the external boundaries $\Gamma_D = \partial \left(\Omega\cup\Omega_P\right)$ of the PMLs.

The perfect conductor layer is attached on the 2-manifold $\Sigma_S$ immersed in $\Omega$. The design variable is used to indicate the no-jump boundary ($\mu_r^{-1} \left \llbracket \nabla \times \mathbf{E} \right\rrbracket \times \mathbf{n}=\mathbf{0}$) and perfect conductor ($\mathbf{n} \times \mathbf{E} = \mathbf{0}$) parts of $\Sigma_S$. The corresponding material interpolation is implemented as
\begin{equation}\label{equ:InterpolatedBNDElectromagnetics2Manifold}
\begin{split}
     & \mu_r^{-1} \left \llbracket \nabla \times \left( \mathbf{E}_s + \mathbf{E}_i \right) \right\rrbracket \times \mathbf{n} = \alpha\left(\bar{\gamma}\right) \left[  \mathbf{n} \times \left(\mathbf{E}_s+\mathbf{E}_i\right) \right] ,~\mathrm{on}~\Sigma_S
\end{split}
\end{equation}
where $\alpha\left(\bar{\gamma}\right)$ is the penalization function in equation \ref{equ:MaterialInterpolationHeatTransfer2Manifold}. In this penalization function, $\alpha_{\max}$ is chosen as a large but finite value to ensure the domination of the term $\mathbf{n} \times \left(\mathbf{E}_s+\mathbf{E}_i\right)$ corresponding to perfect conductor boudanry, when the material density takes the value of $0$; equation \ref{equ:InterpolatedBNDElectromagnetics2Manifold} degenerates into the no-jump boundary condition, when the material density takes the value of $1$. Based on numerical tests, $\alpha_{\max}=1\times10^4$ and $q=1\times10^0$ are chosen in this section.
The perfect conductor layer is optimized to maximize the energy of the scattering field:
\begin{equation}\label{equ:ObjElctromagnetics}
  J = \int_\Omega \mathbf{E}_s \cdot \mathbf{E}_s^* \,\mathrm{d}\Omega.
\end{equation}
Based on the adjoint analysis introduced in \cite{DengJCP2018}, the adjoint derivative is derived with the same form as that in equation \ref{equ:AdjDrvHeatTransfer2ManifoldJ}. The adjoint variables are derived by solving the adjoint equations in the following variational formulations:
\begin{itemize}
  \item find $\mathbf{E}_{sa}$ with $Re\left(\mathbf{E}_{sa}\right) \in \mathcal{V}_{\mathbf{E}}$, $Im\left(\mathbf{E}_{sa}\right) \in \mathcal{V}_{\mathbf{E}}$ and $\mathbf{n}\times\mathbf{E}_{sa}=\mathbf{0}$ on $\Gamma_D$, satisfying
\begin{equation}\label{CombinedWeakAdjTransformedTOOPProblem3DEs}
\begin{split}
& \int_\Omega 2 \mathbf{E}_s^* \cdot \hat{\mathbf{E}}_{sa} + \mu_r^{-1} \left(\nabla \times \mathbf{E}_{sa}^* \right) \cdot \left(\nabla \times \hat{\mathbf{E}}_{sa} \right) - k_0^2 \epsilon_r \mathbf{E}_{sa}^* \cdot \hat{\mathbf{E}}_{sa} \,\mathrm{d}\Omega + \\
& \int_{\Omega_P} \mu_r^{-1} \left( \mathbf{T} \nabla \times \mathbf{E}_{sa}^* \right) \cdot \left( \mathbf{T} \nabla \times \hat{\mathbf{E}}_{sa} \right) \left|\mathbf{T}\right|^{-1} - k_0^2 \epsilon_r \mathbf{E}_{sa}^* \cdot \hat{\mathbf{E}}_{sa} \left|\mathbf{T}\right| \,\mathrm{d}\Omega = 0, \forall \hat{\mathbf{E}}_{sa} \in \mathcal{V}_{\mathbf{E}}
\end{split}
\end{equation}
  \item find $\tilde{\gamma}_a \in \mathcal{H}\left(\Sigma_s\right)$ satisfying
\begin{equation}\label{WeakAdjTransformedTOOPProblem3DEsFilter}
\begin{split}
& \int_{\Sigma_s} r^2\nabla \tilde{\gamma}_a \cdot \nabla \hat{\tilde{\gamma}}_a + \tilde{\gamma}_a \hat{\tilde{\gamma}}_a + {\partial \alpha \over \partial \bar{\gamma} } {\partial \bar{\gamma} \over \partial \tilde{\gamma} } \big[ Re\left( \mathbf{n} \times \left(\mathbf{E}_s+\mathbf{E}_i\right) \right) \cdot Re\left( \mathbf{n} \times \mathbf{E}_{sa}^* \right) + \\
& Im\left( \mathbf{n} \times \left(\mathbf{E}_s+\mathbf{E}_i\right) \right) \cdot Im\left( \mathbf{n} \times \mathbf{E}_{sa}^* \right) \big] \hat{\tilde{\gamma}}_a \,\mathrm{d}\Sigma = 0, ~ \forall \hat{\tilde{\gamma}}_a \in\mathcal{H}\left(\Sigma_s\right)
\end{split}
\end{equation}
\end{itemize}
where $Re$ and $Im$ are operators used to extract the real and imaginary parts of a complex; $\mathbf{E}_{sa}$ is the adjoint variables of $\mathbf{E}_s$; $\mathcal{V}_{\mathbf{E}}$ is the functional space $\big\{\mathbf{u}\in\mathcal{H}\left(\mathrm{curl};\Omega\cup\Omega_P\right) \big| \nabla \cdot \mathbf{u} = 0 \big\}$ with $\mathcal{H} \left( \mathrm{curl}; \Omega\cup\Omega_P \right) = \big\{\mathbf{u}\in\left(\mathcal{L}^2\left(\Omega\cup\Omega_P\right)\right)^3 \big| \nabla \times\mathbf{u}\in\left(\mathcal{L}^2\left(\Omega\cup\Omega_P\right)\right)^3 \big\}$ and  $\mathcal{L}^2\left(\Omega\cup\Omega_P\right)$ denoting the second order Lebesque space for the real functionals defined on $\Omega\cup\Omega_P$.

\begin{figure}[!htbp]
  \centering
  \subfigure[Torus]
  {\includegraphics[width=0.32\columnwidth]{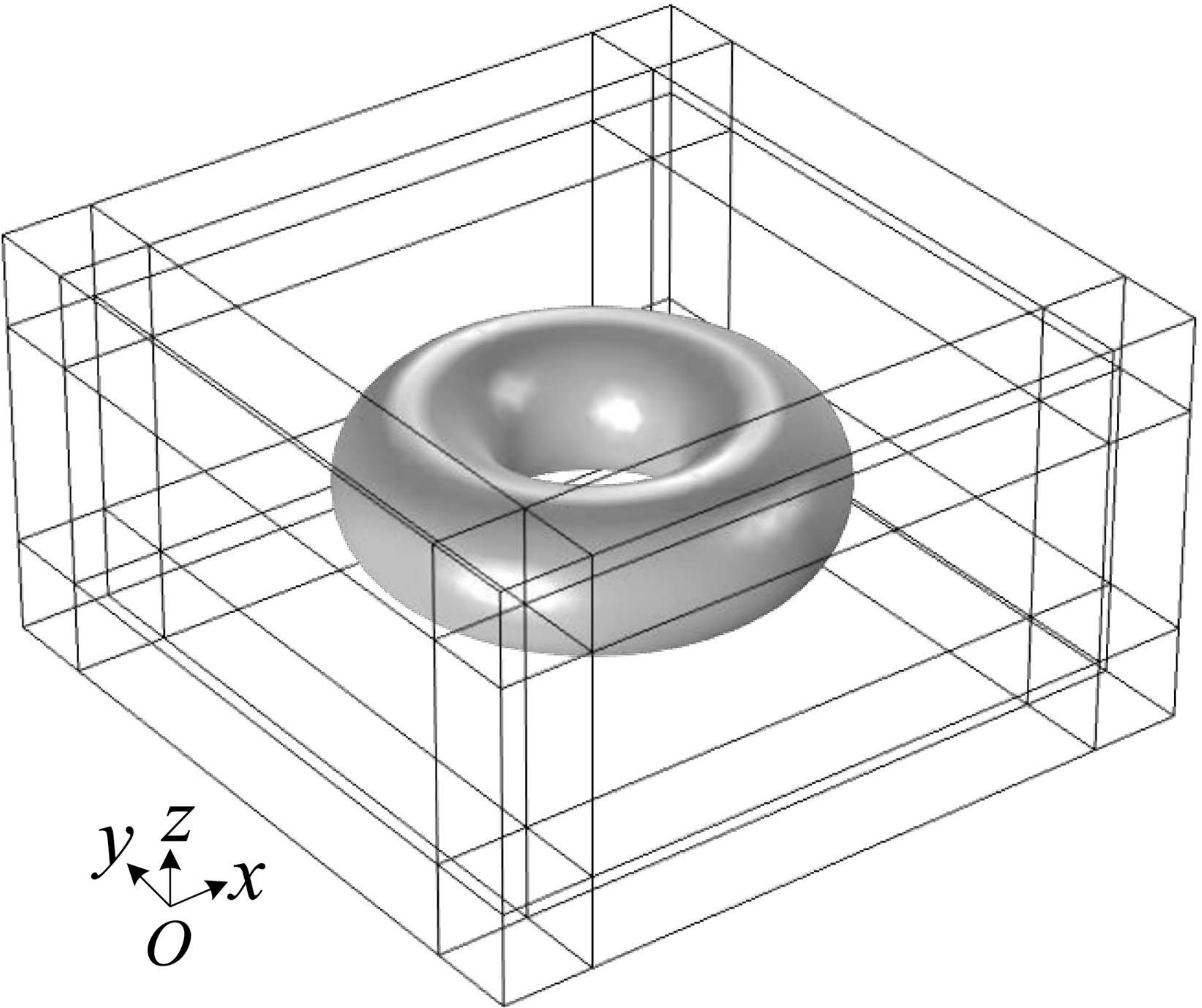}}\hspace{3ex}
  \subfigure[M\"{o}bius strip]
  {\includegraphics[width=0.32\columnwidth]{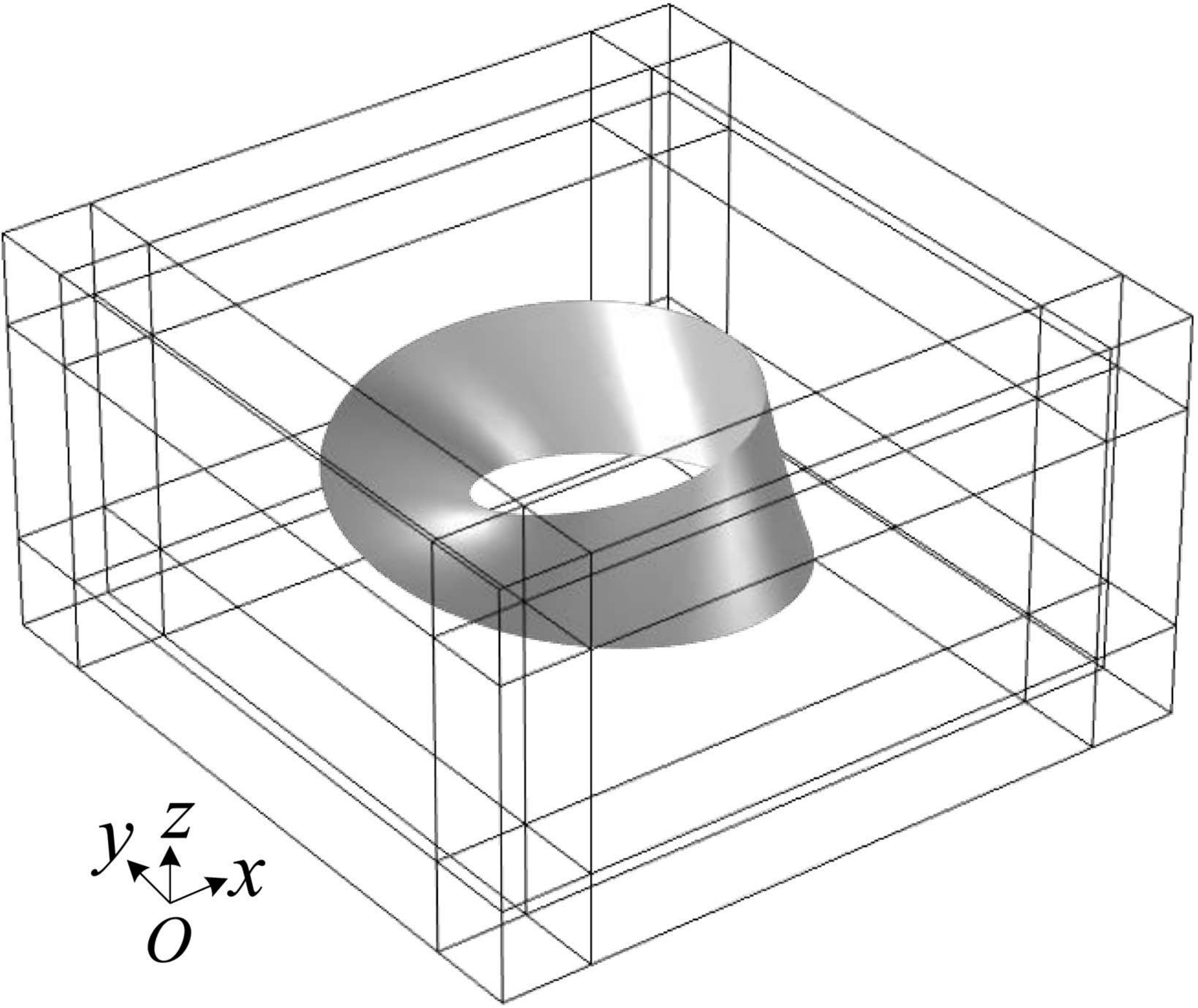}}\\
  \subfigure[$\mathcal{P}=\left\{P_1,P_2\right\}$ on torus]
  {\includegraphics[height=0.15\columnwidth]{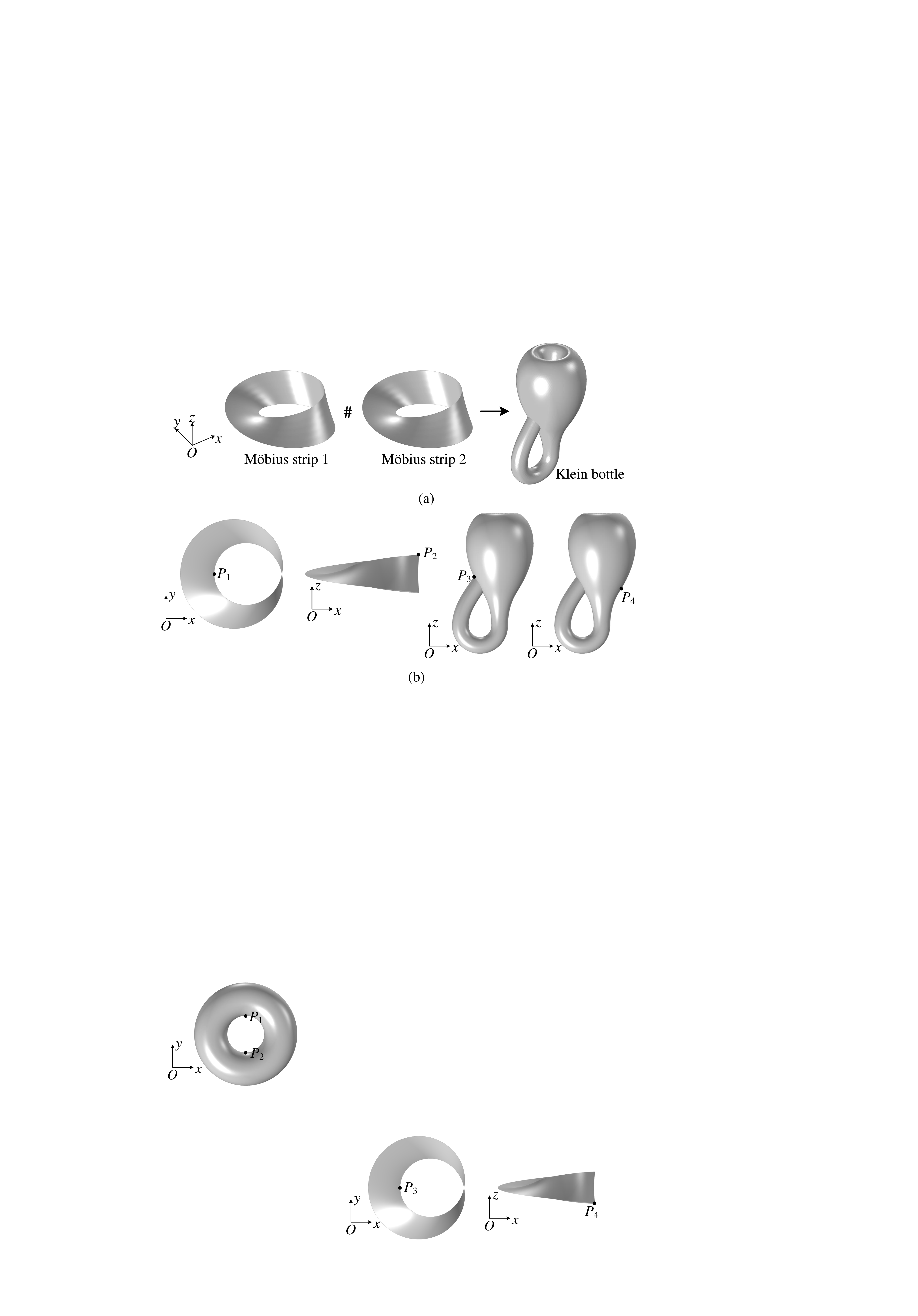}}\hspace{6ex}
  \subfigure[$\mathcal{P}=\left\{P_3,P_4\right\}$ on M\"{o}bius strip]
  {\includegraphics[height=0.15\columnwidth]{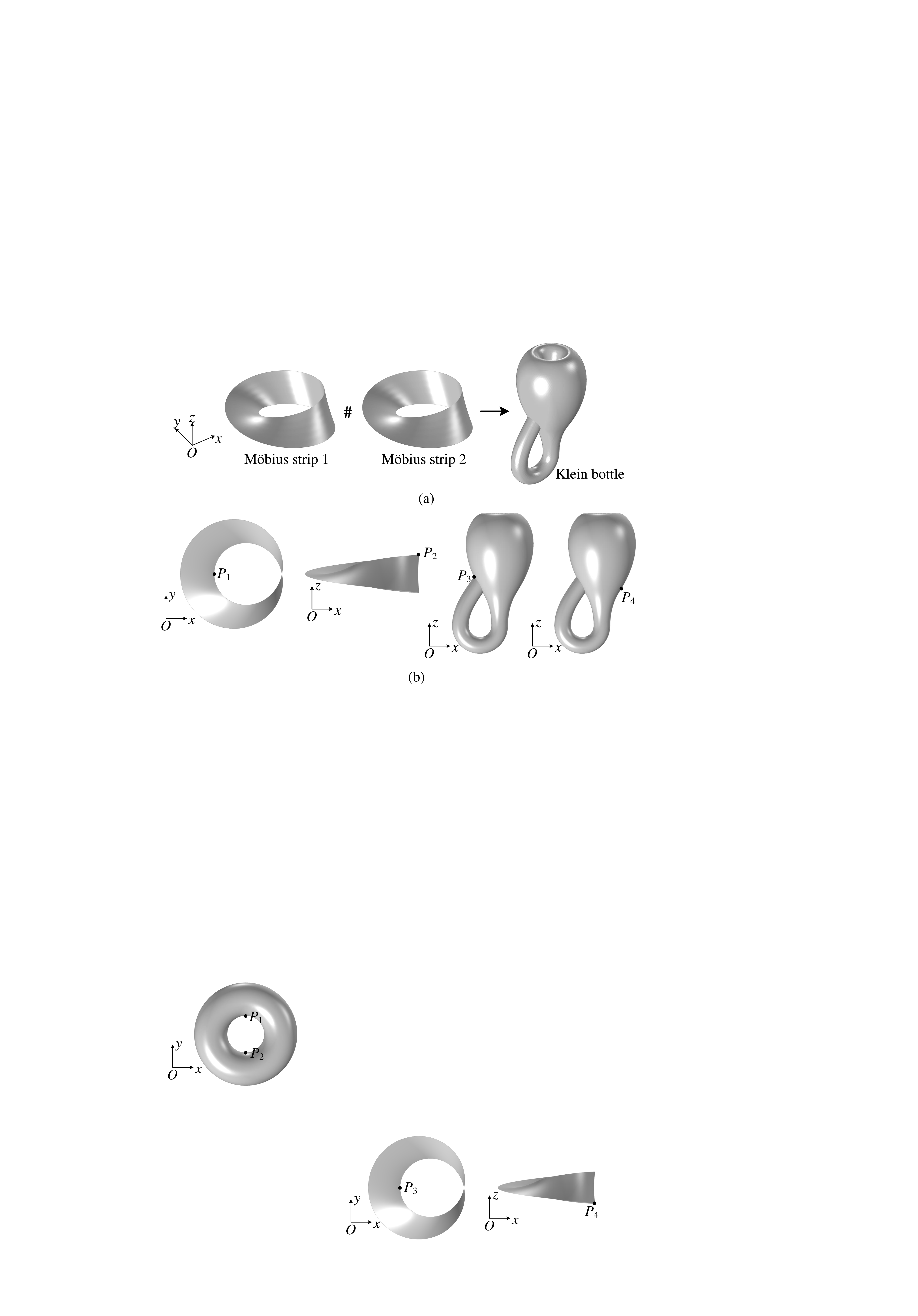}}
  \caption{(a-b) Sketches of the torus and M\"{o}bius strip immersed in the volume domain $\Omega$; (c-d) sketches of the point set $\mathcal{P}$ on the torus and the M\"{o}bius strip.}\label{fig:GeomEM}
\end{figure}

\begin{figure}[!htbp]
  \centering
  \subfigure[$x$-polarization]
  {\includegraphics[width=0.32\columnwidth]{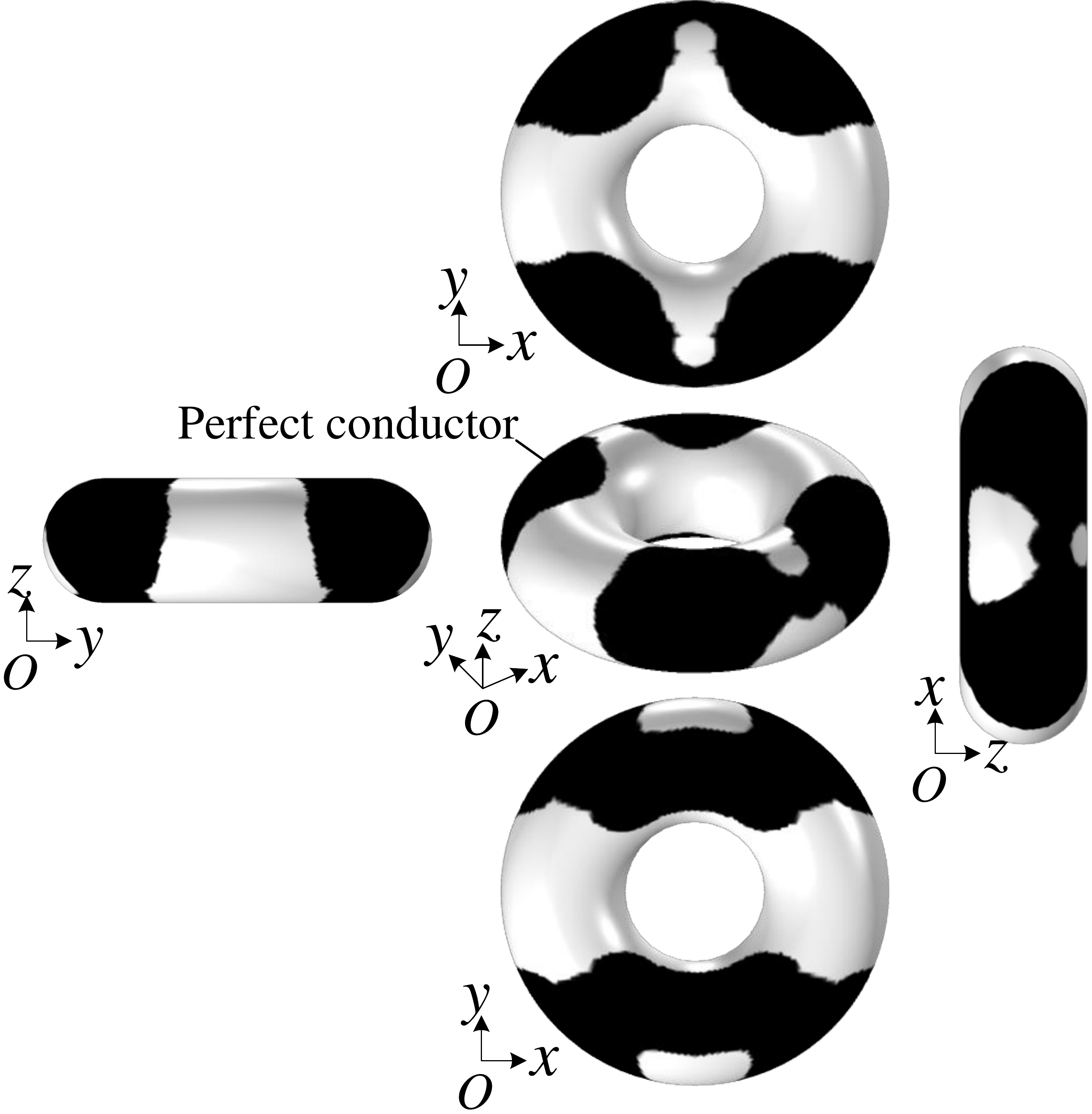}}\hspace{1ex}
  \subfigure[$y$-polarization]
  {\includegraphics[width=0.32\columnwidth]{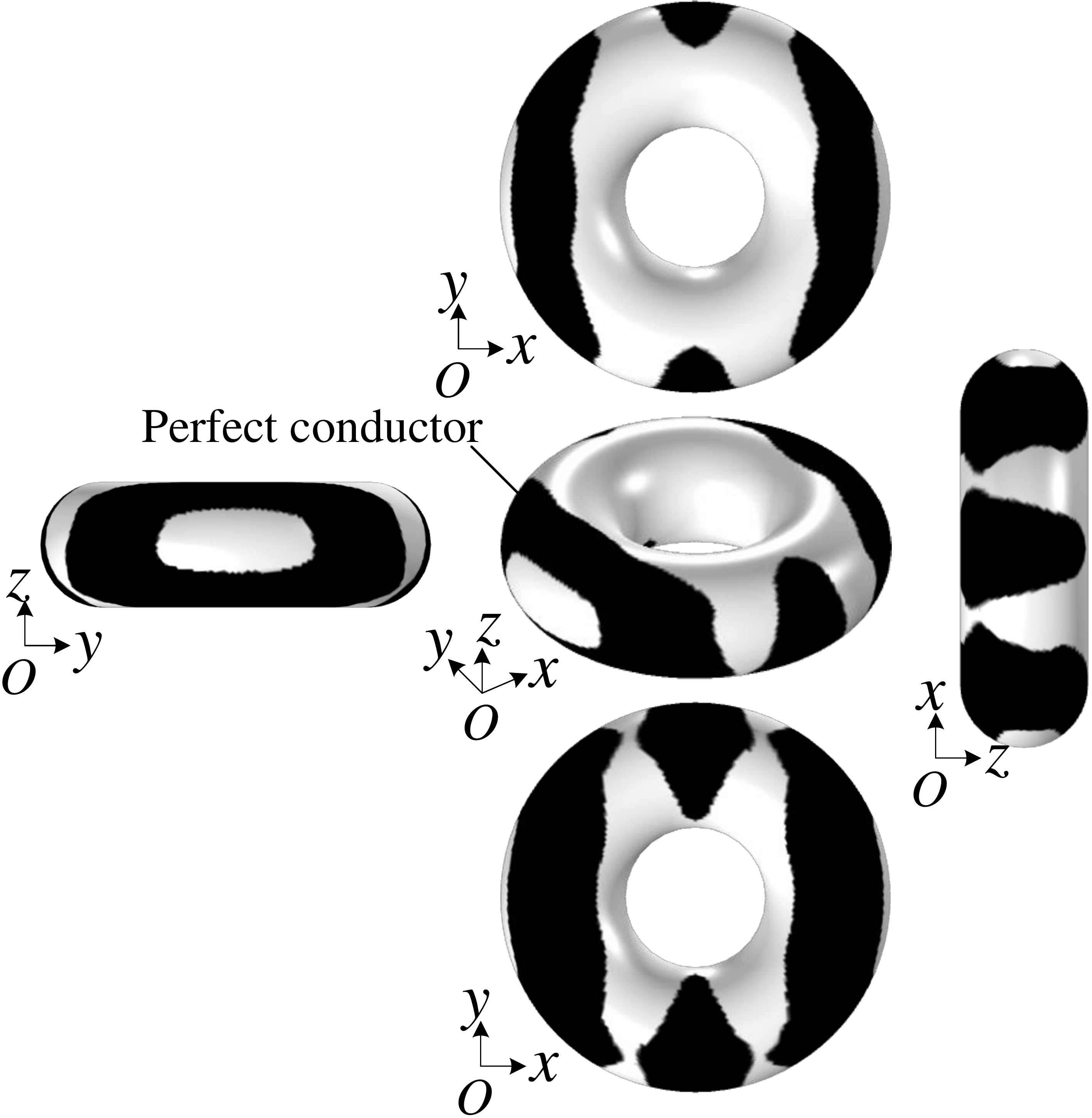}}\hspace{1ex}
  \subfigure[Circular polarization]
  {\includegraphics[width=0.32\columnwidth]{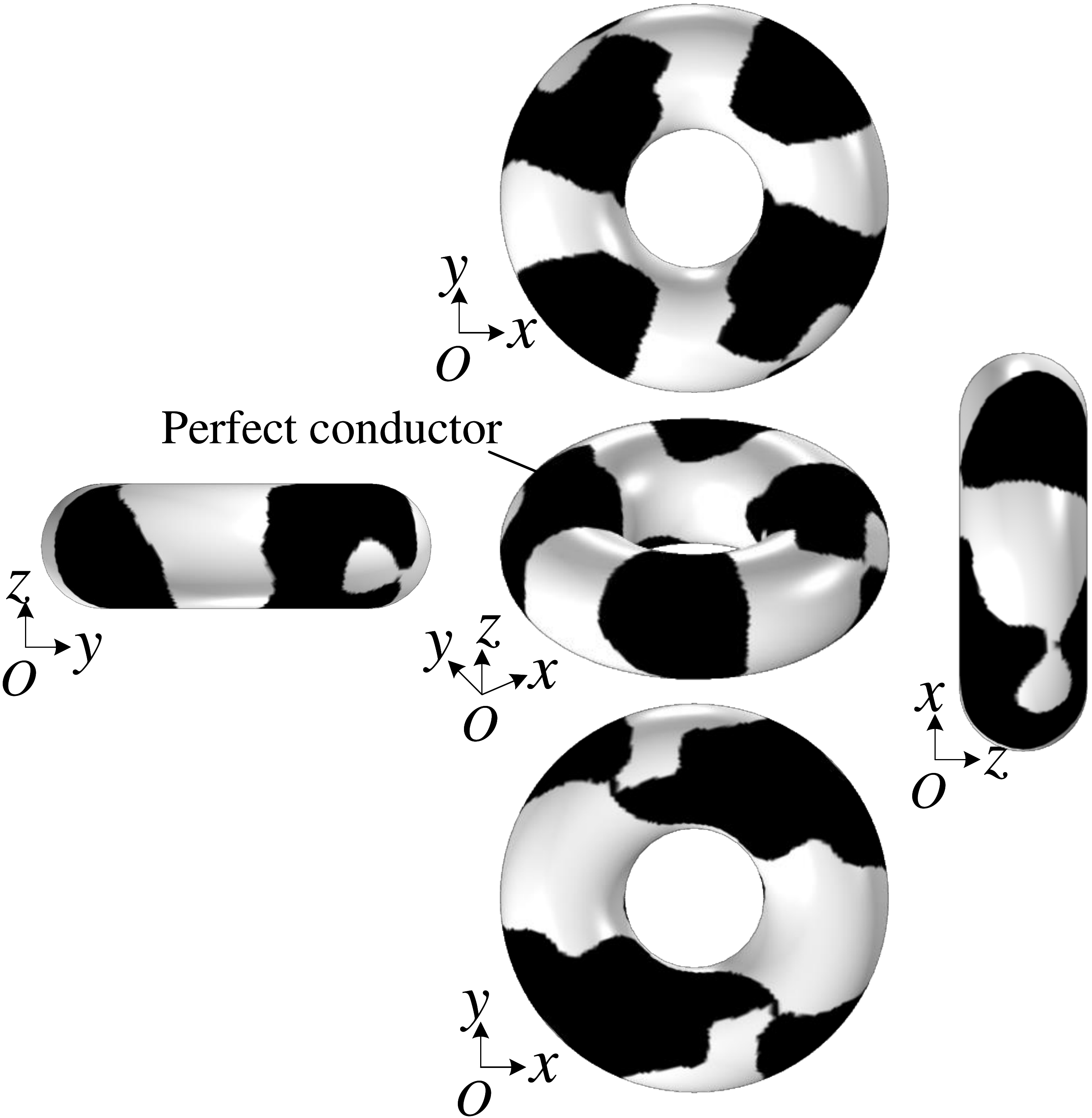}}
  \caption{Derived patterns of the perfect conductor on the torus: (a) for the incident wave polarized in the $x$-axis; (b) for the incident wave polarized in the $y$-axis; (c) for the incident wave right-circularly polarized in the $xOy$-plane.}\label{fig:TorusEM}
\end{figure}

\begin{figure}[!htbp]
  \centering
  \subfigure[Iteration 1]
  {\includegraphics[height=0.115\columnwidth]{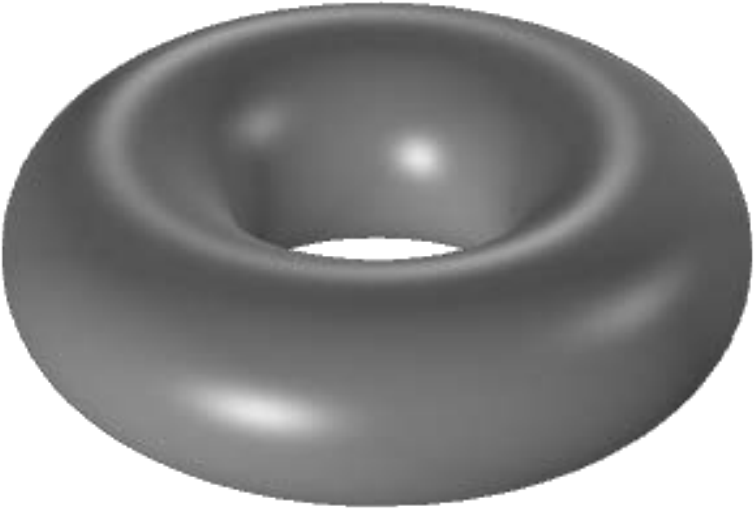}}\hspace{1ex}
  \subfigure[Iteration 5]
  {\includegraphics[height=0.115\columnwidth]{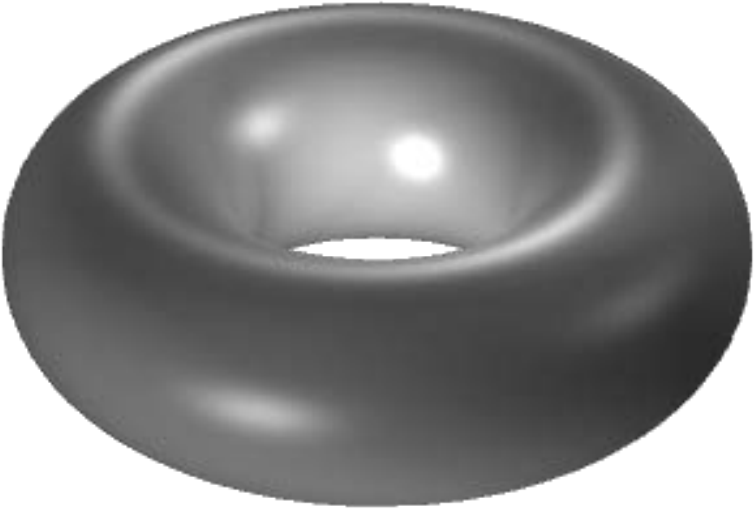}}\hspace{1ex}
  \subfigure[Iteration 20]
  {\includegraphics[height=0.115\columnwidth]{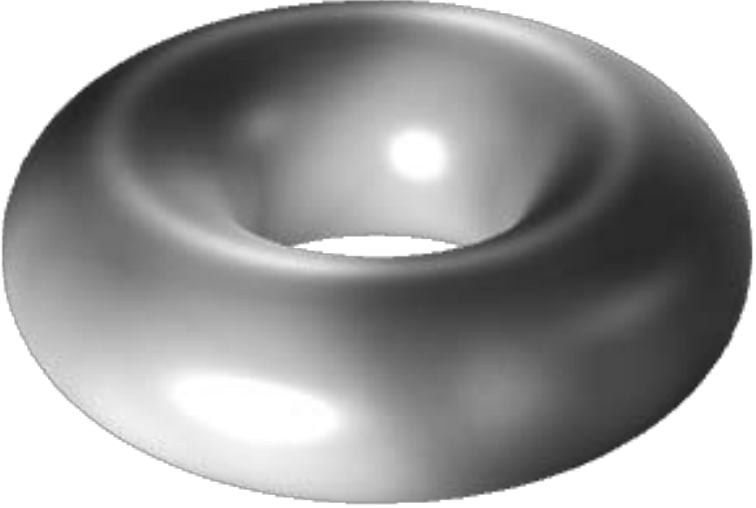}}\hspace{1ex}
  \subfigure[Iteration 70]
  {\includegraphics[height=0.115\columnwidth]{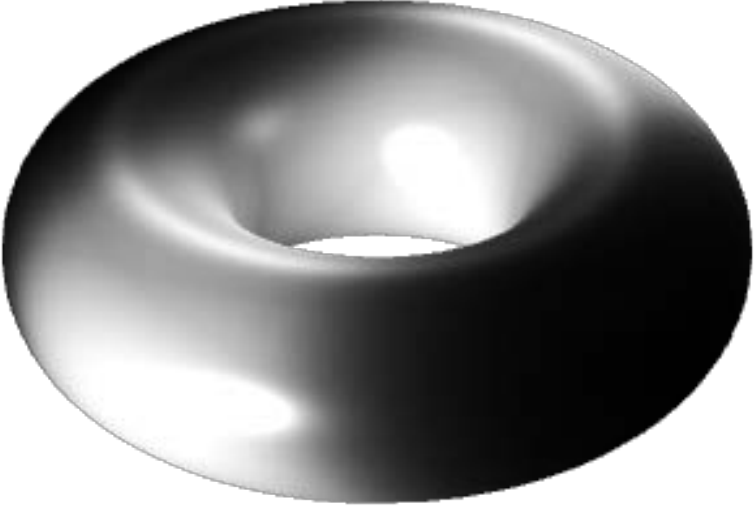}}\hspace{1ex}
  \subfigure[Iteration 100]
  {\includegraphics[height=0.115\columnwidth]{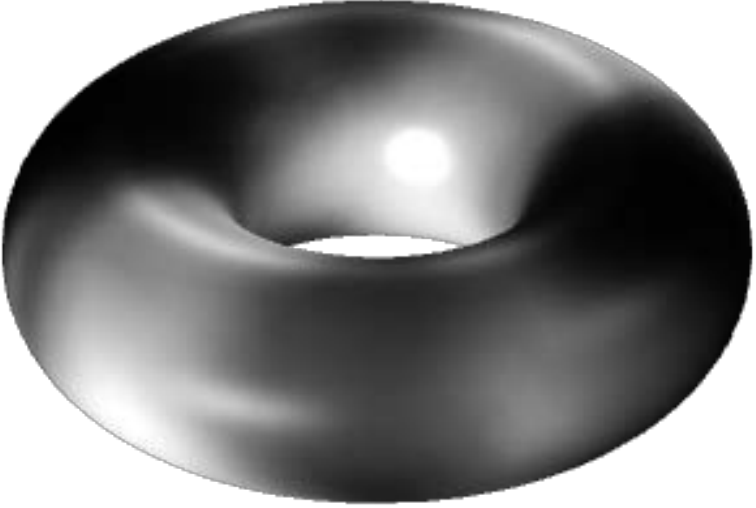}\hspace{1ex}
   \includegraphics[height=0.115\columnwidth]{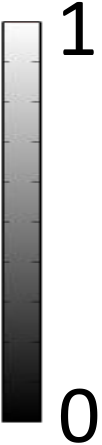}}\\
  \subfigure[Iteration 110]
  {\includegraphics[height=0.115\columnwidth]{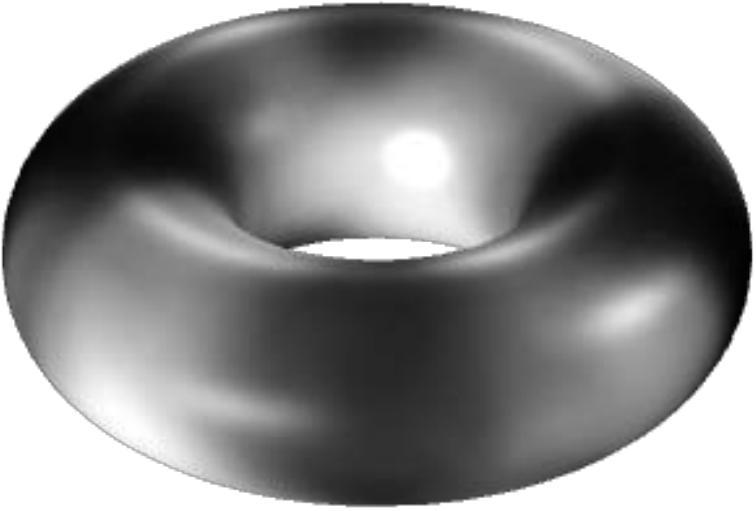}}\hspace{1ex}
  \subfigure[Iteration 150]
  {\includegraphics[height=0.115\columnwidth]{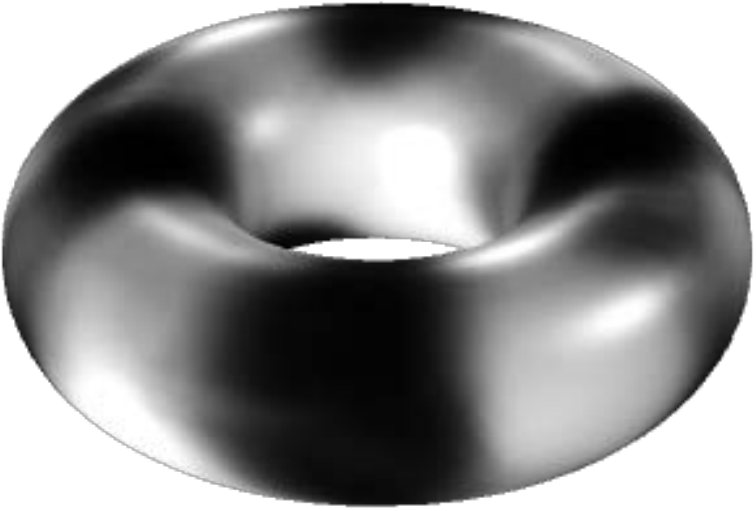}}\hspace{1ex}
  \subfigure[Iteration 200]
  {\includegraphics[height=0.115\columnwidth]{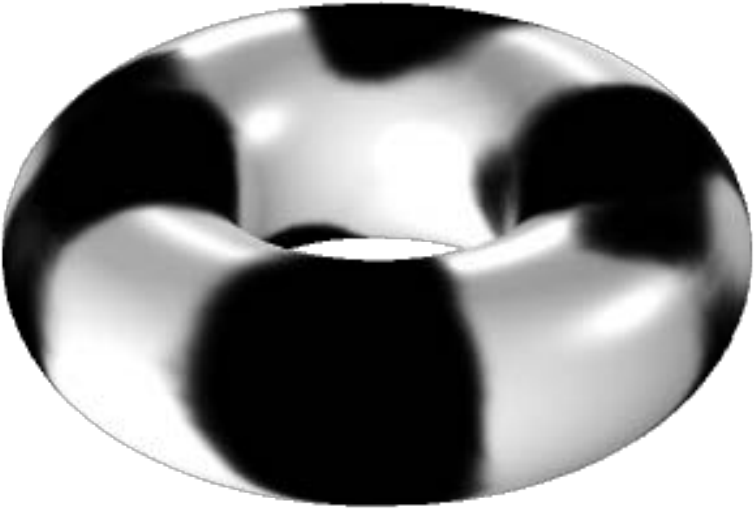}}\hspace{1ex}
  \subfigure[Iteration 260]
  {\includegraphics[height=0.115\columnwidth]{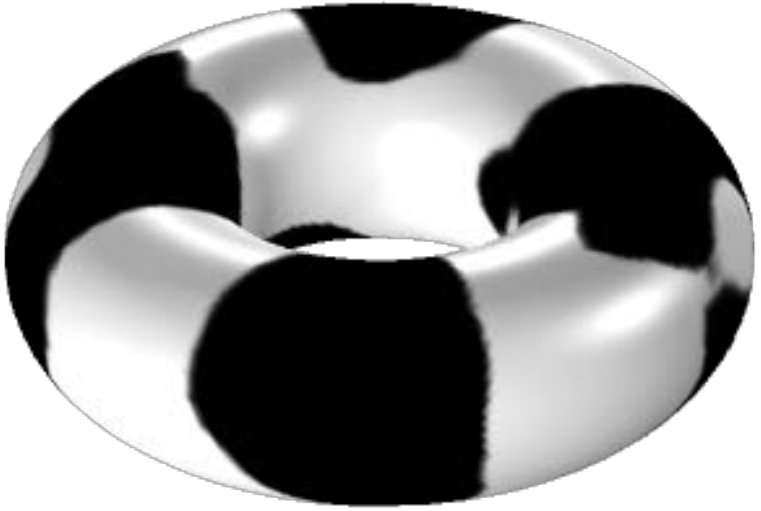}}\hspace{1ex}
  \subfigure[Iteration 310]
  {\includegraphics[height=0.115\columnwidth]{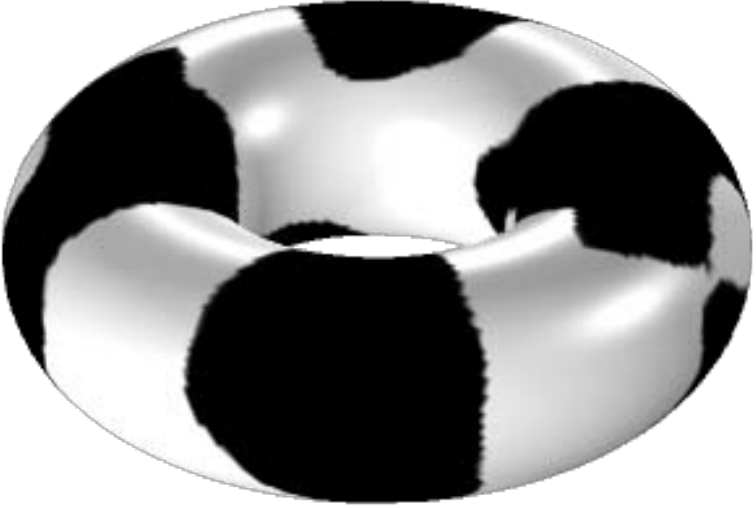}\hspace{1ex}
   \includegraphics[height=0.115\columnwidth]{Figures/LegBarTop.pdf}}\\
  \caption{Snapshots for the evolution of the material density in the topology optimization of the patterns of the perfect conductor on the torus, where the incident wave is right-circularly polarized in the $xOy$-plane.}\label{fig:TorusEMSnapshots}
\end{figure}

\begin{figure}[!htbp]
  \centering
  \subfigure[Iteration 1]
  {\includegraphics[height=0.12\columnwidth]{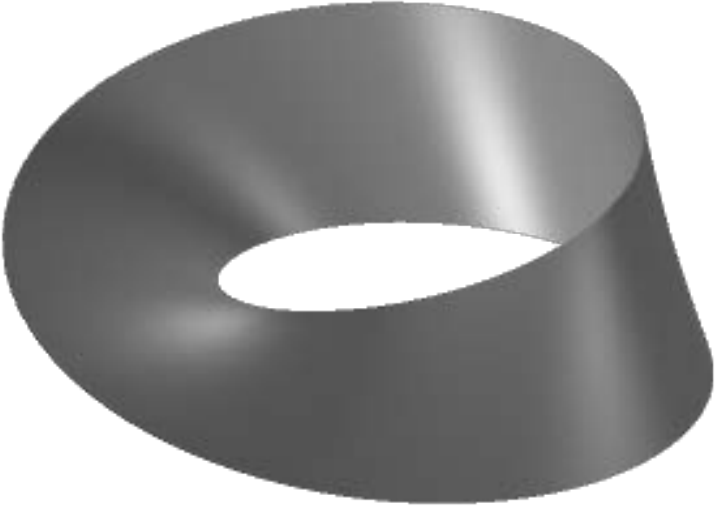}}\hspace{1ex}
  \subfigure[Iteration 5]
  {\includegraphics[height=0.12\columnwidth]{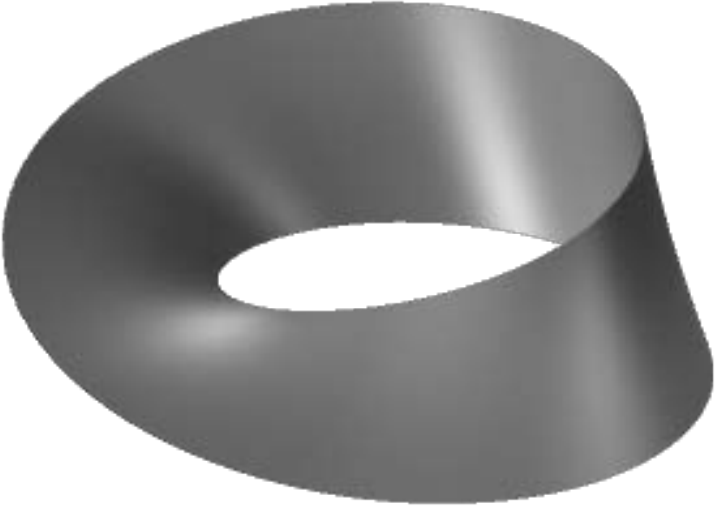}}\hspace{1ex}
  \subfigure[Iteration 20]
  {\includegraphics[height=0.12\columnwidth]{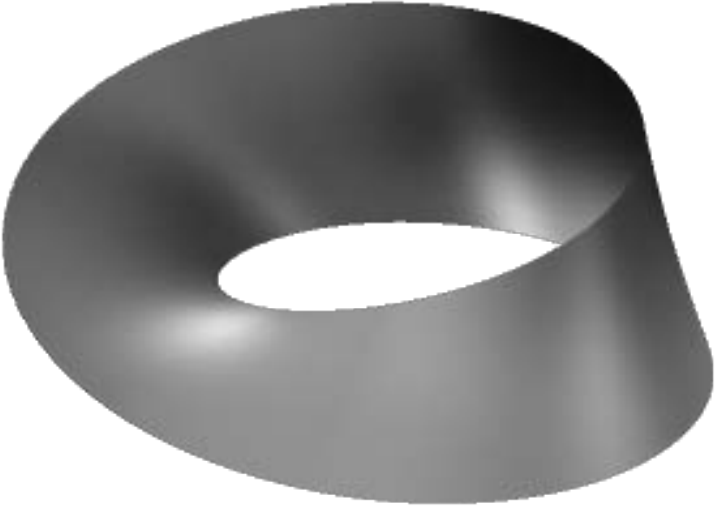}}\hspace{1ex}
  \subfigure[Iteration 40]
  {\includegraphics[height=0.12\columnwidth]{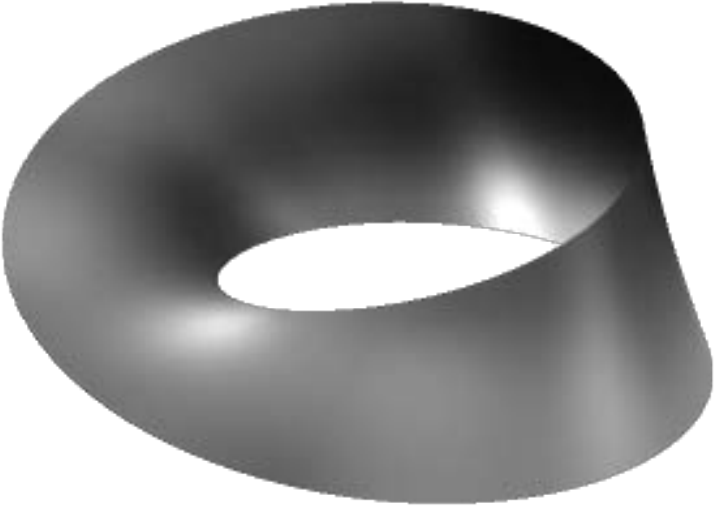}}\hspace{1ex}
  \subfigure[Iteration 70]
  {\includegraphics[height=0.12\columnwidth]{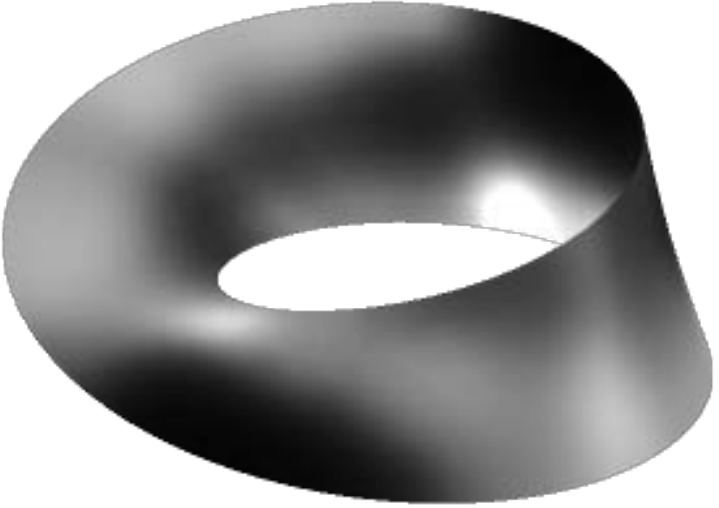}\hspace{1ex}
   \includegraphics[height=0.12\columnwidth]{Figures/LegBarTop.pdf}}\\
  \subfigure[Iteration 110]
  {\includegraphics[height=0.12\columnwidth]{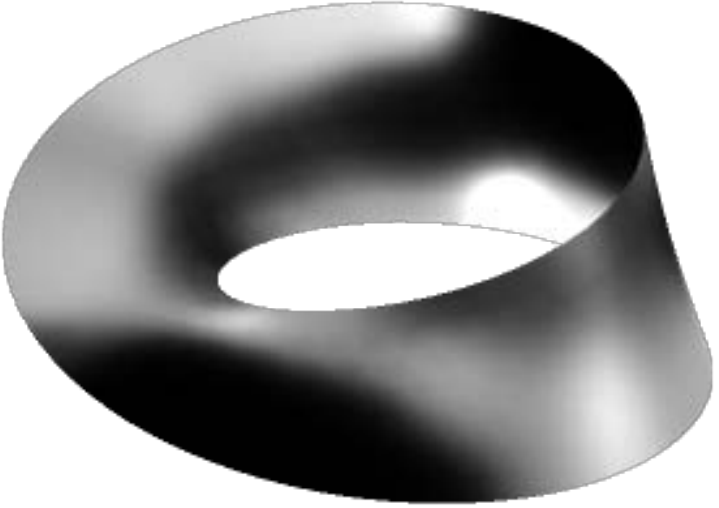}}\hspace{1ex}
  \subfigure[Iteration 150]
  {\includegraphics[height=0.12\columnwidth]{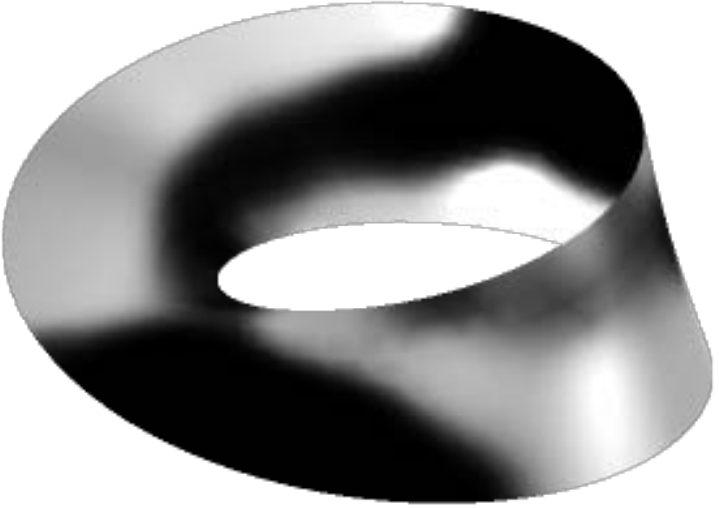}}\hspace{1ex}
  \subfigure[Iteration 200]
  {\includegraphics[height=0.12\columnwidth]{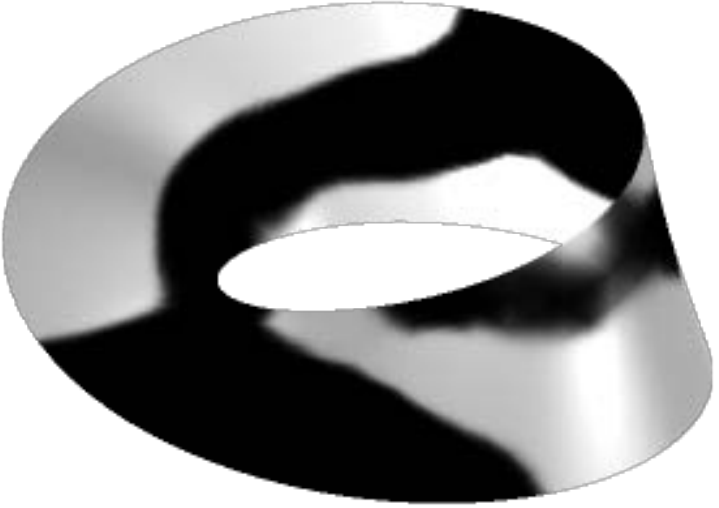}}\hspace{1ex}
  \subfigure[Iteration 260]
  {\includegraphics[height=0.12\columnwidth]{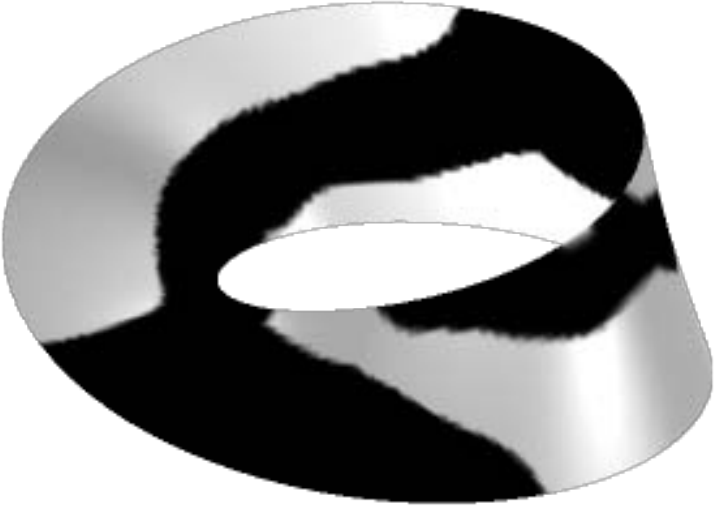}}\hspace{1ex}
  \subfigure[Iteration 310]
  {\includegraphics[height=0.12\columnwidth]{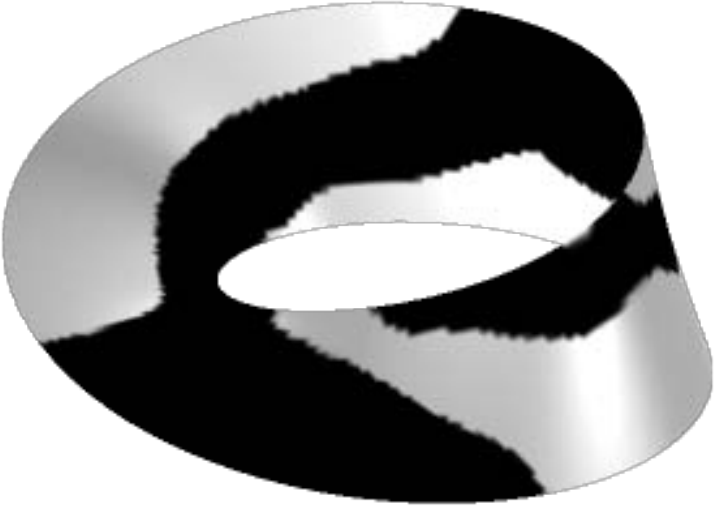}\hspace{1ex}
   \includegraphics[height=0.12\columnwidth]{Figures/LegBarTop.pdf}}\\
  \caption{Snapshots for the evolution of the material density in the topology optimization of the patterns of the perfect conductor on the M\"{o}bius strip, where the incident wave is right-circularly polarized in the $xOy$-plane.}\label{fig:MoebiusEMSnapshots}
\end{figure}

Based on the numerical implementation in Section \ref{sec:NumericalImplementation}, topology optimization for the patterns of the perfect conductor is implemented on the torus and M\"{o}bius strip (Figure \ref{fig:GeomEM}a and \ref{fig:GeomEM}b). These 2-manifolds are immersed into the volume domain $\Omega$, a brick enclosed by the PMLs. The size of $\Omega$ is $2.4\times2.4\times0.96$ in the unit of $\mathrm{m}$. This domain is discretized by the cubic elements with a size of $0.08$. The coordinate origin is set as the center of $\Omega$. For these two 2-manifolds, the area fraction in the area constraint is set as $0.5$. The point set $\mathcal{P}$ is set as shown in Figure \ref{fig:GeomEM}c and \ref{fig:GeomEM}d. The maximal iteration number and the updating interval are set as $n^{sub}_{max}=315$ and $n^{upt}=30$, respectively. The incident waves is set to propagate in the $+z$ with a wavelength of $0.8$m. For three different polarizations of the incident waves, the optimized patterns of the perfect conductor are derived as shown in Figure \ref{fig:TorusEM} and \ref{fig:MoebiusEM}. The evolution snapshots for the material density are shown in Figure \ref{fig:TorusEMSnapshots} and \ref{fig:MoebiusEMSnapshots} for the incident waves with circular polarization. From the evolution snapshots, the performance of the topology optimization procedures can be confirmed for this electromagnetic problem.

\begin{figure}[!htbp]
  \centering
  \subfigure[$x$-polarization]
  {\includegraphics[width=0.32\columnwidth]{Figures/MoebiusEMx.pdf}}\hspace{1ex}
  \subfigure[$y$-polarization]
  {\includegraphics[width=0.32\columnwidth]{Figures/MoebiusEMy.pdf}}\hspace{1ex}
  \subfigure[Circular polarization]
  {\includegraphics[width=0.32\columnwidth]{Figures/MoebiusEMxy.pdf}}
  \caption{Derived patterns of the perfect conductor on the M\"{o}bius strip: (a) for the incident wave polarized in the $x$-axis; (b) for the incident wave polarized in the $y$-axis; (c) for the incident wave right-circularly polarized in the $xOy$-plane.}\label{fig:MoebiusEM}
\end{figure}

\begin{figure}[!htbp]
  \centering
  \subfigure[$x$-polarization]
  {\includegraphics[width=0.32\columnwidth]{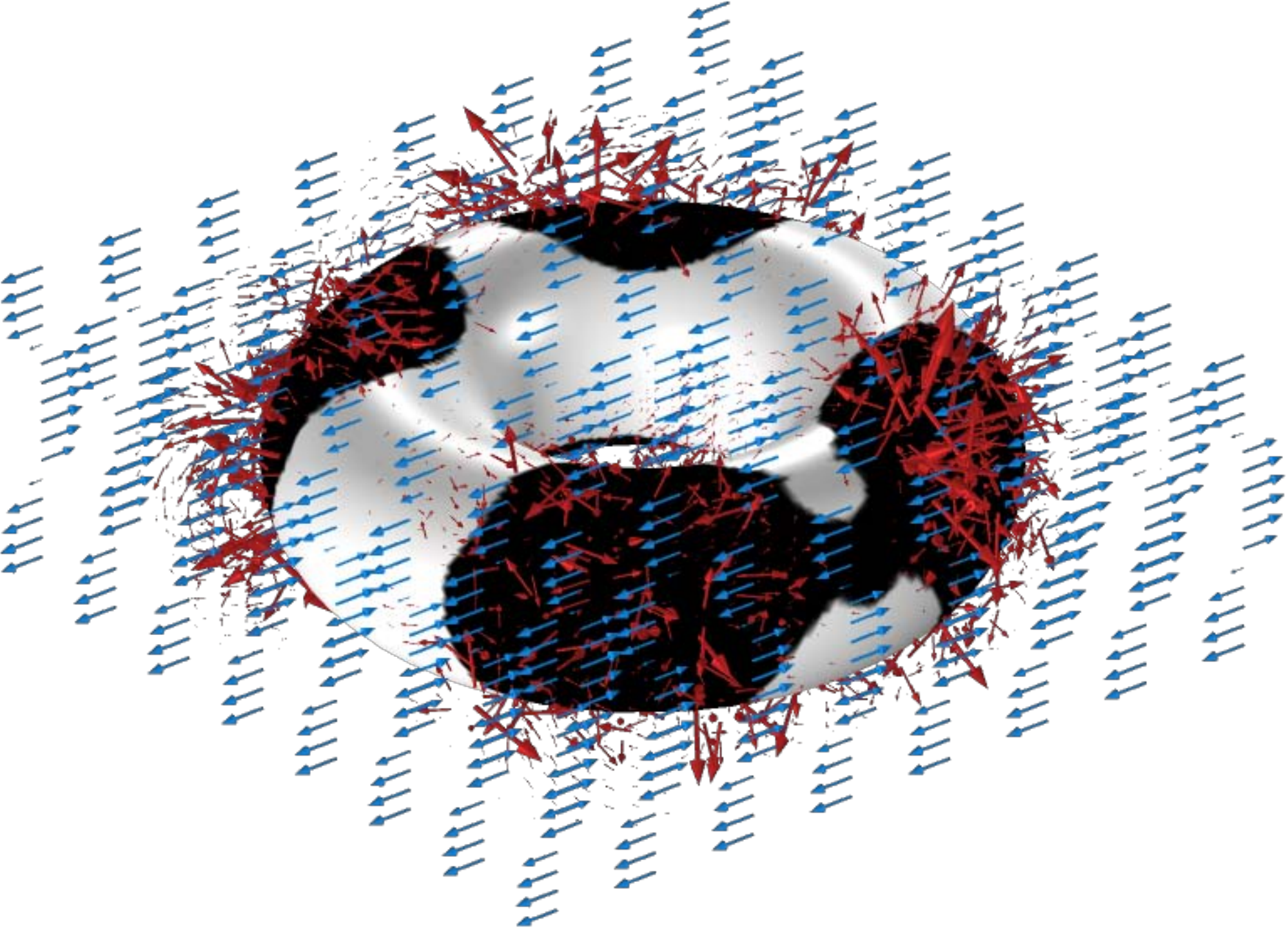}}\hspace{1ex}
  \subfigure[$y$-polarization]
  {\includegraphics[width=0.32\columnwidth]{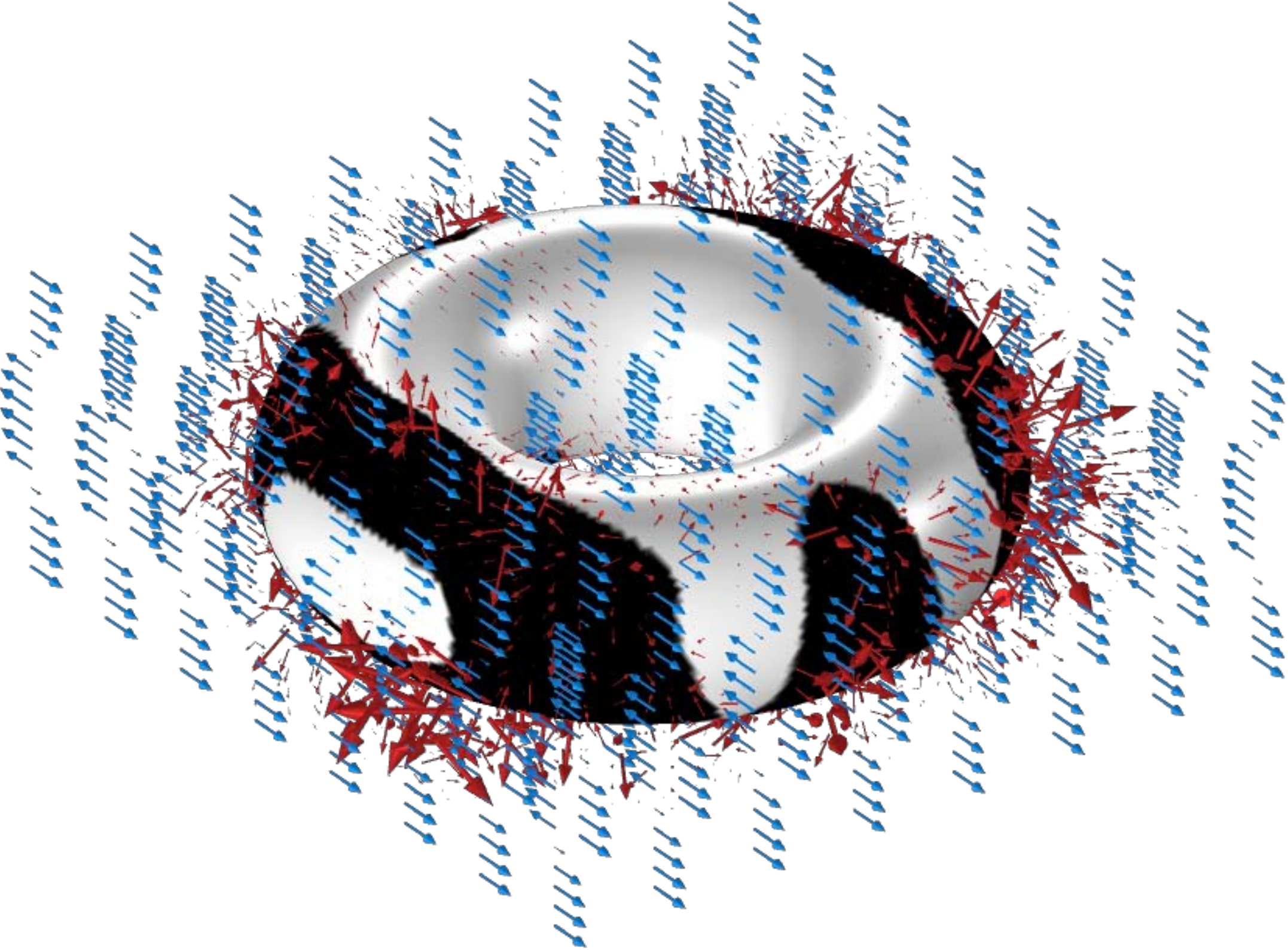}}\hspace{1ex}
  \subfigure[Circular polarization]
  {\includegraphics[width=0.32\columnwidth]{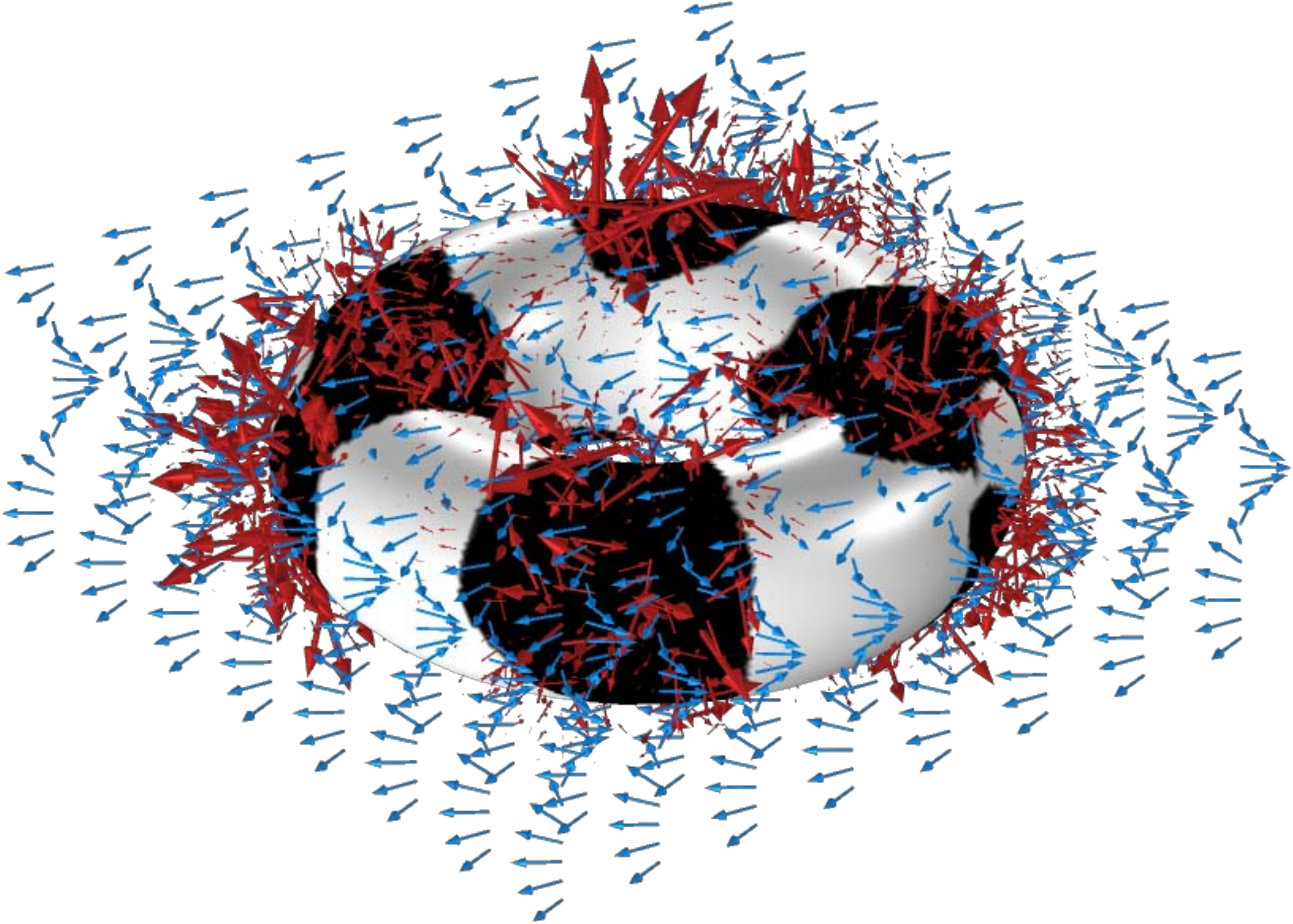}}
  \caption{Distributions of the vectors of the scattering field (red arrows) and the incident field (blue arrows) correspond to the three cases in Figure \ref{fig:TorusEM}, respectively.}\label{fig:ScatteringFieldVectorsTorus}
\end{figure}

\begin{figure}[!htbp]
  \centering
  \subfigure[$x$-polarization]
  {\includegraphics[width=0.32\columnwidth]{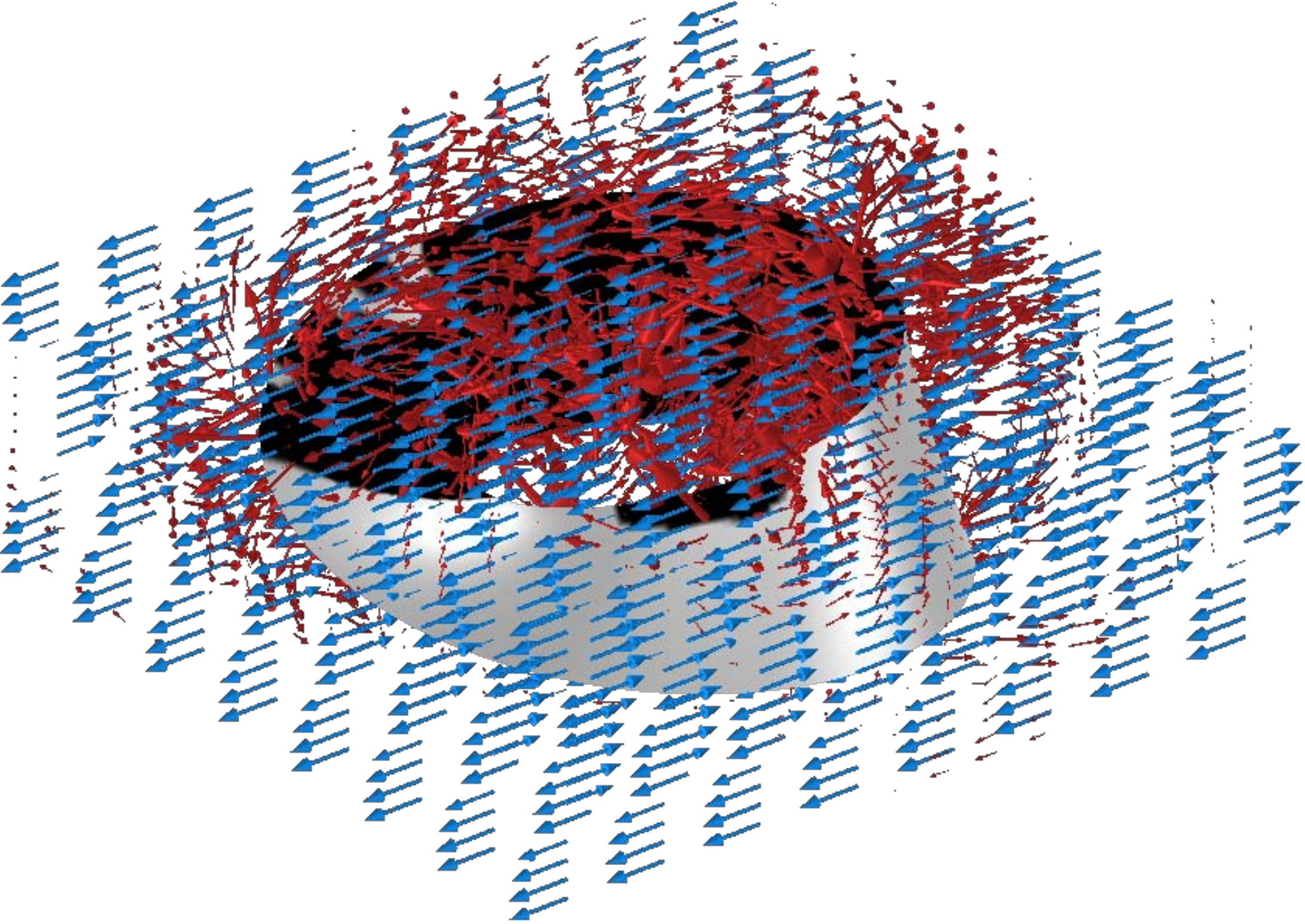}}\hspace{1ex}
  \subfigure[$y$-polarization]
  {\includegraphics[width=0.32\columnwidth]{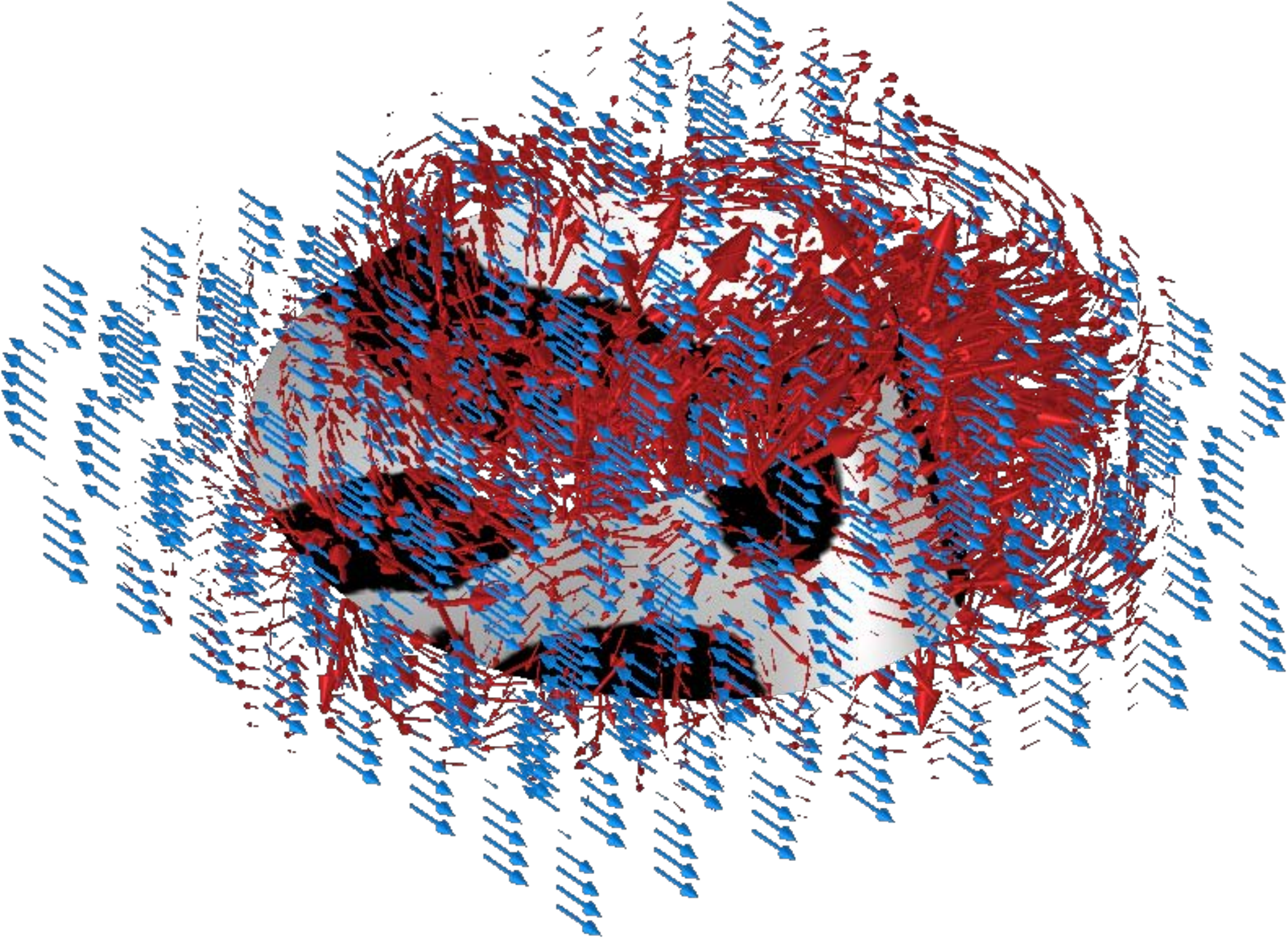}}\hspace{1ex}
  \subfigure[Circular polarization]
  {\includegraphics[width=0.32\columnwidth]{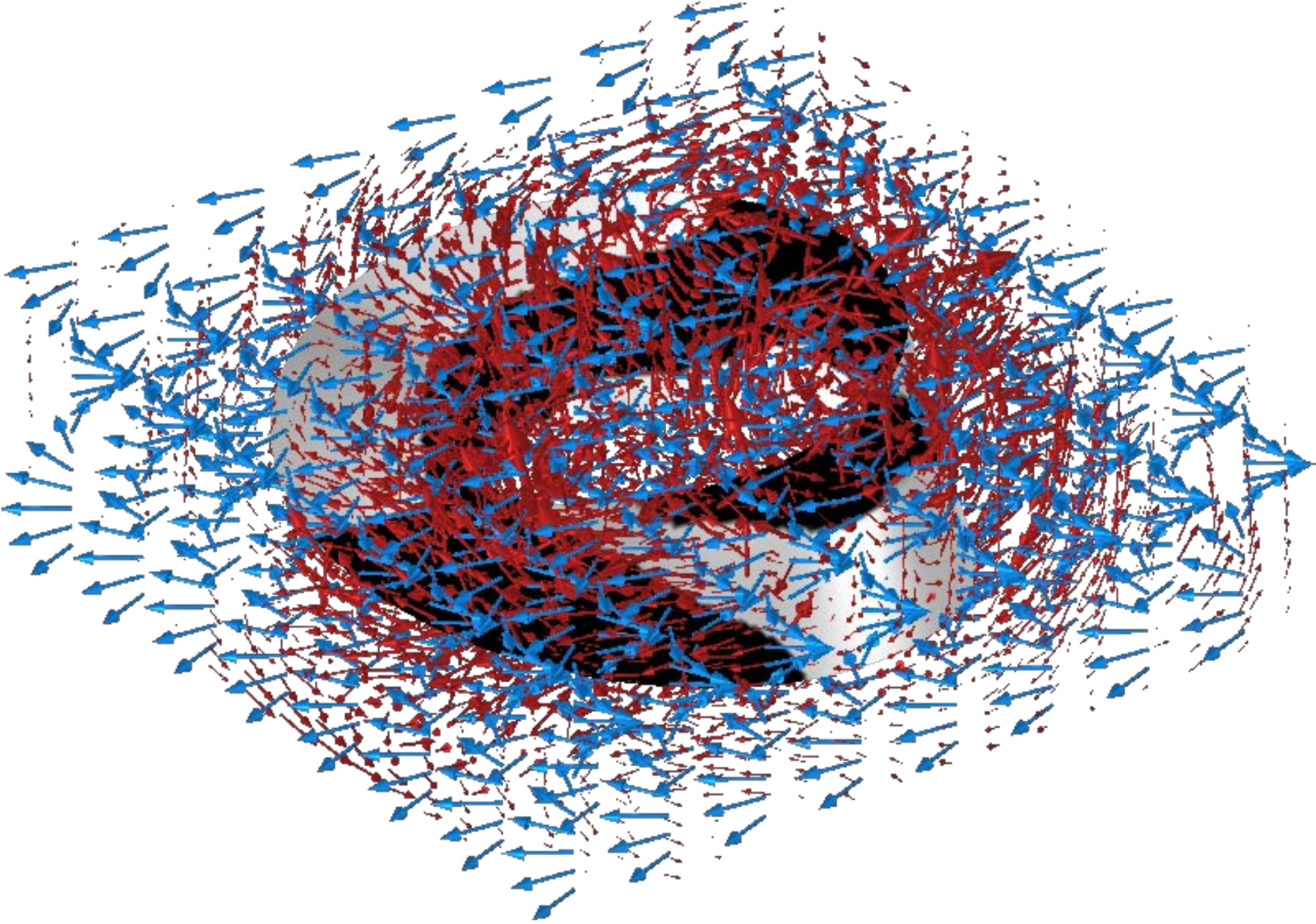}}
  \caption{Distributions of the vectors of the scattering field (red arrows) and the incident field (blue arrows) correspond to the three cases in Figure \ref{fig:MoebiusEM}, respectively.}\label{fig:ScatteringFieldVectorsMoebius}
\end{figure}

The performance of the derived patterns of the perfect conductor on maximizing the energy of the scattering fields can be confirmed from the vector distributions of the scattering field compared with those of the incident field (Figure \ref{fig:ScatteringFieldVectorsTorus} and \ref{fig:ScatteringFieldVectorsMoebius}). The comparison of the vector distributions shows that the perfect layers can enhance the energy of the scattering field by effectively reflecting and disturbing the incident waves.

To check the optimality, the optimized patterns of the perfect conductor in Figure \ref{fig:TorusEM} and \ref{fig:MoebiusEM} are used to scatter the incident waves with different polarizations and the energy of the scattering fields is cross-compared, where the other parameters are kept without change. From the cross comparison of the values in every row of the sub-tables in Table \ref{tab:PerfectConductorLayerPatternOptimalityConfirmation}, the scattering performance of the derived patterns of the perfect conductor can be confirmed.

\begin{table}[!htbp]
\centering
\subtable[Torus]{
\begin{tabular}{c|ccc}
  \toprule
        & Figure \ref{fig:TorusEM}(a) & Figure \ref{fig:TorusEM}(b) & Figure \ref{fig:TorusEM}(c) \\
  \midrule
  $x$-polarization & $\mathbf{107.8}$ & $81.7$ & $97.9$ \\
  \midrule
  $y$-polarization & $58.2$ & $\mathbf{110.1}$ & $68.5$ \\
  \midrule
  Circular polarization & $54.3$ & $102.1$ & $\mathbf{125.5}$ \\
  \bottomrule
\end{tabular}}\hspace{2ex}
\subtable[M\"{o}bius strip]{
\begin{tabular}{c|ccc}
  \toprule
        & Figure \ref{fig:MoebiusEM}(a) & Figure \ref{fig:MoebiusEM}(b) & Figure \ref{fig:MoebiusEM}(c) \\
  \midrule
  $x$-polarization & $\mathbf{22.8}$ & $12.9$ & $10.1$ \\
  \midrule
  $y$-polarization & $18.3$ & $\mathbf{19.9}$ & $8.4$ \\
  \midrule
  Circular polarization & $16.0$ & $15.8$ & $\mathbf{16.7}$ \\
  \bottomrule
\end{tabular}}
\caption{Energy of the scattering field corresponds to the patterns of the perfect conductor optimized on the torus and the M\"{o}bius strip for the incident waves with different polarizations. The optimized entries are noted in bold.}\label{tab:PerfectConductorLayerPatternOptimalityConfirmation}
\end{table}

\section{Conclusions}\label{sec:Conclusions}

This paper discussed the topology optimization implemented on 2-manifolds. When a physical field is defined on a 2-manifold, the topology optimization is implemented by interpolating a material parameter in the PDE used to describe this physical field; this case has been demonstrated by the topology optimization of the micro-textures for the wetting behaviors in the Cassie-Baxter mode. When the physical field is defined on a 3D domain and its boundary conditions are defined on a 2-manifold, the material density is used to formulate a mixed boundary condition of the physical field and implement the penalization between two different types of boundary conditions; this case has been demonstrated by the topology optimization of the patterns of the heat sinks for heat transfer and the patterns of the perfect conductor for electromagnetics. Typical 2-manifolds, e.g., sphere, torus, M\"{o}bius strip and Klein bottle, have been included in the numerical examples.
Based on the homeomorphic property of 2-manifolds, it can be concluded that this topology optimization can be implemented on any compact 2-manifolds.



\section*{Acknowledgements}\label{Acknowledgements}


Y.\ Deng acknowledges the support from a Humboldt Research Fellowship for Experienced Researchers (Humboldt-ID: 1197305), the National Natural Science Foundation of China (No. 51875545), the Youth Innovation Promotion Association of the Chinese Academy of Sciences (No. 2018253), and the Open Fund of SKLAO; Z. \ Liu acknowledges the support from the National Natural Science Foundation of China (No. 51675506); J.G.\ Korvink acknowledges support from an EU2020 FET grant (TiSuMR, 737043), the DFG under grant KO 1883/20-1 Metacoils, in the framework of the German Excellence Intitiative under grant EXC 2082 "3D Matter Made to Order", and by the VirtMat initiative "Virtual Materials Design". The authors are grateful to Prof.\ K.\ Svanberg for supplying the MMA codes.


\begin{thebibliography}{}

\bibitem{Bendsoe2003}
M. P. Bends{\o}e, O. Sigmund, Topology optimization---theory methods and applications, Springer, Berlin, 2003.

\bibitem{Michell1904}
A. G. M. Michell, The limit of economy of material in frame-structures, \textit{Phil. Mag.} 1904, \textbf{8}, 589-597.

\bibitem{Bendsoe1988}
M. Bends{\o}e, N. Kikuchi, Generating optimal topologies in optimal design using a homogenization method, \textit{Comput. Methods Appl. Mech. Eng.} 1988, \textbf{71}, 197-224.

\bibitem{Sigmund2001}
O. Sigmund, A 99-line topology optimization code written in Matlab, Struct. Multidisc. Optim. 21 (2001) 120-127.

\bibitem{Sigmund1997}
O. Sigmund, On the design of compliant mechanisms using topology optimization, Mech. Struct. Mach. 25 (1997) 495-526.

\bibitem{Rozvany2001}
G. I. N. Rozvany, Aims scope methods history and unified terminology of
computer-aided optimization in structural mechanics, \textit{Struct. Multidisc. Optim.}, 2001, \textbf{21}, 90-108.

\bibitem{Bendsoe1999}
M. P. Bends{\o}e, O. Sigmund, Material interpolations in topology optimization, Arch. Appl. Mech. 69 (1999) 635-654.

\bibitem{Gersborg-Hansen2006}
A. Gersborg-Hansen, M.P. Bends{\o}e, O. Sigmund, Topology optimization of heat conduction problems using the finite volume method, \textit{Struct. Multidisc.
Optim.} 2006, \textbf{31}, 251-259.

\bibitem{Akl2008}
W. Akl, A. El-Sabbagh, K. Al-Mitani, A. Baz, Topology optimization of a plate coupled with acoustic cavity, \textit{Int. J. Solids Struct.} 2008, \textbf{46}, 2060-2074.

\bibitem{Borrvall2003}
T. Borrvall, J. Petersson, Topology optimization of fluid in Stokes flow, \textit{Int. J. Numer. Methods Fluids} 2003, \textbf{41}, 77-107.

\bibitem{Sigmund2008}
O. Sigmund, K. G. Hougaard, Geometric properties of optimal photonic crystals, \textit{Phys. Rev. Lett.} 2008, \textbf{100}, 153904.

\bibitem{Nomura2007}
T. Nomura, K. Sato, K. Taguchi, T. Kashiwa, S. Nishiwaki, Structural topology optimization for the design of broadband dielectric resonator antennas using the finite difference time domain technique, \textit{Int. J. Numer. Methods Eng.} 2007, \textbf{71}, 1261-1296.

\bibitem{Duhring2008}
M. B. Duhring, J. S. Jensen, O. Sigmund, Acoustic design by topology optimization, \textit{J. Sound Vibr.} 2008, \textbf{317}, 557-575.

\bibitem{Kreissl2011}
S. Kreissl, G. Pingen, K. Maute, An explicit level-set approach for generalized shape optimization of fluids with the lattice Boltzmann  method, \textit{Int. J. Numer. Meth. Fluids}, 2011,
\textbf{65},496-519.

\bibitem{Zhou2008}
S. Zhou, Q. Li, A variational level set method for the topology optimization of steady-state Navier-Stokes flow, \textit{J. Comput. Phys.}, 2008, \textbf{227}, 10178-10195.

\bibitem{Guest2006}
J. K. Guest, J. H. Prevost, Topology optimization of creeping fluid flows using a Darcy-Stokes finite element, \textit{Int. J. Numer. Methods Eng.}, 2006, \textbf{66}, 461-484.


\bibitem{Takezawa2014}
A. Takezawa, M. Haraguchi. T. Okamoto, M. Kitamura, Cross-sectional optimization of whispering-gallery mode sensor with high electric field intensity in the detection domain, \textit{IEEE. J. Sel. Top. Quant. Electron.} 2014, \textbf{20}(6), 1-10.



\bibitem{Deng2011}
Y. Deng, Z. Liu, P. Zhang, Y. Liu, Y. Wu, Topology optimization of unsteady incompressible Navier-Stokes flows, J. Comput. Phys., 230 (2011), 6688-6708.

\bibitem{Andkjaer2011}
J. Andkj{\ae}r, O. Sigmund, Topology optimized low-contrast all-dielectric optical cloak, \textit{Appl. Phys. Lett.} 2011, \textbf{98}, 021112.

\bibitem{Andkjaer2012}
J. Andkj{\ae}r, N. A. Mortensen, O. Sigmund, Towards all-dielectric, polarization-independent optical cloaks, \textit{Appl. Phys. Lett.} 2012, \textbf{100}, 101106.

\bibitem{Fujii2013}
G. Fujii, H. Watanabe, T. Yamada, T. Ueta, M. Mizuno, Level set based topology optimization for optical cloaks, \textit{Appl. Phys. Lett.} 2013, \textbf{102}, 251106.

\bibitem{Diaz2010}
A. R. Diaz, O. Sigmund, A topology optimization method for design of negative
permeability metamaterials, \textit{Struct Multidisc Optim} 2010, \textbf{41}, 163-177.

\bibitem{Zhou2010Meta}
S. Zhou, W. Li, G. Sun, Q. Li, A level-set procedure for the design of electromagnetic metamaterials, \textit{Optics Express} 2010, \textbf{18}, 6693-6702.

\bibitem{Zhou2011}
S. Zhou, W. Li, Y. Chen, G. Sun, Q. Li, Topology optimization for negative permeability metamaterials using level-set algorithm, \textit{Acta Materialia} 2011, \textbf{59}, 2624-2636.

\bibitem{Andkjaer20101}
J. Andkj{\ae}r, S. Nishiwaki, T. Nomura, O. Sigmund, Topology optimization of grating couplers for the efficient excitation of surface plasmons, \textit{JOSA B} 2010, \textbf{27}, 1828-1832.

\bibitem{Zhou20102}
S. Zhou, W. Li, Q. Li, Level-set based topology optimization for electromagnetic dipole antenna design, \textit{Journal of Computational Physics} 2010, \textbf{229}, 6915-6930.

\bibitem{Hassan2014}
E. Hassan, E. Wadbro, M. Berggren, Topology optimization of metallic antennas, \textit{IEEE Transactions on Antennas And Propagations} 2014, \textbf{62}, 2488-2500.

%

\bibitem{Shim2009}
H. Shim, V. T. T. Ho, S. Wang, D. A. Tortorelli, Level set-based topology optimization for electromagnetic systems, \textit{IEEE Transactions on Magnetics} 2009, \textbf{45}, 1582-1585.

\bibitem{Feichtner2012}
T. Feichtner, O. Selig, M. Kiunke, B. Hecht, Evolutionary optimization of optical antennas, \textit{Phys. Rev. Lett.} 2012, \textbf{109}, 127701.



\bibitem{Wang2003}
M. Y. Wang, X. Wang, D. Guo, A level set method for structural optimization, Comput. Methods Appl. Mech. Eng. 192 (2003) 227-246.

\bibitem{Allaire2004}
G. Allaire, F. Jouve, A. Toader, Structural optimization using sensitivity analysis and a level-set method, J. Comput. Phys. 194 (2004) 363-393.

\bibitem{Liu2008}
Z. Liu, J.G. Korvink, Adaptive moving mesh level set method for structure optimization, Eng. Optim. 40 (2008) 529-558.

\bibitem{Xing2010}
X. Xing, P. Wei, M.Y. Wang, A finite element-based level set method for structural optimization, Int. J. Numer. Methods Eng. 82 (2010) 805-842.

\bibitem{XieStevenSpringer1997}
Y. M. Xie, G. P. Steven, Evolutionary structural optimization, Springer, 1997.

\bibitem{StevenComputMech2000}
G. P. Steven, Q. Li, Y. M. Xie, Evolutionary topology and shape design for physical field problems, Comput. Mech., 26 (2000), 129-139.

\bibitem{TanskanenCMAME2002}
P. Tanskanen, The evolutionary structural optimization method: theoretical aspects, Comput.
Methods Appl. Mech. Engrg., 191 (2002), 47-48.

\bibitem{GuoJAM2014}
Guo X, Zhang W, Zhong W (2014) Doing topology optimization explicitly and geometrically --- a new moving morphable components based framework. J Appl Mech 81:081009.

\bibitem{GuoCMAME2016}
Guo X, Zhang W, Zhang J, Yuan J (2016) Explicit structural topology optimizationbased onmoving morphable components (MMC) with curved skeletons. Comput Methods Appl Mech Eng, 310:711-748.

\bibitem{TakezawaJCP2010}
Akihiro Takezawa, ShinjiNishiwaki, MitsuruKitamura (2010) Shape and topology optimization based on the phase field method and sensitivity analysis. Journal of Computational Physics, 229:2697-2718.

\bibitem{QianYe2019}
Qian Ye, Yang Guo, Shikui Chen, Na Lei, Xianfeng Gu, Topology optimization of conformal structures on manifolds using extended level set methods (X-LSM) and conformal geometry theory, Computer Methods in Applied Mechanics and Engineering, 344 (2019): 164-185.

\bibitem{PanagiotisVogiatzis2018}
Panagiotis Vogiatzis, Ming Ma, Shikui Chen and Xianfeng Gu, Computational design and additive manufacturing of periodic conformal metasurfaces by synthesizing topology optimization with conformal mapping, Computer Methods in Applied Mechanics and Engineering, 328 (2018): 477-497.

\bibitem{AllaireESAIM2014}
G. Allaire, C. Dapogny, G. Delgado, G. Michailidis. Multi-phase structural optimization via a level set method, \textit{ESAIM: Control, Optimisation and Calculus of Variations}, 2014, 20, 576-611.

\bibitem{VermaakSMO2014}
N. Vermaak, G. Michailidis, G. Parry, R. Estevez, G. Allaire, Y. Br\'{e}chet. Material interface effects on the topology optimization of multi-phase structures using a level set method, \textit{Struct. Multidisc. Optim.} 2014, 50, 623-644.

\bibitem{SigmundJMPS1997}
O. Sigmund, S. Torquato, Design of materials with extreme thermal expansion using a three-phase
topology optimization method, \textit{J. Mech. Phys. Solids} 1997, 45, 1037-1067.

\bibitem{GibianskyJMPS2000}
L.V. Gibiansky, O. Sigmund, Multiphase composites with extremal bulk modulus, \textit{J. Mech. Phys. Solids} 2000, 48, 461-498.

\bibitem{GaoZhangIJNME2011}
T. Gao, W. Zhang, A mass constraint formulation for structural topology optimization with multiphase materials, \textit{Int. J. Numer. Meth. Engng.} 2011, 88, 774-796.

\bibitem{LuoAiaaJ2012}
Y.J. Luo, Z. Kang, Z.F. Yue, Maximal stiffness design of two-material structures by topology
optimization with nonprobabilistic reliability, \textit{AIAA J.} 2012, 50, 1993-2003.

\bibitem{YinSMO2001}
L. Yin, G.K. Ananthasuresh, Topology optimization of compliant mechanisms with multiple
materials using a peak function material interpolation scheme, \textit{Struct. Multidisc. Optim.} 2011, 23, 49-62.

\bibitem{WangCMAME2004}
M.Y. Wang, X.M. Wang, "Color" level sets: a multi-phase method for structural topology
optimization with multiple materials, \textit{Comput. Methods Appl. Mech. Eng.} 2004, 193, 469-496.

\bibitem{ZhouWangSMO2007}
S. W. Zhou, M.Y. Wang, Multimaterial structural topology optimization with a generalized Cahn-Hilliard model of multiphase transition, \textit{Struct. Multidisc. Optim.} 2007, 33, 89-111.

\bibitem{KrogOlhoffComputersStructures1996}
L. Krog, N. Olhoff, Optimum topology and reinforcement design of disk and plate structures with multiple stiffness
and eigenfrequency objectives, \textit{Computers and Structures} 1996, 72, 535-63.

\bibitem{AnsolaComputersStructures2002}
R. Ansola, J. Canales, J. A. T\'{a}rrago, J. Rasmussen, An integrated approach for shape and topology optimization of shell structures, \textit{Computers and Structures} 2002, 80, 449-458.

\bibitem{HassaniSMO2013}
B. Hassani, S. M. Tavakkoli, H. Ghasemnejad, Simultaneous shape and topology optimization of shell structures, \textit{Struct. Multidisc. Optim.} 2013, 48, 221-233.

\bibitem{LochnerAldingerSchumacher2014}
I. Lochner-Aldinger, A. Schumacher, Homogenization method. In: S. Adriaenssens, P. Block, D. Veenendaal, C. Williams (eds) \textit{Shell structures for architecture-form finding and optimization}, Routledge, New York, 2014.

\bibitem{ClausenActaMechanicaSinica2017}
A. Clausen, E. Andreassen, O. Sigmund, Topology optimization of 3D shell structures with porous infill, \textit{Acta Mechanica Sinica} 2017, 33, 778-791.

\bibitem{YoonFSIIJNME2010}
G. H. Yoon, Topology optimization for stationary fluid每structure interaction problems using a new monolithic formulation, \textit{Int. J. Numer. Meth. Engng} 2010, 82, 591-616.

\bibitem{LundgaardFSISMO2018}
C. Lundgaard, J. Alexandersen, M. Zhou, C. S. Andreasen, O. Sigmund, Revisiting density-based topology optimization for fluid-structure-interaction problems, \textit{Struct. Multidisc. Optim.} 2018, 58, 969-995.

\bibitem{AndreasenFSISMO2019}
C. S. Andreasen, A framework for topology optimization of inertial microfluidic particle manipulators, \textit{Struct. Multidisc. Optim.} 2019, under review.

\bibitem{AuligUlm2012}
N. Aulig, I. Lepenies, \textit{A topology optimization interface for LS-DYNA}. In: 11. LS-DYNA Forum, Ulm, 2012.

\bibitem{MauteIJNME2017}
R. Behrou, M. Lawry, K. Maute, Level set topology optimization of structural problems with interface cohesion, \textit{Int. J. Numer. Meth. Engng.} 2017, 112, 990-1016.

\bibitem{MauteSMO2005}
M. Raulli, K. Maute, Topology optimization of electrostatically actuated microsystems, \textit{Struct. Multidisc. Optim.}  2005, 30: 342-359.

\bibitem{Reshetnyak2Manifolds1993}
Y. G. Reshetnyak (1993) Two-Dimensional Manifolds of Bounded Curvature. In: Reshetnyak Y.G. (eds) Geometry IV. Encyclopaedia of Mathematical Sciences, vol 70. Springer, Berlin, Heidelberg.

\bibitem{LazarovIntJNumerMethodsEng2011}
B. Lazarov, O. Sigmund, Filters in topology optimization based on Helmholtz type differential equations, \textit{Int. J. Numer. Methods Eng.} 2011, 86, 765-781.

\bibitem{WangStructMultidiscipOptim2011}
F. Wang, B. S. Lazarov, O. Sigmund, On projection methods, convergence and robust formulations in topology optimization, \textit{Struct. Multidiscip. Optim.} 2011, 43, 767-784.

\bibitem{GuestIntJNumerMethodsEng2004}
J. Guest, J. Prevost, T. Belytschko, Achieving minimum length scale in topology optimization using nodal design variables and projection functions, \textit{Int. J. Numer. Methods Eng.} 2004, 61, 238-254.


\bibitem{GilbargTrudingerSpringer1988}
D. Gilbarg, N.S. Trudinger, \textit{Elliptic partial differential equations of second order}, Springer,
1988.

\bibitem{ChernWorldScientificPublishing1999}
S. S. Chern, W. H. Chen, K. S. Lam, \textit{Lectures on differential geometry}, World Scientific Publishing, 1999.

\bibitem{ZeidlerSpringer1986}
E. Zeidler, Nonlinear Functional Analysis and Its Applications. I, Fixed-Point Theorems,
Springer, Berlin, 1986.

\bibitem{HinzeSpringer2009}
M. Hinze, R. Pinnau, M. Ulbrich, S. Ulbrich, Optimization with PDE Constraints, Springer,
Berlin, 2009.

\bibitem{DziukActaNumerica2013}
G. Dziuk, C. M. Elliott, Finite element methods for surface PDEs, Acta Numerica. (2013) 289-396.

\bibitem{SvanbergIJNME1987}
K. Svanberg, The method of moving asymptotes: a new method for structural optimization,
Int. J. Numer. Methods Eng. 24 (1987) 359-373.

\bibitem{FengAdvMater2006}
X. Feng, L. Jiang, Design and creation of superwetting/antiwetting surfaces. \textit{Adv. Mater.} 2006, 18, 3063-3078.


\bibitem{DengCMAME2018}
Y. Deng, D. Mager, Y. Bai, T. Zhou, Z. Liu, L. Wen, Y. Wu, J. G. Korvink, Inversely designed micro-textures for robust Cassie-Baxter mode of super-hydrophobicity. \textit{Comput. Methods Appl. Mech. Engrg.} 2018, 341, 113-132.

\bibitem{BicoEurophysLett1999}
J. Bico, C. Marzolin, D. Qu\'{e}r\'{e}. \textit{Europhys. Lett.} 1999, 47, 220.

\bibitem{LafumaNatMater2003}
A. Lafuma, D. Qu\'{e}r\'{e}. \textit{Nat. Mater.} 2003, 2, 457.

\bibitem{YoungPhilTrans1805}
T. Young. An essay on the cohesion of fluids. \textit{Phil. Trans.} 1805, 65.

\bibitem{LaplaceMechaniqueCeleste1806}
P. Laplace. Supplement to the tenth edition. M\'{e}chanique c\'{e}leste 10, 1806.

\bibitem{LiIJHMT2004}
Q. Li, G.P. Steven, Y.M. Xie, O.M. Querin, Evolutionary topology optimization for temperature reduction of heat conducting fields, Int. J. Heat Mass Transf. 47 (23) (2004) 5071-5083.

\bibitem{GersborgHansenSMO2006}
A. Gersborg-Hansen, M.P. Bendsoe, O. Sigmund, Topology optimization of heat conduction problems using the finite volume method, Struct. Multidiscip. Optim. 31 (2006) 251-259.

\bibitem{ZhuangCMAME2007}
C. Zhuang, Z. Xiong, H. Ding, A level set method for topology optimization of heat conduction problem under multiple load cases, Comput. Methods Appl. Mech. Eng. 196 (2007) 1074-1084.

\bibitem{DengJCP2014}
Y. Deng, Z. Liu, Y. Liu, Y. Wu, Combination of topology optimization and optimal control method, Journal of Computational Physics 257 (2014) 374-399.

\bibitem{KrausCarverElectromagnetics1973}
J. D. Kraus, K. R. Carver, Electromagnetics, 1973, McGraw-Hill edition, in English, 2nd ed.

\bibitem{JinWiley2002}
J. Jin, The Finite Element Method in Electromagnetics, 2nd edition, John Wiley \& Sons, New York, 2002.

\bibitem{BerengerJCP1994}
J. Berenger, A perfectly matched layer for the absorption of electromagnetic waves, J. Comput. Phys. 114 (1994) 185-200.


\bibitem{DengJCP2018}
Y. Deng, J. G. Korvink, Self-consistent adjoint analysis for topology optimization of electromagnetic waves, J. Comput. Phys. 361 (2018) 353-376.

%
%
%
%
%
%
%
%
%




%
%
%
%
%
%
%
%
%
%
%
%
%
%
%
%
%
%
%
%
%
%
%
%
%
%
%
%
%
%
%
%
%
%
%
%
%
%

\end{thebibliography}
\end{document}